\documentclass[10pt,a4paper]{article}

\usepackage[round]{natbib}
\usepackage{amsfonts}
\usepackage{dsfont}
\usepackage{physics}
\usepackage{xcolor}
\usepackage{algorithm}
\usepackage{indentfirst}
\usepackage{booktabs}
\usepackage{graphicx}
\usepackage{subcaption}
\newtheorem{theorem}{Theorem}
\usepackage{authblk}
\usepackage{parskip}
\usepackage{xurl}
\usepackage[bottom=2.5cm,top=2.5cm,left=2.5cm,right=2.5cm]{geometry}

\usepackage[utf8]{inputenc}  
\usepackage[T1]{fontenc}

\begin{document}

\providecommand{\keywords}[1]{\textbf{\textit{Keywords --}} #1}

\title{Neural Hawkes: Non-Parametric Estimation in High Dimension and Causality Analysis in Cryptocurrency Markets}
\author[1,2]{Timothée Fabre}
\author[1]{Ioane Muni Toke}
\affil[1]{Laboratoire MICS, CentraleSupélec, Université Paris-Saclay, France.}
\affil[2]{SUN ZU Lab, Paris, France.}
\date{\today}

\maketitle

\begin{abstract}
\noindent We propose a novel approach to marked Hawkes kernel inference which we name the moment-based neural Hawkes estimation method. Hawkes processes are fully characterized by their first and second order statistics through a Fredholm integral equation of the second kind. Using recent advances in solving partial differential equations with physics-informed neural networks, we provide a numerical procedure to solve this integral equation in high dimension. Together with an adapted training pipeline, we give a generic set of hyperparameters that produces robust results across a wide range of kernel shapes. We conduct an extensive numerical validation on simulated data. We finally propose two applications of the method to the analysis of the microstructure of cryptocurrency markets. In a first application we extract the influence of volume on the arrival rate of BTC-USD trades and in a second application we analyze the causality relationships and their directions amongst a universe of 15 cryptocurrency pairs in a centralized exchange.

\vspace{0.5cm}

\noindent \keywords{Hawkes process, Non-parametric estimation, Physics-informed neural network, Cryptocurrency}
\end{abstract}

\setcounter{tocdepth}{2}
\tableofcontents

\section{Introduction}

Hawkes processes are a class of point process that models a sequence of arrivals of events that can cluster over time. Their properties can display self-exciting or self-inhibiting behaviors in the unidimensional case, and mutual-exciting or mutual-inhibiting behaviors for multidimensional processes. In the recent years, their application has been explored in many areas such as earthquakes \citep{wang2012markov, kwon2023flexible}, neuroscience \citep{reynaud2013inference}, criminality \citep{olinde2020self} and in finance \citep{toke2012modelling, bacry2013modelling, bacry2015hawkes, rambaldi2017role, gao2018transform, jusselin2020no}. A $D$-variate linear Hawkes process $N=(N^1,\ldots,N^D)$ is a counting process with conditional intensity vector $\lambda_t=(\lambda^1_t,\ldots,\lambda^D_t)$ such that for any $i=1,\ldots, D$, 
\begin{equation}
    \lambda_t^i\,=\,\mu^i+\sum_{j=1}^D\int_{(-\infty,t)}\phi^{ij}(t-s)\,\mathrm{d}N_s^j,
\end{equation}
where $\mu=(\mu^1,\ldots,\mu^D)$ is the baseline intensity which corresponds to the exogenous component and $\Phi=(\phi^{ij})_{1\leq i,j\leq D}$ is the kernel matrix. For any $i=1,\ldots, D$, $j=1,\ldots, D$, the element $\phi^{ij}$ is a function that gives the impact each past event has over the instantaneous probability of observing a new event at time $t$. This interpretability of Hawkes processes justifies their widespread use especially in causality analysis \citep{xu2016learning, achab2017uncovering}. In the common framework where $\phi$ is positive we say that the processes $N^1,\ldots,N^D$ are self- and mutually exciting. The use of Hawkes processes that exhibit kernels with negative parts provides more flexibility and better fits to inhibitory behaviors that are often observed in the data, at the price of a lesser mathematical tractability \citep{lu2018high, duval2022interacting}.

In general the estimation of such processes is carried on by specifying a parametric form of the influence kernel and then calibrating the parameters via likelihood maximization \citep[][among recent contributions]{lee2017marked, lu2018high, bonnet2023inference}. Specifying an exponentially decreasing kernel preserves the Markovian property and is therefore a common choice. But even the estimation of exponential kernels can sometimes be tricky --- especially in large dimension --- since the likelihood is strongly non-convex. Moreover, much more specific forms of Hawkes kernels are required to fit the data in practice. For example, latency effects may cause the kernel to be non-monotonic when dealing with high-frequency financial data. Furthermore, the dynamic of the point process may depend on an additional component such as the past realizations of marks which makes the kernel even more complex to specify. The application of a non-parametric method is thus essential in these cases.

\subsection{Literature review of non-parametric estimation methods}

Non-parametric inference has been explored through different methodologies. An expectation-maximization (EM) algorithm was designed using the cluster representation of \cite{hawkes1974cluster} for univariate kernel estimation \citep{lewis2011nonparametric} and was later extended to the multivariate case \citep{zhou2013learning}. Nevertheless this method has three drawbacks which are particularly exacerbated in the context of high-frequency financial data, as outlined in \citet{bacry2016first}:
\begin{itemize}
    \item high computational cost: the evaluation of the log-likelihood has a quadratic time complexity in the number of observations ;
    \item tail-modulated speed of convergence: the EM algorithm slowly converges when the kernel is power-law decaying, and such kernels are often observed in financial data ;
    \item removal of inhibition effects: as the kernel is constrained to take positive values, inhibitory effects cannot be quantified.
\end{itemize}

The exploitation of an infinite autoregressive representation has been proposed as well in \citet{eichler2017graphical}. Similarly, the INAR representation of Hawkes processes is used in \citet{kirchner2017estimation}. Note that other kernel estimation methods have been proposed in the literature and either explore least squares estimation \citep{bacry2020sparse} or Bayesian inference perspectives \citep{zhang2018efficient, donnet2020nonparametric, sulem2024bayesian}. However, these methods are unsuitable to general multivariate marked kernels. Other methods focus on the estimation of the norm of the kernels rather than their shape as in \citet{achab2017uncovering}.

The use of neural networks has been explored in \citet{joseph2020shallow, joseph2023neural}. The authors train a neural network model that outputs the kernel matrix and aims at maximizing the associated log-likelihood over event time observations. This procedure solves the problem of high dimension but shares the drawbacks that are directly linked to a Hawkes likelihood optimization, \textit{i.e.} the potentially large number of local minima and a high computational cost.

To overcome these disadvantages, another class of non-parametric methods encompassing moment-based procedures has been made popular in the literature. Based on the Laplace transform of the autocovariance function, the inference of symmetric kernels is developed in \citet{bacry2012non}. This methodology is extended in \citet{bacry2016first} by relaxing the symmetry assumption. The authors demonstrate a characterization equation that links the kernel matrix of any stationary linear Hawkes process to its second order statistics. The resulting system of equations involves Fredholm integral equations of the second kind that are numerically solved as a Wiener-Hopf system by discretizing the time domain with quadratures.

Although the Wiener-Hopf (WH) method of \citet{bacry2016first} seems to provide a convenient framework as both power-laws and inhibitory effects, as well as marked kernels are inferred, its practical application can be laborious as it is prone to the curse of dimensionality when $D$ increases. Solving such a system involves the inversion of a matrix that can be arbitrarily large depending on the dimension of the Hawkes process, on the size of the support of the mark process and on the number of quadrature points. Moreover estimation issues involving oscillatory behaviours have been reported --- see for example \citet{cartea2021gradient} --- and the optimal number of quadratures might be highly problem-dependent. 

The problem of the Wiener-Hopf approach does not lie in the methodology itself since solving the characterization equation should lead to the desired solution, but rather in the numerical solving procedure. Supported by recent advances in the literature about physics-informed neural networks (PINNs), we propose in this work to infer the solution by training a neural network to solve the characterization integral equation. Originally developed in \citet{psichogios1992hybrid, dissanayake1994neural, lagaris1998artificial} for differential equations and initial/boundary value problems that arise in physics, this idea has gained much more attention in the past years due to the rising availability of computational resources.

\subsection{Contribution and organization of the paper}

Our main contributions are the following.
\begin{itemize}
    \item We introduce a novel non-parametric estimation method for the kernel of marked Hawkes processes in high dimension. Our method leverages some of the latest advances in PINNs training to train a robust kernel learner that solves the characterization equation of \citet{bacry2016first}.
    \item We provide a general configuration of hyperparameters and network architecture that approximates the solution with excellent accuracy with respect to the characterization equation residual.
    \item Using our estimation method on high-frequency cryptocurrency data, we extract the impact of volume over the arrival rate of BTC-USD trades and we analyze spillover effects within a high-dimensional universe of cryptocurrency pairs. We define two ratios that quantify the direction of causal relationships.
\end{itemize}

The structure of this paper is as follows. Section \ref{methodology_section} first defines our framework and notation, then introduces the characterization equation of \citet{bacry2016first} and finally presents the neural Hawkes non-parametric estimation method. A fully detailed training procedure is provided. Section \ref{validation_section} validates the estimation method on simulated data and provides insights about hyperparameter tuning and robustness. Section \ref{application_section} finally applies the estimation method to high-frequency cryptocurrency data.

\section{Moment-based non-parametric neural Hawkes estimation}
\label{methodology_section}

\subsection{Framework and notation}

Let $N$ be a $D$-dimensional linear marked Hawkes process such that for $i\in\{1,\dots,D\}$, $(N_t^i)_{t\geq0}$ has the intensity process $(\lambda_t^i)_{t\geq0}$ and mark process $(\xi_t^i)_{t\geq0}$. We assume the non-zero marks are i.i.d. and $\{t,\xi_t^i\neq0\}=\{t,N_t^i-N_{t^-}^i=1\}$. We denote by $p^i$ the probability density function of the random variable $\xi_t^i$ conditionally on $\{N_t^i-N_{t^-}^i=1\}$. We furthermore denote by $\mathcal{X}_i\subset(0,+\infty)$ the domain of $p^i$.

We denote by $\Phi:=(\phi^{ij})_{(i,j)\in\{1,\dots,D\}}$ the kernel matrix of $N$ and by $\mu^i$ its baseline intensity and we assume that for any $i=1,\ldots,D$, $j=1,\ldots,D$, $\phi^{ij}$ takes values in $\mathbb{R}_+$. For any $i\in\{1,\dots,D\}$, the arrival intensity of events of type $i$ at time $t\geq0$ is
\begin{equation}\label{eq:intensity_hawkes}
    \lambda_t^i\,=\,\mu^i+\sum_{j=1}^D\int_{(-\infty,t)}\phi^{ij}(t-s,\,\xi_s^j)\,\mathrm{d}N_s^j.
\end{equation}

Let $\| \Phi \rVert:=\left(\| \phi^{ij} \rVert\right)_{(i,j)\in\{1,\dots,D\}^2}$ be the matrix of $L^1$-norms of the kernel components, i.e. for all $(i,j)\in\{1,\dots,D\}^2$,
\begin{equation}
    \| \phi^{ij} \rVert:=\iint_{(0,+\infty)\times\mathcal{X}_j} \phi^{ij}(s,z) \,p^j(z)\mathrm{d}s\mathrm{d}z.
\end{equation}

The branching ratio of the process $N$ is
\begin{equation}
    \mathcal{R}:=\max_{\lambda\in\mathcal{S}(\Phi)}\abs{\lambda},
\end{equation}
where $\mathcal{S}(\Phi)$ is the spectrum of the matrix $\| \Phi \rVert$. The condition $\mathcal{R}<1$ ensures the stationarity of $N$ and consequently, the vector of first order statistics $\Lambda:=\mathbb{E}(\lambda_t)=(\Lambda^i)_{i\in\{1,\dots,D\}}$ is
\begin{equation}\label{eq:order_1_characterization}
    \Lambda=\left(\mathbb{I}-\| \Phi \rVert\right)^{-1}\mu,
\end{equation}
where $\mathbb{I}$ is the identity matrix of size $D$.

Let us now define two quantities that will be of interest in our paper.  For all $(i,j)\in\{1,\dots,D\}^2$, we call \emph{aggregated time kernel}, denoted $\varphi^{ij}$, the expectation of the kernel component $(i,j)$ with respect to the mark distribution :
\begin{equation}\label{eq:aggregated_time_kernel}
    \varphi^{ij}(t):=\mathbb{E}\left(\phi^{ij}\left(t,\xi^j\right)\right)=\int_{\mathcal{X}_j}\phi^{ij}(t,z)\,p^j(z)\mathrm{d}z.
\end{equation}
Similarly, we call \emph{aggregated mark kernel}, denoted $f^{ij}$, the norm of the kernel component $(i,j)$ along the time coordinate
\begin{equation}\label{eq:aggregated_mark_kernel}
    f^{ij}(x):=\frac{1}{\| \phi^{ij} \rVert}\int_{(0,+\infty)}\phi^{ij}(s,x)\mathrm{d}s.
\end{equation}
Note the normalization such that $\mathbb{E}\left(f^{ij}(\xi^j)\right)=1$. The aggregated kernels are such that if the kernel is multiplicative in the sense that $\phi^{ij}(t,x)=\psi(t)h(x)$ for some functions $\psi$ and $h$ with $\mathbb{E}(h(\xi))=1$, then $\psi\equiv \varphi^{ij}$ and $h\equiv f^{ij}$.

The second order statistics characterization equation of a point process $N$ with intensity defined in Equation \eqref{eq:intensity_hawkes} is given by the following result.

\begin{theorem}[\cite{bacry2016first}]\label{th:characterization_theorem}
If $G$ is the second order statistics matrix defined by
\begin{equation}
    G^{ij}(t,x)\mathrm{d}t\mathrm{d}x:=\mathbb{E}\left[\mathrm{d}N_t^i\,\Big|\,\mathrm{d}N_0^j=1,\,\xi_0^j\in[x,x+\mathrm{d}x]\right]-\mathds{1}_{\{i=j,\,t=0\}}\mathrm{d}x-\Lambda^i\mathrm{d}t\mathrm{d}x,
    \label{eq:def_G_statistics}
\end{equation}
then for all $i\in\{1,\dots,D\}$, the cross-excitation kernels of the $i$-th event type $\phi^{i.}:=(\phi^{ij})_{j\in\{1,\dots,D\}}$ are the solutions of the following $D$-system of equations
\begin{equation}\label{eq:characterization_equation}
    G^{ij}(t,x)=\phi^{ij}(t,x)+\sum_{k=1}^D\iint_{(0,+\infty)\times\mathcal{X}_k}\phi^{ik}(s,z)H^{kj}(t-s,x,z)p^k(z)\,\mathrm{d}s\mathrm{d}z, \hspace{0.3cm} 1\leq j \leq D
\end{equation}
where
\begin{equation}
    H^{kj}(t,x,z):=G^{kj}(t,x)\mathds{1}_{\{t>0\}}+\frac{\Lambda^k}{\Lambda^j}G^{jk}(-t,z)\mathds{1}_{\{t<0\}}.
\end{equation}
\end{theorem}

The process is fully characterized by this result in the sense that once the kernel matrix is obtained from Equation \eqref{eq:characterization_equation}, the baseline intensity $\mu$ is deduced using Equation \eqref{eq:order_1_characterization}.

In practice the mark universe is discretized which turns Equation \eqref{eq:characterization_equation} into a Fredholm equation of the second kind. The Wiener-Hopf method consists in discretizing the time domain using quadratures and inverting the corresponding matrix to deduce the values taken by the solution at the quadrature points. Thus, the entire kernel matrix $\Phi$ is obtained by proceeding to the inversion of $D$ matrices of size $Q\times M \times D$, where $Q$ is the number of quadratures and $M$ the size of the discretized mark universe. For a large $Q$ --- 100 to 1000 points ---, a reasonably large $M$ --- 3 to 10 points --- and a large dimension --- 10 event types --- the inversion becomes both computationally costly and noisy with a potentially high approximation error.

Note that depending on the sign of the second order statistics, inhibitory effects can be observed, \textit{i.e.} a kernel that explores negative values. As discussed in \citet{bacry2016first, rambaldi2017role} this does not invalidate the non-parametric estimation method as long as $\mathbb{P}\left(\lambda_t^i<0\right)$ is negligible for any $i=1,\ldots,D$. In this case, the authors have shown the method leads to reliable results.

\subsection{Physics-Informed Neural Networks}

Assume $f:[0,T]\times\Omega\rightarrow\mathbb{R}$ is a solution to a partial differential equation (PDE) with representation
\begin{equation} \label{eq:pde_pinn}
    \partial_tf+\mathcal{D}f=0, \hspace{0.3cm} t\in[0,T], \hspace{0.1cm}\,x\in\Omega\subset\mathbb{R^d},
\end{equation}
where $\mathcal{D}$ is a differential operator, $T$ the time horizon and $\Omega$ the spatial domain. Suppose that some initial and boundary conditions ensure the uniqueness of solution $f$:
\begin{equation}
    f(0,x)=g(x), \quad x\in\Omega,
\end{equation}
\begin{equation}
    \mathcal{B}f(t,x)=0, \quad t\in[0,T], \, x\in\partial\Omega,
\end{equation}
where $\mathcal{B}$ is the boundary operator associated to Equation \eqref{eq:pde_pinn} and $\partial\Omega$ is the boundary of $\Omega$.

A data-driven approach to solve this PDE is to specify the solution $f$ as a neural network $f_\theta$ where $\theta$ is the set of trainable parameters. Since $f$ is assumed continuous on $[0,T]\times\Omega$, the existence of this approximation is guaranteed by the universal approximation theorem \citep{hornik1989multilayer} provided there are enough hidden units and the activation function is continuous, unbounded. This physics-informed neural network (PINN) specification was formulated in \cite{raissi2019physics} and have motivated many researchers in the past years. The idea is to train $f_\theta$ such that it minimizes the following loss function
\begin{equation}
    \mathcal{L}(\theta,\mathcal{E}):=\frac{1}{{N^c}}\sum_{n=1}^{N^c}\left|\partial_tf_\theta(t_n,x_n)+\mathcal{D}f_\theta(t_n,x_n)\right|^2+\frac{1}{N^i}\sum_{n=1}^{N^i}\left|f_\theta(0,x_n^i)-g(x_n^i)\right|^2+\frac{1}{N^b}\sum_{n=1}^{N^b}\mathcal{B}f_\theta(t_n^b,x_n^b)^2,
\end{equation}
where $\mathcal{E}:=\{\{(t_n,x_n)\}_{1\leq n\leq N^c}, \{x_n^i\}_{1\leq n\leq N^i}, \{(t_n^b,x_n^b)\}_{1\leq n\leq N^b}\}$ are either a reference grid used for validation or randomly generated data points at each iteration of a gradient descent algorithm. The notation $N^c$, $N^i$ and $N^b$ holds respectively for the number of collocation points, initial condition points and boundary condition points. The gradients and other derivatives are computed via automatic differentiation.

At the beginning of its development, the method presented a lot of training issues that have been tackled since, encompassing unbalanced gradients \citep{wang2021understanding, deguchi2023dynamic}, the reproduction of time causality \citep{wang2022respecting}, the complexity of the PINNs loss landscapes \citep{krishnapriyan2021characterizing} and the difficulty of optimizing the sum of multiple competitive loss functions \citep{bischof2021multi}. Although the literature about PINNs is still burgeoning, several good practices have already emerged and now allow researchers to train robust models.

This data-driven approach can be transposed from PDEs to integral equations (IEs). If $f$ is a solution to an integral equation with representation
\begin{equation}\label{eq:integral_pinn}
    f+\mathcal{I}f=0,\hspace{0.3cm} t\in[0,T], \hspace{0.1cm} x\in\Omega\subset\mathbb{R^d},
\end{equation}
where $\mathcal{I}$ is an integral operator, $T$ the time horizon and $\Omega$ the spatial domain, then the loss function of the associated PINN $f_\theta$ is written
\begin{equation}
    \mathcal{L}(\theta,\mathcal{E}):=\frac{1}{N}\sum_{n=1}^N\left|f_\theta(t_n,x_n)+\mathcal{I}f_\theta(t_n,x_n)\right|^2.
\end{equation}
For example, if $f$ is the solution of the following Fredholm integral equation of the second kind
\begin{equation}
    g(t)=f(t)+\int_0^Tg(t-s)f(s)\mathrm ds,\hspace{0.3cm} t\in[0, T],
\end{equation}
where $g$ is a continuous function satisfying $\underset{0\leq t\leq T}{\sup}\int_0^T|g(t-s)|\mathrm{d}s<1$ which ensures the uniqueness of the solution, we write for all $t\in[0,T]$
\begin{equation}
    \mathcal{I}f(t):= - g(t) + \int_0^Tg(t-s)f(s)\mathrm ds.
\end{equation}
This time, the training of a PINN $f_\theta$ that approximates the solution of Equation \eqref{eq:integral_pinn} does not involve the estimation of gradients, but the computation of an integral via a quadrature method. The exploitation of neural networks in such frameworks has been proposed in e.g., \cite{effati2012neural, zappala2022neural, guan2022solving, yuan2022pinn}. Inspired by the best practices from the literature concerning PINNs applied to PDEs and IEs, we now design a PINN model to solve the Hawkes characterization equation.

\subsection{Moment-based neural Hawkes estimation procedure}

Let $u_\theta:(t,x)\mapsto (u_\theta^{ij}(t,x))_{1\leq i,j\leq D}$ be a physics-informed neural network with trainable parameters $\theta$. Our goal is to find $\theta^*$ that minimizes a loss function $\mathcal{L}$ over the domain of Equation \eqref{eq:characterization_equation}. We truncate the time domain and define the kernels for $t\in[0,T]$ instead of $[0,+\infty)$. Without any loss of generality we assume that for all $i$, $\mathcal{X}_i=\mathcal{X}\subset(0,+\infty)$ and that $\mathcal{X}$ is bounded. In practice, it simplifies the computation of the loss function and we can always extend the sets $\mathcal{X}_i$ such that all $D$ probability density functions $p^i$ share the same domain $\mathcal{X}$.

\subsubsection{Loss function}

Consider a set of data points $\mathcal{E}:=\{(t_n,x_n),\,1\leq n\leq N^c\}$ sampled over the domain $(0,T)\times\mathcal{X}$. We assume that the second order statistics $G^{ij}(t_n,x_n)$ defined at Equation \eqref{eq:def_G_statistics} are non-zero for all $i,j$ and $n$. From the characterization equation of Theorem \ref{th:characterization_theorem}, we define the $(i,j)$-equation residual 
at the $n$-th data point by
\begin{equation}
    \varepsilon_n^{ij}(\theta):=G^{ij}(t_n,x_n)-u_\theta^{ij}(t_n,x_n)-\sum_{k=1}^D\iint_{(0,T)\times\mathcal{X}}u_\theta^{ik}(s,z)H^{kj}(t_n-s,x_n,z)p_k(z)\,\mathrm{d}s\mathrm{d}z,
\end{equation}
and define the quadratic loss function $\mathcal{L}$ with
\begin{equation}\label{eq:unweighted_loss_function_nn}
    \mathcal{L}(\theta,\mathcal{E}):=\sum_{n=1}^{N^c}\sum_{1\leq i,j\leq D}\varepsilon_n^{ij}(\theta)^2.
\end{equation}

Following the recommendations of \cite{wang2023expert} on the training of PINNs, we use a loss weighting procedure to better infer the time causality of the kernel matrix. This weighting procedure was introduced in \cite{wang2022respecting}. The authors noticed that neural networks may encounter issues in approximating the solution of PDEs over the time component $t$ due to the propensity of the model to minimize all loss components at the same time. They proposed to create temporal weights that will not give importance to a loss component evaluated at any time $t$ unless all the previous components --- for all time $s<t$ --- are sufficiently small. This can be particularly relevant for Hawkes kernels when the accuracy of estimation at short times is crucial.

We revisit this methodology by introducing a normalized weighting procedure that scales the loss and controls the weights' range of values. Assume the set $\mathcal{E}$ is built such that for all $n\in\{1,\dots,N^c\}$, $t_{n-1}<t_{n}$ and define the cumulative loss at time $t_n$:
\begin{equation}
    \mathcal{S}_n^{ij}:=\sum_{k=1}^n\varepsilon_k^{ij}(\theta)^2.
\end{equation}
We define the $(i,j)$-temporal weight evaluated at the $n$-th data point
\begin{equation}
    \omega_n^{ij}(\theta):=
    \begin{cases}
        1 & \text{if } n=1,\\
        \displaystyle \exp{-\epsilon\frac{\mathcal{S}_{n-1}^{ij}}{\mathcal{S}_{N^c}^{ij}}}  & \text{otherwise},
    \end{cases}
\end{equation}
where $\epsilon$ is a a hyperparameter that controls the minimum value taken by the weights as they all take values in $]e^{-\epsilon},1]$.
We finally define the weighted loss function as
\begin{equation}
    \mathcal{L}^{\omega}(\theta,\mathcal{E}):=\sum_{n=1}^{N^c}\mathcal{L}_n^{\omega}(\theta),
\end{equation}
where
\begin{equation}\label{eq:weighted_loss_function_nn}
    \mathcal{L}_n^{\omega}(\theta,\mathcal{E}):=\sum_{1\leq i,j\leq D}\omega_n^{ij}(\theta)\,\varepsilon_n^{ij}(\theta)^2.
\end{equation}

\subsubsection{Sampling procedure}

We now detail the sampling procedure of the set of data points $\mathcal{E}:=\{(t_n,x_n),\,1\leq n\leq N^c\}$. Our method uses a generic randomized sampling. The model did not show any significant improvement by using residual-based adaptive sampling such as the ones described in \cite{wu2023comprehensive} as standard sampling already provides accurate estimates of the kernel matrix.

As our practical applications are usually done for short time scales -- such as high-frequency data -- we are particularly concerned by the inference of kernels that may have large variations for $t<10^{-3}$, corresponding to a millisecond. Therefore we adapt the randomized time sampling by generating more data points for small times. Let $S$ be the proportion of time points taking values under a short-time scale threshold $\tau\ll T$. For a training set of size $N^c$, we sample $\lfloor SN^c \rfloor$ data points using a uniform law on the interval $(0,\tau)$, and $N^c-\lfloor SN^c \rfloor$ data points using a uniform law on the interval $(\tau,T)$, $\lfloor . \rfloor$ being the floor function. In the rest of the paper, we set $S=0.3$. The choice of the threshold $\tau$ highly depends on the problem and the time scale of the analysis. Other sampling distributions of time points have been tested, but did not result in any significant improvement to the estimation. Later, we introduce an analog threshold $h$ for the estimation grid of second order statistics and we set $\tau=h$. The value of $h$ is specified at each numerical experiment.

Finally, an integer uniform grid sampling is used for the mark data points. 

\subsubsection{Architecture and detailed training pipeline}

The choice of the neural network architecture can be crucial when designing a PINN. For example in \cite{wang2021understanding}, the authors propose a novel architecture to help reducing gradient flow issues during training. Furthermore, neural networks inspired from the long-short term memory cell may lead to better performance as argued in \cite{sirignano2018dgm}. In the rest of the paper, we use the Deep Galerkin Method (DGM) network architecture introduced in \cite{sirignano2018dgm}. We make a slight modification to the original architecture by using ReLU activation functions in place of hyperbolic tangent. Indeed, in our numerical experiments, we found that discontinuous kernels are much better reproduced with this configuration\footnote{We also explored the idea of using the $L^\infty$-norm instead of the $L^2$-norm in the training process as \cite{wang20222} suggests that it might lead to more accurate solutions in some problems. We finally chose to stick to the original norm as the validation results presented in the next section did not improve with this change.}.

As discussed in \cite{bacry2016first}, it is noteworthy that the system of $D^2$ Fredholm integral equations can be formulated as $D$ independent systems of $D$ Fredholm integral equations. Thus, the approximated solution can be splitted into $D$ neural networks each predicting an output of size $D$ --- the kernels $\phi^{i.}$ or in other words, the $i$-th row of the kernel matrix $\Phi$. This splitting procedure thus involves the training of $D$ models and has the additional benefit that the $D$ models are suitable for parallel computing, leading to a much faster training. We use this methodology to learn the entire kernel matrix $\Phi$.
The loss functions of Equations \eqref{eq:unweighted_loss_function_nn} and \eqref{eq:weighted_loss_function_nn} are decomposed as $D$ loss functions such that for all $i$,
\begin{equation}\label{eq:unweighted_loss_function_nn_i}
    \mathcal{L}_i(\theta,\mathcal{E}):=\sum_{n=1}^{N^c}\sum_{j=1}^D\varepsilon_n^{ij}(\theta)^2,
\end{equation}
and
\begin{equation}\label{eq:weighted_loss_function_nn_i}
    \mathcal{L}_{i,n}^{\omega}(\theta,\mathcal{E}):=\sum_{j=1}^D\omega_n^{ij}(\theta)\,\varepsilon_n^{ij}(\theta)^2.
\end{equation}
The detailed training pipeline is described in Algorithm \ref{algo:pinn_training}.

\begin{algorithm}[htb]
    \caption{Training procedure of the moment-based neural Hawkes}
    \label{algo:pinn_training}
    \begin{enumerate}
        \item
        Estimate the first order statistics $\Lambda$ and second order statistics $G$ given by Equation \eqref{eq:def_G_statistics} over the time and mark domains.

        \item
        For $i=1$ to $D$:
        \begin{enumerate}
            \item
            Represent the solution of the $i$-th characterization equation of Theorem \ref{th:characterization_theorem} by a DGM neural network $u_\theta^i$ and initialize its weights $\theta$ with a Glorot scheme.
            
            \item
            Set the number of epochs $E$, the size $N^c$ of the training set $\mathcal{E}$, the batch size $B$ such that $N^c=0\,[B]$, a decreasing sequence of learning rates $(\gamma_e)_{e\geq 1}$ and a random sampling scheme, \textit{e.g.} the procedure described above. We use $\gamma_e=\gamma_0100^{-\frac{e}{E}}$.
            
            \item
            For $e=1$ to $E$:
            \begin{enumerate}
                \item Sample a training set $\mathcal{E}=\{(t_n,x_n), 1\leq n\leq N^c\}$ and a validation set $\bar{\mathcal{E}}$.
                \item Apply a standardization and log-scaling to the data points.
                \item Over the elements of $\mathcal{E}$: compute the equation residuals $(\varepsilon_n^{ij}(\theta))_{1\leq j\leq D}$ and the temporal weights $(\omega_n^{ij}(\theta))_{1\leq j\leq D}$.
                \item For $s = 1$ to $\frac{N^c}{B}$:
                \begin{itemize}
                    \item Using Equation \eqref{eq:weighted_loss_function_nn_i}, update the weights via a mini-batch gradient descent
                    
                    \begin{equation}
                      \theta \leftarrow \theta - \gamma_e\frac{1}{B}\sum_{n=(s-1)B+1}^{sB}\nabla_\theta\mathcal{L}_{i,n}^{\omega}(\theta).
                    \end{equation}
                \end{itemize}
                \item Using Equation \eqref{eq:unweighted_loss_function_nn_i}, compute the unweighted validation loss $\mathcal{L}_i(\theta, \bar{\mathcal{E}})$.
            \end{enumerate}
            
            \item \textbf{Outputs:} An estimate $\theta^i$ of the optimal weights for the $i$-th model and a list of validation losses over the epochs.
        \end{enumerate}
        \item \textbf{Outputs:} $D$ models $(u_{\theta^i}^i)_{1\leq i\leq D}$, the $i$-th model representing the $i$-th row of the kernel matrix.
    \end{enumerate}
\end{algorithm}

\subsection{The practitioner's corner}

The aim of this section is to give a practical advice regarding training and potential questions that arise during the implementation step.

\paragraph{On the non-dimensionalization of the problem.} In order to standardize the time and mark inputs of the neural network, we recommend the following transformations.
\begin{itemize}
    \item \textbf{Time}: Apply a logarithmic transformation $t\mapsto \log_p{t}$, with $p=10$ for example. This allows the network to better infer kernels that have large fluctuations over small time windows. In practice, as this is commonly observed in financial microstructure when studying the order flow at short time scales, we advise the practitioners to use this scaling by default. Furthermore, our numerical experiments have shown significant improvements by using this transformation.
    \item \textbf{Mark}: A good standardization methodology for the marks should depend on the data generating process. By default, we applied a z-score transformation $x\mapsto \frac{x-m}{\sigma}$ where $m$ and $\sigma$ are respectively the empirical average and empirical standard deviation computed over a sample of the training data.
\end{itemize}

\paragraph{On the estimation of the integral.} The integral component of Equation \eqref{eq:loss_function_nn} is numerically estimated with time quadratures.  Although a large number of quadratures --- larger than 1000 --- may provide slightly better results, we did find that 500 quadratures were amply sufficient to obtain a good performance. One should keep in mind that the training procedure becomes costly if this number is too large as $D^2M$ integrals need to be computed at each evaluation of the loss. In the numerical experiments, we use a quadrature scheme with a logarithmic grid.

\paragraph{On the estimation of the second order statistics.} A correct choice of the time grid over which the statistics $G$ are estimated is crucial. As the number of observations increases, we can expect to get a more precise estimation and thus we can set a finer mesh. As the choice of the sample-grid may be highly problem-dependent, we will use a grid that is well suited for our scope of application. To this extent, we shall use a scheme inspired by the work of \citet{bacry2016estimation} and \citet{rambaldi2017role} for the estimation of kernels with short time horizon. It consists in a linear scheme for very short times up to an intermediate step $h$, and then a logarithmic scheme from $h$ to the time horizon $T$. This allows to account for the structure of high-frequency data as explained by the authors. We thus use the following rule. We set $h<T$, the number of points in the linear part $n_{\text{lin}}$ and the number of points in the logarithmic part $n_{\text{log}}$ and estimate the statistics on the following grid
\begin{equation}\label{eq:g_grid}
    \left[t_{\text{min}}, t_{\text{min}}+\frac{1}{n_{\text{lin}}}(h-t_{\text{min}}),t_{\text{min}}+\frac{2}{n_{\text{lin}}}(h-t_{\text{min}}),\dots,h,h\left(\frac{T}{h}\right)^{\frac{1}{n_{\text{log}}}},h\left(\frac{T}{h}\right)^{\frac{2}{n_{\text{log}}}},\dots,T\right].
\end{equation}
A linear interpolation method is used to obtain the values for any $t\in[t_{\text{min}},T]$. In the numerical experiments, we set $t_{\text{min}}=\frac{h}{n_{\text{lin}}}$.

\paragraph{On the continuity of the kernel at time $T^-$.} Since the kernel matrix is defined for $t\leq T$, we may impose a continuity condition $u(T,x)=0$, for all $x\in\mathcal{X}$ depending on the practical application. In the PINN literature, this boundary constraint is simply introduced to the training by adding an additional error term to the quadratic loss. A regularization weight $\omega_c$ is then applied to this loss term.

\paragraph{On kernels with different orders of magnitude.} Approximation issues may arise when the values taken by each kernel component of the matrix explore very different orders of magnitude. In this case, the kernels with high values will be correctly estimated but the relative error made over the other kernels might be high. This inference problem is similar to the causality issue that was studied in \cite{wang2022respecting} in the sense that the neural network averages the errors made over the statistics $G$. Thus, statistics that take high values will overshadow the ones taking small values. An idea to mitigate this problem is to use a weighted version of the loss function defined in Equation \eqref{eq:loss_function_nn}. To this extent, we propose the following weighted loss function
\begin{equation}\label{eq:loss_function_nn}
    \mathcal{L}^{\omega}(u,G,\mathcal{E}):=\sum_{n=1}^{N^c}\sum_{1\leq i,j\leq D}\zeta_{ij}(x_n)\omega_n^{ij}(\theta)\varepsilon_n^{ij}(\theta)^2,
\end{equation}
with
\begin{equation}
    \zeta_{ij}(x):=\frac{\displaystyle\left(\int_{(0,T)}\abs{G^{ij}(t,x)}\,\mathrm{d}t\right)^{-1}}{\displaystyle\sum_{1\leq i',j'\leq D}\left(\int_{(0,T)}\abs{G^{i'j'}(t,x)}\,\mathrm{d}t\right)^{-1}}, \hspace{0.25cm} 1\leq i, j \leq D.
\end{equation}
This weighting procedure puts more weight to kernels that exhibit smaller absolute second order statistics $G$ with respect to other kernels. Obviously, this technique becomes irrelevant if all the components of the kernel matrix $\Phi$ take similar values.

\paragraph{On the training time.} As the Wiener-Hopf approach, our non-parametric method has the benefit of having a computation time that is independent from the number of observations. In fact, the number of observations influences the computation time of second order statistics but not the training time itself. This strongly contrasts with common likelihood approaches.

\section{Numerical validation}
\label{validation_section}

In this section we present several numerical experiments conducted over a wide range of simulated data. We show that the moment-based non-parametric neural Hawkes estimation method detailed in Section \ref{methodology_section} provides accurate kernel estimates in various configurations of dimensions, kernel types and mark distributions. We measure the rate of convergence of the method with respect to the sample size and provide some elements regarding the robustness of the method with respect to hyperparameters.

\subsection{Numerical validation for various shapes of kernels}
\label{subsec:VariousKernels}
We set up six numerical experiments. In each experiment, we use a thinning algorithm to simulate a sequence of timestamps generated by a multidimensional linear Hawkes process with kernel configurations that may appear in financial applications, e.g., with slowly decreasing tails, delayed modes and inhibition patterns. For each simulated event, a mark $\xi$ is independently generated from a discrete uniform law taking values in $\mathcal{X}=\{1,\dots,M\}$ where $M$ is specified for each experiment.

More precisely, we use in the first five experiments a multiplicative kernel matrix $\Phi$ with components satisfying
\begin{equation}
    \phi^{ij}(t,x):=\varphi^{ij}(t)f^{ij}(x),\hspace{0.3cm} 1\leq i,j \leq D,\hspace{0.1cm} t\in(0,+\infty),\hspace{0.1cm} x\in\mathcal{X}.
\end{equation}
Each experiment tests a specific form of aggregated time kernels $\varphi^{ij}$, namely exponentially decreasing kernels, slowly-decreasing kernels, kernels with latencies, kernels with inhibition and kernels with multiple Gaussian modes. 
As for marks, aggregated mark kernels in the first five experiments can be either a constant mark kernel
\begin{equation}
    f_0(x)=1,
\end{equation}
or a linear mark kernel
\begin{equation}
    f_1(x)=\frac{2}{M+1}\,x,
\end{equation}
or a quadratic mark kernel
\begin{equation}
    f_2(x)=\frac{6}{(M+1)(2M+1)}\,x^2.
\end{equation}
In each case $\mathbb{E}\left(f(X)\right)=1$. 

Observe that we use in the first five experiments a decoupled kernel for the sake of simplicity and illustration but that the estimation method is designed for a non-multiplicative kernel, as it is clear in Sections  \ref{methodology_section} and \ref{application_section}. In the last (sixth) experiment, we test the estimation method on a non-multiplicative kernel in order to validate the general case.

In all experiments, we train a neural network following Algorithm \ref{algo:pinn_training} with the set of hyperparameters displayed in Table \ref{table:hyperparameters_config} and we estimate the kernel over the truncated time domain $(0,T)$.
\begin{table}[!htb]
   \small
   \centering
   \caption{Hyperparameters configuration in the numerical experiment.}
   \begin{tabular}{cccccccc}
   \toprule\toprule
   Neurons & DGM depth & Learning rate & Quadratures & Batch size & Training size & Validation size & Epochs\\ 
   \midrule
   64 & 1 & $10^{-3}$ & 250 & 8 & 1024 & 128 & 1000\\ 
   \bottomrule
   \end{tabular}
   \label{table:hyperparameters_config}
\end{table}

\paragraph{Exponentially decreasing kernel.} Exponentially decreasing functions are amongst the most popular kernels found in the literature. The Markov property that results from such a specification provides mathematical tractability and a greatly reduced computation cost when simulating. The non-parametric estimation of these kernels from simulated data constitutes the first step of our validation procedure.

We set $D=4$, $M=10$ and simulate $10^7$ events of a linear marked Hawkes process with kernel functions
\begin{equation}\label{eq:exponentially_decreasing_kernel}
    \phi^{ij}(t,x):=\alpha_{ij} e^{-\beta_{ij} t}f^{ij}(x), \hspace{0.3cm} 1\leq i,j\leq D, \hspace{0.1cm}t\in[0,+\infty), \hspace{0.1cm}x\in\mathcal{X}.
\end{equation}

The parameters of the time component and the mark kernels are given in Table \ref{table:config_exponentially_decreasing_kernel}. The second order statistics are estimated using the grid of Equation \eqref{eq:g_grid} with parameters $h=0.1$, $n_{\text{lin}}=10$, $n_{\text{log}}=50$. The baseline intensity is $\mu=[0.5, 0.25, 0.75, 0.15]'$.
\begin{table}[!htb]
    \caption{\textit{Exponentially decreasing kernel} --- Parameter configuration of the simulation.}
    \begin{subtable}{.33\linewidth}
      \centering
        \caption{$\alpha$}
        \begin{tabular}{c|cccc}
           \toprule
           $\alpha_{ij}$ & 1 & 2 & 3 & 4 \\ 
           \midrule
            1 & 1.5 & 1 & 2 & 0.75 \\
            2 & 1 & 2 & 2 & 1 \\
            3 & 1 & 1 & 2 & 1.5 \\ 
            4 & 2 & 1 & 2 & 0.5 \\ 
           \bottomrule
       \end{tabular}
    \end{subtable}%
    \begin{subtable}{.33\linewidth}
      \centering
        \caption{$\beta$}
        \begin{tabular}{c|cccc}
           \toprule
           $\beta_{ij}$ & 1 & 2 & 3 & 4 \\ 
           \midrule
            1 & 8 & 5 & 10 & 8 \\
            2 & 5 & 15 & 8 & 5 \\
            3 & 8 & 10 & 8 & 8 \\ 
            4 & 5 & 5 & 8 & 4 \\ 
           \bottomrule
       \end{tabular}
    \end{subtable}%
    \begin{subtable}{.33\linewidth}
      \centering
        \caption{$f$}
        \begin{tabular}{c|cccc}
           \toprule
           $f^{ij}$ & 1 & 2 & 3 & 4 \\ 
           \midrule
            1 & $f_1$ & $f_2$ & $f_1$ & $f_1$\\
            2 & $f_1$ & $f_2$ & $f_2$ & $f_0$\\
            3 & $f_0$ & $f_0$ & $f_2$ & $f_0$\\ 
            4 & $f_2$ & $f_1$ & $f_2$ & $f_1$\\ 
           \bottomrule
       \end{tabular}
    \end{subtable}%
    \label{table:config_exponentially_decreasing_kernel}
\end{table}

\begin{figure}
    \centering
    \includegraphics[width=1.\linewidth]{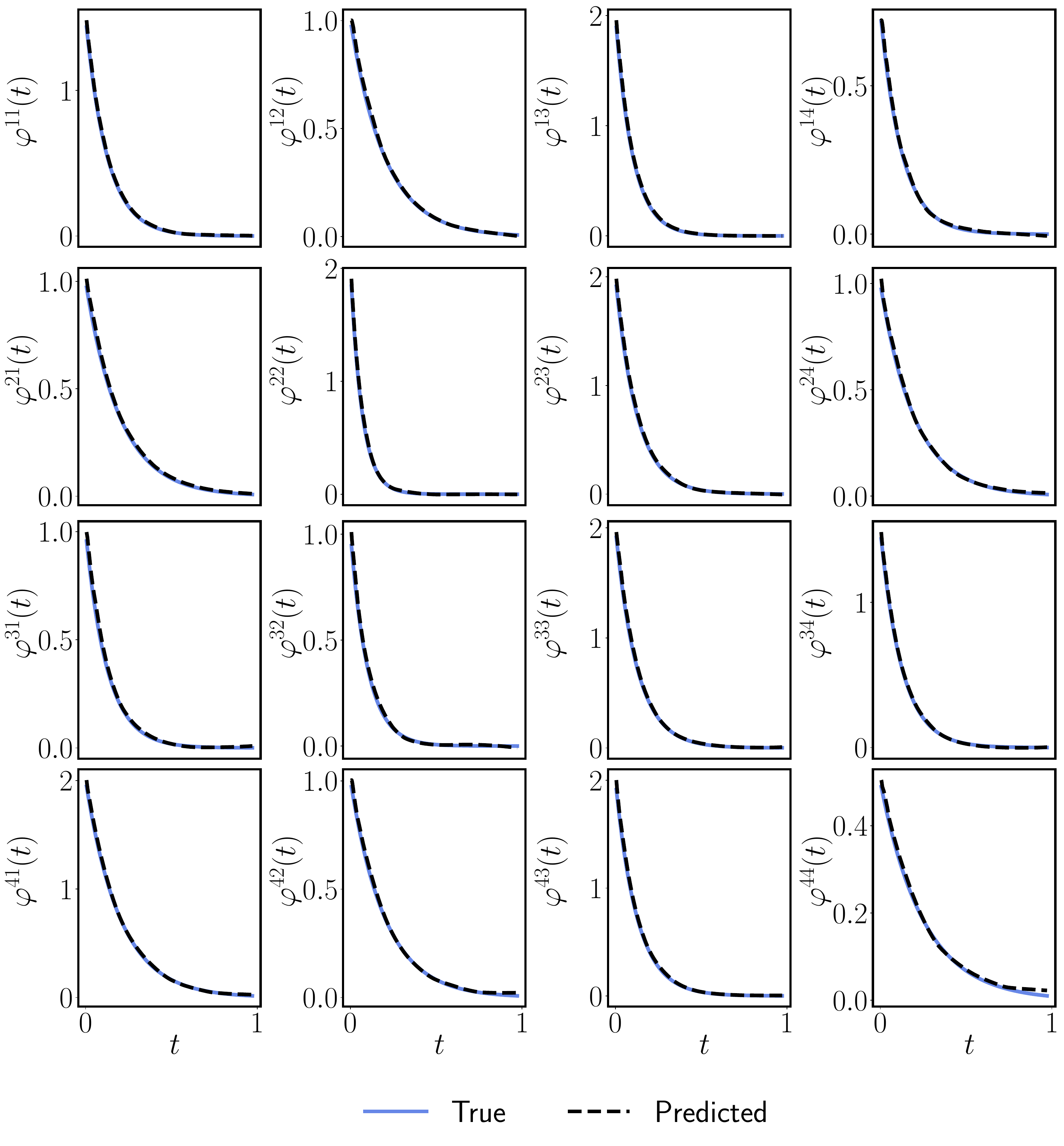}
    \caption{\textit{Exponentially decreasing kernel} --- Comparison of the predicted and exact aggregated time kernel matrix $(\varphi^{ij})_{1\leq i,j\leq D}$ defined in Equation \eqref{eq:aggregated_time_kernel}.}
    \label{fig:exponential_4d_10m_time_kernel}
\end{figure}

\begin{figure}
    \centering
    \includegraphics[width=1.\linewidth]{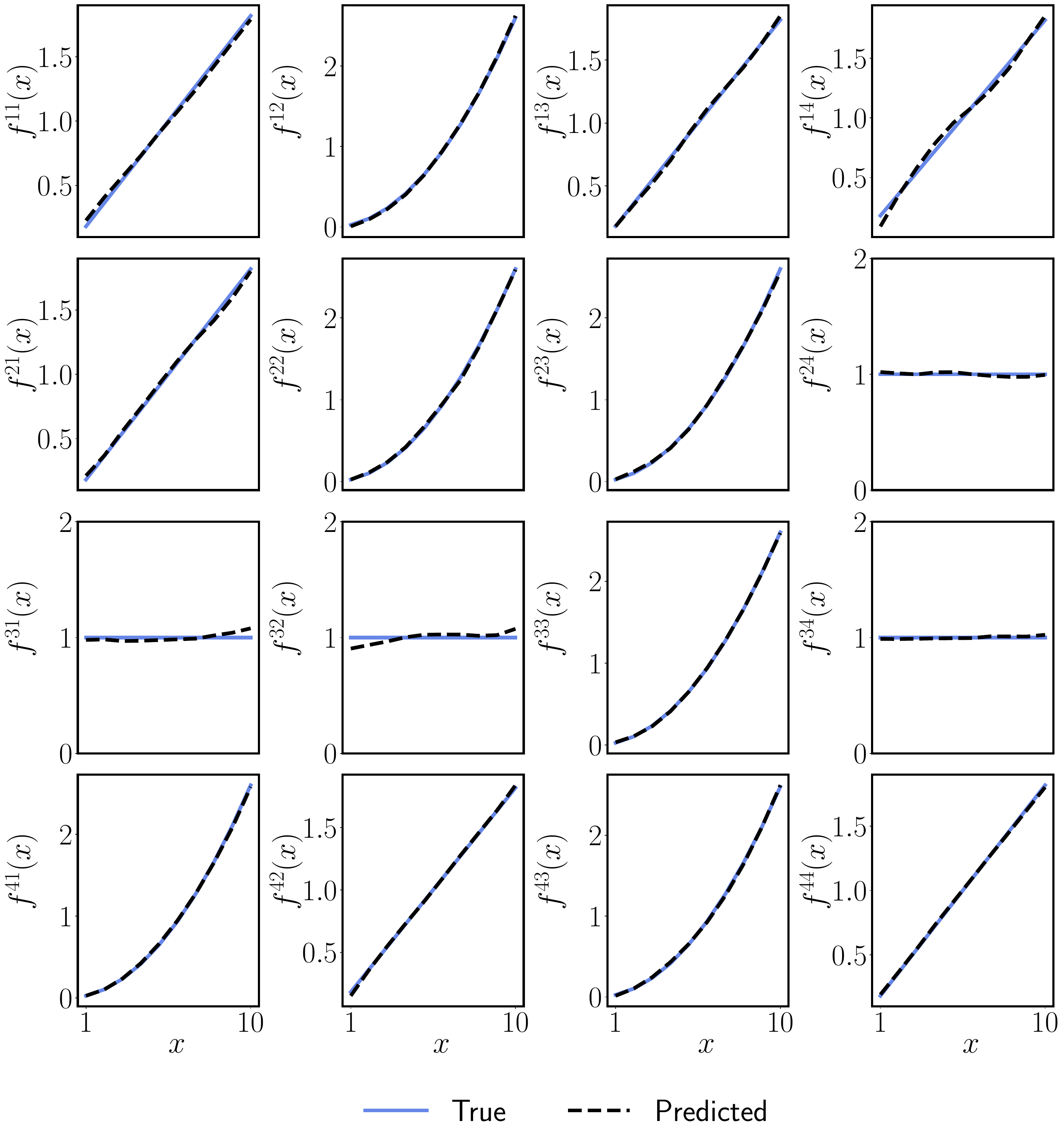}
    \caption{\textit{Exponentially decreasing kernel} --- Comparison of the predicted and exact aggregated mark kernel matrix $(f^{ij})_{1\leq i,j\leq D}$ defined in Equation \eqref{eq:aggregated_mark_kernel}.}
    \label{fig:exponential_4d_10m_mark_kernel}
\end{figure}

The fits of aggregated time kernels and mark kernels are displayed in Figure \ref{fig:exponential_4d_10m_time_kernel} and \ref{fig:exponential_4d_10m_mark_kernel}. We observe a great fit for both components and the model is able to reproduce the various mark kernel functions even when $f\equiv 1$.

\paragraph{Slowly decreasing kernel.} Slowly decreasing kernels are often observed on high-frequency financial data after applying non-parametric estimation methods for Hawkes processes --- see \cite{bacry2012non}, \cite{bacry2016estimation}, \cite{fosset2022non}. Identifying the rates of decrease over time of self-excitation and cross-excitation components gives a better understanding of the market microstructure and the memory of order flow.

We set $D=2$, $M=5$ and simulate $2.10^6$ events of a linear marked Hawkes process with kernel functions
\begin{equation}\label{eq:slowly_decreasing_kernel}
    \phi^{ij}(t,x):=\alpha_{ij}(\gamma_{ij}+t)^{-\beta_{ij}}f^{ij}(x), \hspace{0.3cm} 1\leq i,j\leq D, \hspace{0.1cm}t\in[0,+\infty), \hspace{0.1cm}x\in\mathcal{X},
\end{equation}
with $f^{ij}(x)=f_2(x)$. The parameters of the time component are given in Table \ref{table:config_slowly_decreasing_kernel}. The second order statistics are estimated using the grid of Equation \eqref{eq:g_grid} with parameters $h=10^{-3}$, $n_{\text{lin}}=25$, $n_{\text{log}}=75$. The baseline intensity is $\mu=[0.05, 0.05]'$.
\begin{table}[!htb]
    \caption{\textit{Slowly decreasing kernel} --- Parameter configuration of the simulation.}
    \begin{subtable}{.33\linewidth}
      \centering
        \caption{$\alpha$}
        \begin{tabular}{c|cc}
           \toprule
           $\alpha_{ij}$ & 1 & 2 \\ 
           \midrule
            1 & 0.01 & 0.006 \\
            2 & 0.005 & 0.012 \\
           \bottomrule
       \end{tabular}
    \end{subtable}%
    \begin{subtable}{.33\linewidth}
      \centering
        \caption{$\beta$}
        \begin{tabular}{c|cc}
           \toprule
           $\beta_{ij}$ & 1 & 2 \\ 
           \midrule
            1 & 1.05 & 1.25 \\
            2 & 1.3 & 1.025 \\
           \bottomrule
       \end{tabular}
    \end{subtable}%
    \begin{subtable}{.33\linewidth}
      \centering
        \caption{$\gamma$}
        \begin{tabular}{c|cc}
           \toprule
           $\gamma_{ij}$ & 1 & 2 \\ 
           \midrule
            1 & 0.0005 & 0.00075 \\
            2 & 0.001 & 0.0006 \\
           \bottomrule
       \end{tabular}
    \end{subtable}%
    \label{table:config_slowly_decreasing_kernel}
\end{table}

\begin{figure}
    \centering
    \subfloat[Time kernel $(\varphi^{ij})_{1\leq i,j\leq D}$]{%
        \includegraphics[width=0.5\linewidth]{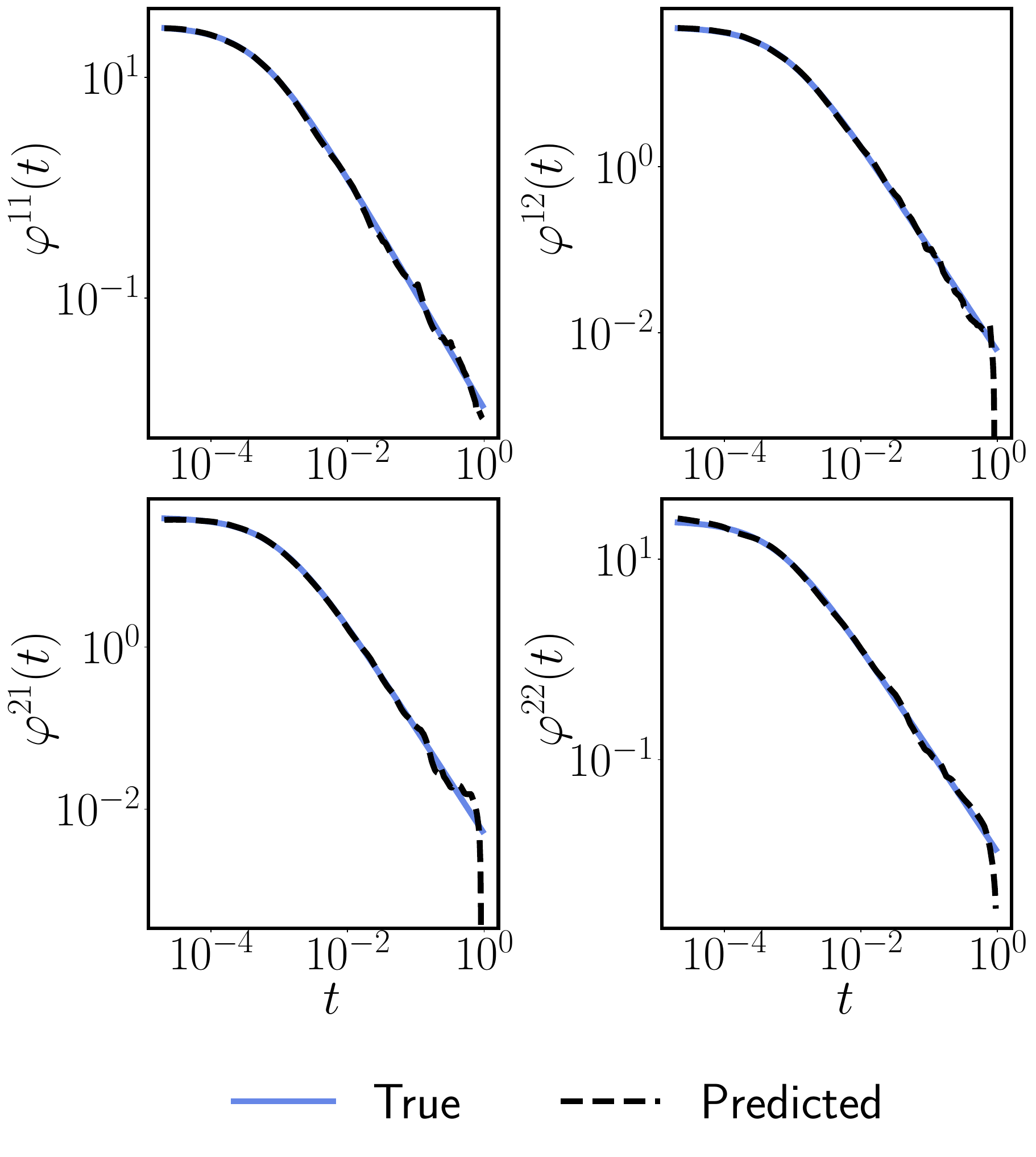}%
    }
    \subfloat[Mark kernel $(f^{ij})_{1\leq i,j\leq D}$]{%
        \includegraphics[width=0.5\linewidth]{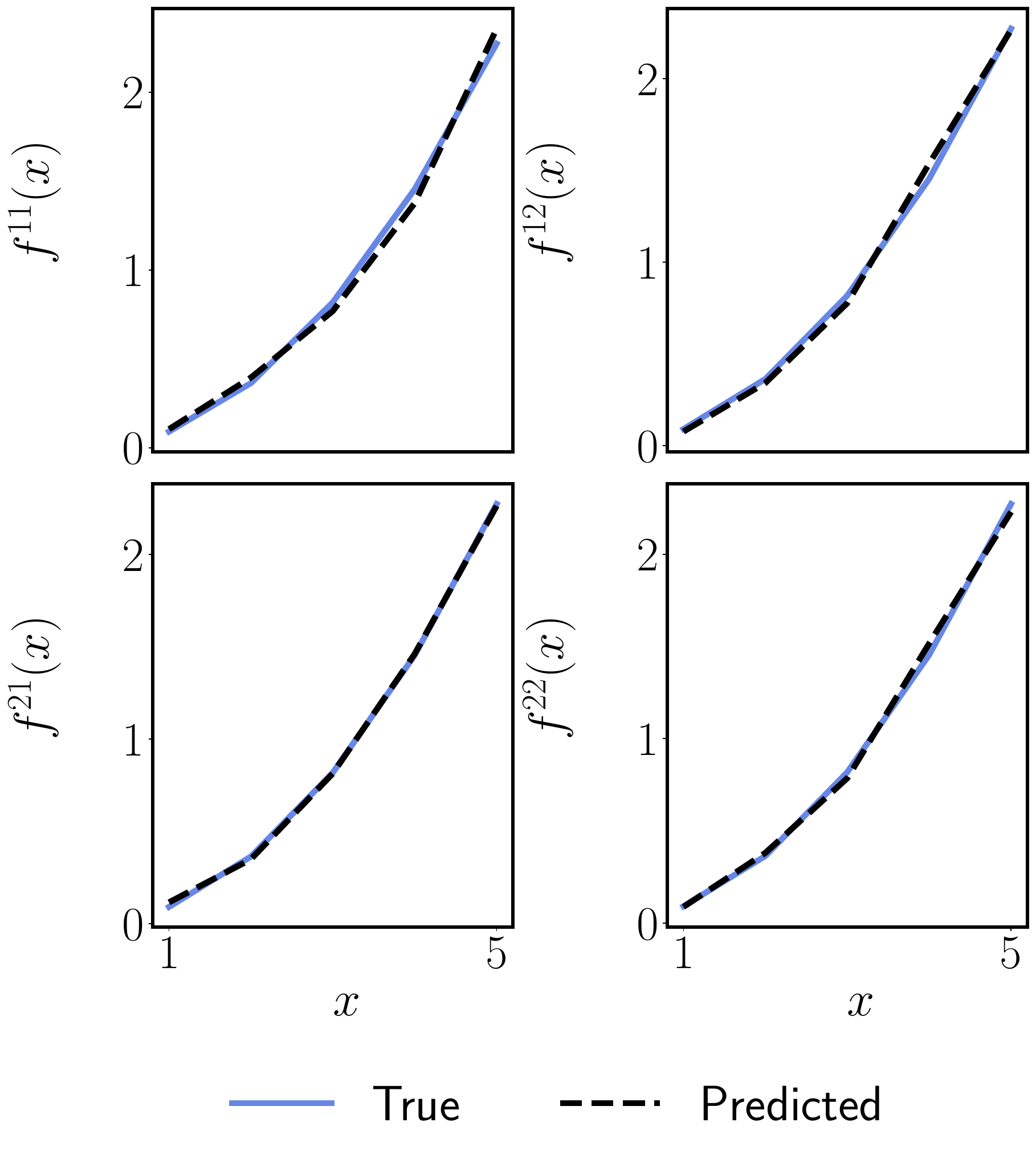}%
    }
    \caption{\textit{Slowly decreasing kernel} --- Comparison of the predicted and exact aggregated kernel matrices defined in Equations \eqref{eq:aggregated_time_kernel} and \eqref{eq:aggregated_mark_kernel}.}
    \label{fig:power_2d_5m_kernels}
\end{figure}

The fits of aggregated time kernels and mark kernels are displayed in Figure \ref{fig:power_2d_5m_kernels}. The model reproduces the power law over 5 decades and fits the mark component as well.

\paragraph{Delayed exponentially decreasing kernel.} In practice, trading algorithms are exposed to transmission delays and latency. This phenomenon passes through to the formation of lagged kernels --- see for example \citet{carreira2021exponential}.

We set $D=2$, $M=5$ and simulate $2.10^6$ events of a linear marked Hawkes process with kernel functions
\begin{equation}
    \phi^{ij}(t,x):=\alpha_{ij} e^{-\beta_{ij} (t-\ell_{ij})}f^{ij}(x)\,\mathds{1}_{\{t\geq \ell_{ij}\}}, \hspace{0.3cm} 1\leq i,j\leq D, \hspace{0.1cm}t\in[0,+\infty), \hspace{0.1cm}x\in\mathcal{X},
\end{equation}
with $f^{ij}(x)=f_2(x)$.

The parameters of the time component are given in Table \ref{table:config_delayed_exponential_kernel}. The second order statistics are estimated using the grid of Equation \eqref{eq:g_grid} with parameters $h=0.1$, $n_{\text{lin}}=10$, $n_{\text{log}}=50$. The baseline intensity is $\mu=[0.05, 0.05]'$.
\begin{table}[!htb]
    \caption{\textit{Delayed exponentially decreasing kernel} --- Parameter configuration of the simulation.}
    \begin{subtable}{.33\linewidth}
      \centering
        \caption{$\alpha$}
        \begin{tabular}{c|cc}
           \toprule
           $\alpha^{ij}$ & 1 & 2 \\ 
           \midrule
            1 & 1.25 & 0.35 \\
            2 & 0.6 & 1.15 \\
           \bottomrule
       \end{tabular}
    \end{subtable}%
    \begin{subtable}{.33\linewidth}
      \centering
        \caption{$\beta$}
        \begin{tabular}{c|cc}
           \toprule
           $\beta^{ij}$ & 1 & 2 \\ 
           \midrule
            1 & 5 & 8 \\
            2 & 8 & 10 \\
           \bottomrule
       \end{tabular}
    \end{subtable}%
    \begin{subtable}{.33\linewidth}
      \centering
        \caption{$\ell$}
        \begin{tabular}{c|cc}
           \toprule
           $\ell^{ij}$ & 1 & 2 \\ 
           \midrule
            1 & 0.1 & 0.2 \\
            2 & 0.25 & 0.4 \\
           \bottomrule
       \end{tabular}
    \end{subtable}%
    \label{table:config_delayed_exponential_kernel}
\end{table}

\begin{figure}
    \centering
    \subfloat[Time kernel $(\varphi^{ij})_{1\leq i,j\leq D}$]{%
        \includegraphics[width=0.5\linewidth]{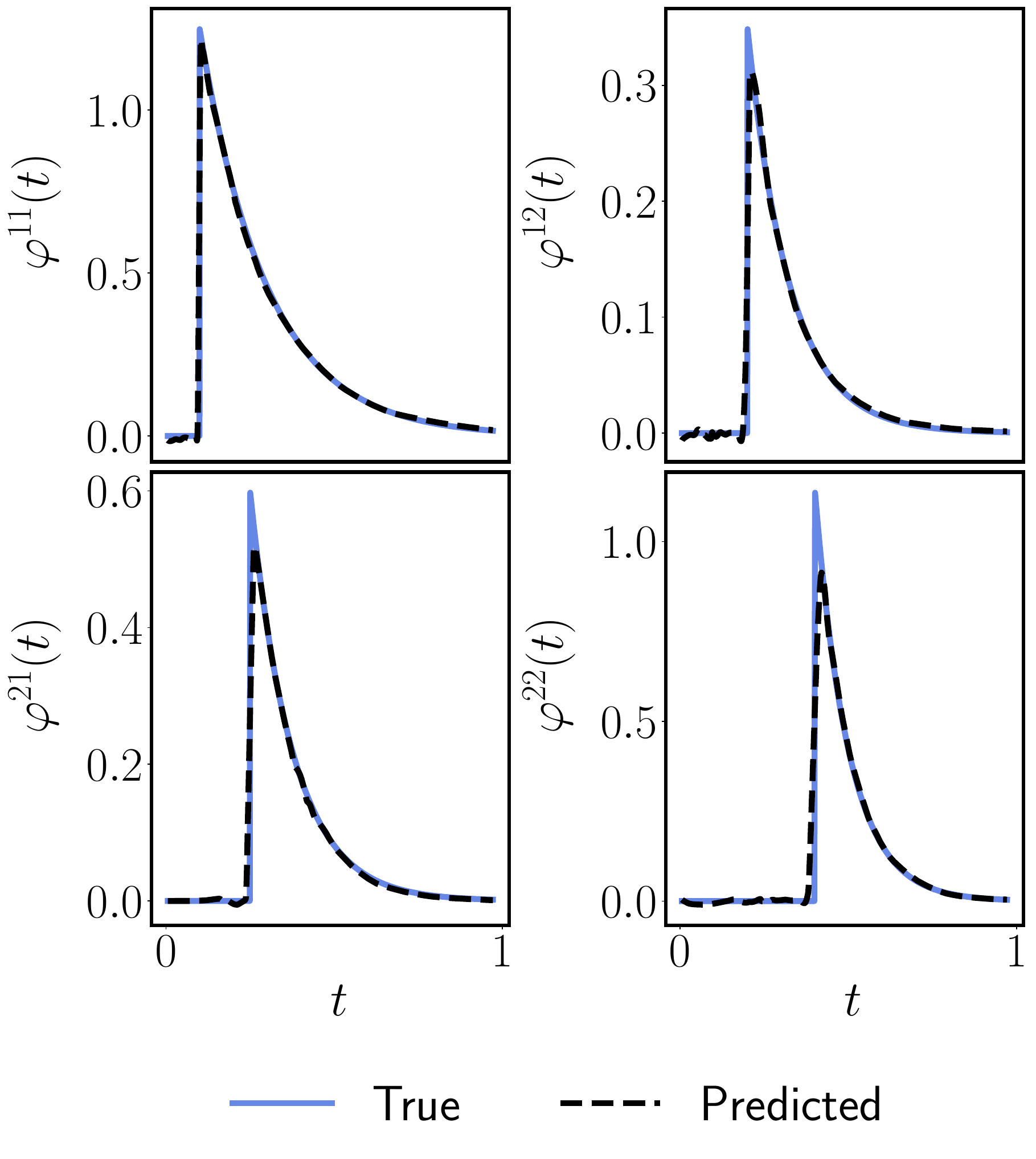}%
    }
    \subfloat[Mark kernel $(f^{ij})_{1\leq i,j\leq D}$]{%
        \includegraphics[width=0.5\linewidth]{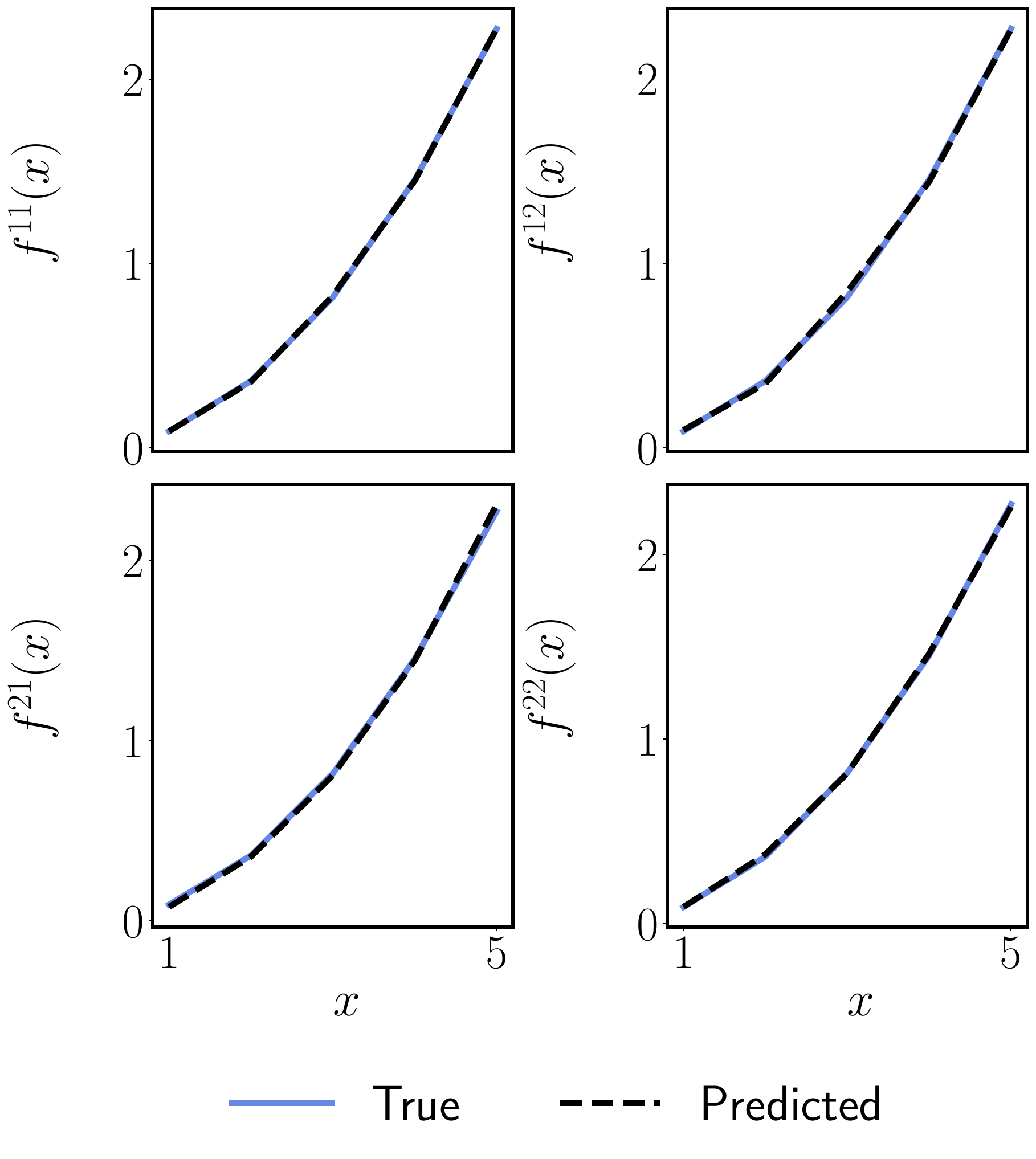}%
    }
    \caption{\textit{Delayed exponentially decreasing kernel} --- Comparison of the predicted and exact aggregated kernel matrices defined in Equations \eqref{eq:aggregated_time_kernel} and \eqref{eq:aggregated_mark_kernel}.}
    \label{fig:delayed_exp_2d_5m_kernels}
\end{figure}

The fits of aggregated time kernels and mark kernels are displayed in Figure \ref{fig:delayed_exp_2d_5m_kernels}. All different 4 latencies are correctly captured by the model and the mark kernels are correctly estimated as well.

\paragraph{Kernel with inhibition.} Inhibition effects have been documented several times in the market microstructure literature - see for example \cite{rambaldi2017role}, \cite{lu2018high}. In this experiment, we combine both latency and inhibition effects in one single kernel.

We set $D=2$, $M=5$ and simulate $2.10^6$ events of a linear marked Hawkes process with kernel functions
\begin{equation}
    \phi^{ij}(t,x):=\left(\underbar{$\alpha$}_{ij} e^{-\underbar{$\beta$}_{ij} t}\,\mathds{1}_{\{t< \ell_{ij}\}}+\overline{\alpha}_{ij} e^{-\overline{\beta}_{ij} (t-\ell_{ij})}\,\mathds{1}_{\{t\geq \ell_{ij}\}}\right)f^{ij}(x), \hspace{0.3cm} 1\leq i,j\leq D, \hspace{0.1cm}t\in[0,+\infty), \hspace{0.1cm}x\in\mathcal{X},
\end{equation}
with $f^{ij}(x)=f_2(x)$ and we set the instantaneous self/cross-impact elements $\underbar{$\alpha$}_{ij}$ and $\overline{\alpha}_{ij}$ such that
\begin{equation}
    \text{sign}(\underbar{$\alpha$}_{ij})=-\text{sign}(\overline{\alpha}_{ij}).
\end{equation}

The parameters of the time component are given in Table \ref{table:config_inhibition_kernel}. The second order statistics are estimated using the grid of Equation \eqref{eq:g_grid} with parameters $h=0.1$, $n_{\text{lin}}=25$, $n_{\text{log}}=75$. The baseline intensity is $\mu=[3, 2.5]'$.
\begin{table}[!htb]
    \caption{\textit{Kernel with inhibition} --- Parameter configuration of the simulation.}
    \begin{subtable}{.2\linewidth}
      \centering
        \caption{$\underbar{$\alpha$}$}
        \begin{tabular}{c|cc}
           \toprule
           $\underbar{$\alpha$}_{ij}$ & 1 & 2 \\ 
           \midrule
            1 & 1 & -0.25 \\
            2 & -0.2 & 1.2 \\
           \bottomrule
       \end{tabular}
    \end{subtable}%
    \begin{subtable}{.2\linewidth}
      \centering
        \caption{$\underbar{$\beta$}$}
        \begin{tabular}{c|cc}
           \toprule
           $\underbar{$\beta$}_{ij}$ & 1 & 2 \\ 
           \midrule
            1 & 3 & 3 \\
            2 & 2 & 2 \\
           \bottomrule
       \end{tabular}
    \end{subtable}%
    \begin{subtable}{.2\linewidth}
      \centering
        \caption{$\overline{\alpha}$}
        \begin{tabular}{c|cc}
           \toprule
           $\overline{\alpha}_{ij}$ & 1 & 2 \\ 
           \midrule
            1 & -0.3 & 1.5 \\
            2 & 1 & -0.25 \\
           \bottomrule
       \end{tabular}
    \end{subtable}%
    \begin{subtable}{.2\linewidth}
      \centering
        \caption{$\overline{\beta}$}
        \begin{tabular}{c|cc}
           \toprule
           $\overline{\beta}_{ij}$ & 1 & 2 \\ 
           \midrule
            1 & 2 & 5 \\
            2 & 3 & 10 \\
           \bottomrule
       \end{tabular}
    \end{subtable}%
    \begin{subtable}{.2\linewidth}
      \centering
        \caption{$\ell$}
        \begin{tabular}{c|cc}
           \toprule
           $\ell_{ij}$ & 1 & 2 \\ 
           \midrule
            1 & 0.25 & 0.5 \\
            2 & 0.15 & 0.6 \\
           \bottomrule
       \end{tabular}
    \end{subtable}%
    \label{table:config_inhibition_kernel}
\end{table}

\begin{figure}
    \centering
    \subfloat[Time kernel $(\varphi^{ij})_{1\leq i,j\leq D}$]{%
        \includegraphics[width=0.5\linewidth]{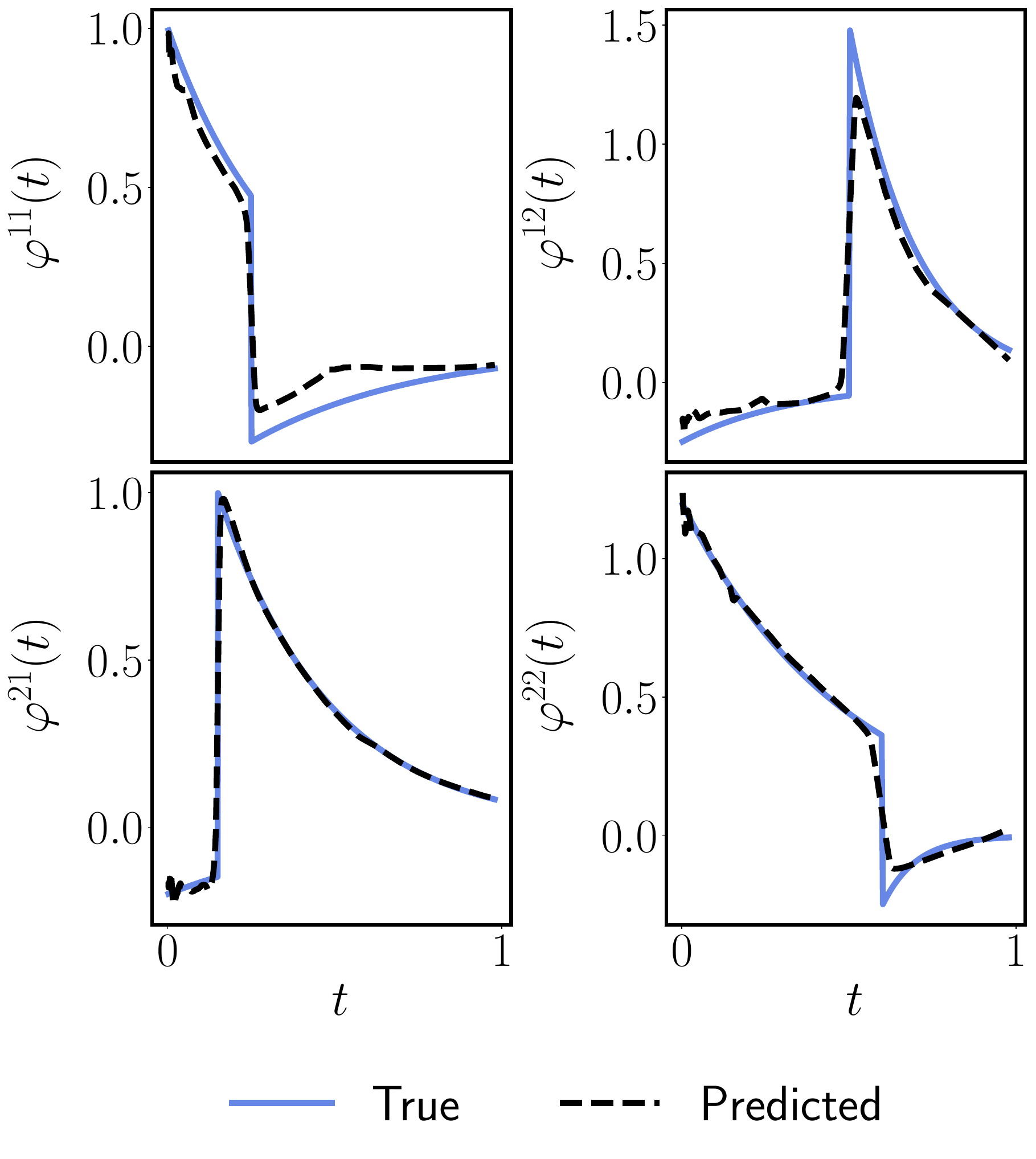}%
    }
    \subfloat[Mark kernel $(f^{ij})_{1\leq i,j\leq D}$]{%
        \includegraphics[width=0.5\linewidth]{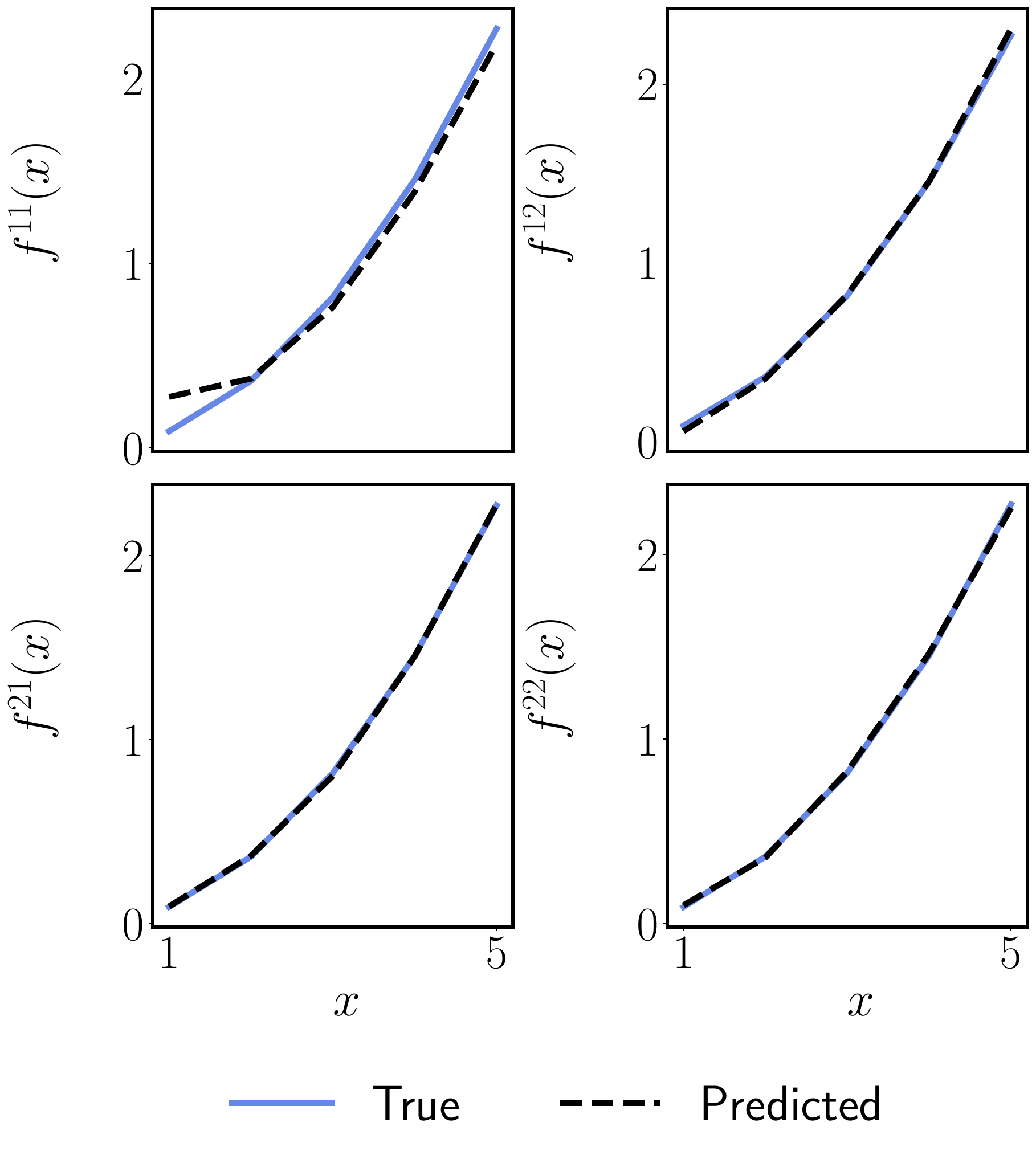}%
    }
    \caption{\textit{Kernel with inhibition} --- Comparison of the predicted and exact aggregated kernel matrices defined in Equations \eqref{eq:aggregated_time_kernel} and \eqref{eq:aggregated_mark_kernel}.}
    \label{fig:inhibition_2d_5m_kernels}
\end{figure}

The fits of aggregated time kernels and mark kernels are displayed in Figure \ref{fig:inhibition_2d_5m_kernels}. Note that the model is able to correctly capture the inhibition effects together with the latencies. 

\paragraph{High-dimensional bimodal Gaussian kernel.} In order to validate the estimation method in high-dimension in the presence of non-Markovian kernels, we design the following experiment.

We set $D=15$, $M=1$ and simulate $5.10^7$ events of a linear marked Hawkes process with bimodal Gaussian kernel that we define, for $1\leq i,j\leq D, \hspace{0.1cm}t\in[0,+\infty), \hspace{0.1cm}x\in\mathcal{X}$, as
\begin{equation}
    \phi^{ij}(t,x):=\frac{\alpha_{ij}}{2\sqrt{2\pi}}\left(\frac{1}{\underbar{$\sigma$}_{ij}}\exp\left(-\frac{1}{2}\left(\frac{t-\underbar{$\mu$}_{ij}}{\underbar{$\sigma$}_{ij}}\right)^2\right)+\frac{1}{\overline{\sigma}_{ij}}\exp\left(-\frac{1}{2}\left(\frac{t-\overline{\mu}_{ij}}{\overline{\sigma}_{ij}}\right)^2\right)\right).
\end{equation}

The parameters of the time component are given in Table \ref{table:config_inhibition_kernel}.  The second order statistics are estimated using a linear grid $\{\frac{k}{n}\}_{1\leq k\leq n}$ over $[0,1]$, with $n=50$. We generate the parameters randomly such that the Hawkes process is stable, \textit{i.e.} the branching ratio $\mathcal{R}<1$. 

\begin{figure}[!ht]
    \centering
    \includegraphics[width=1.\linewidth]{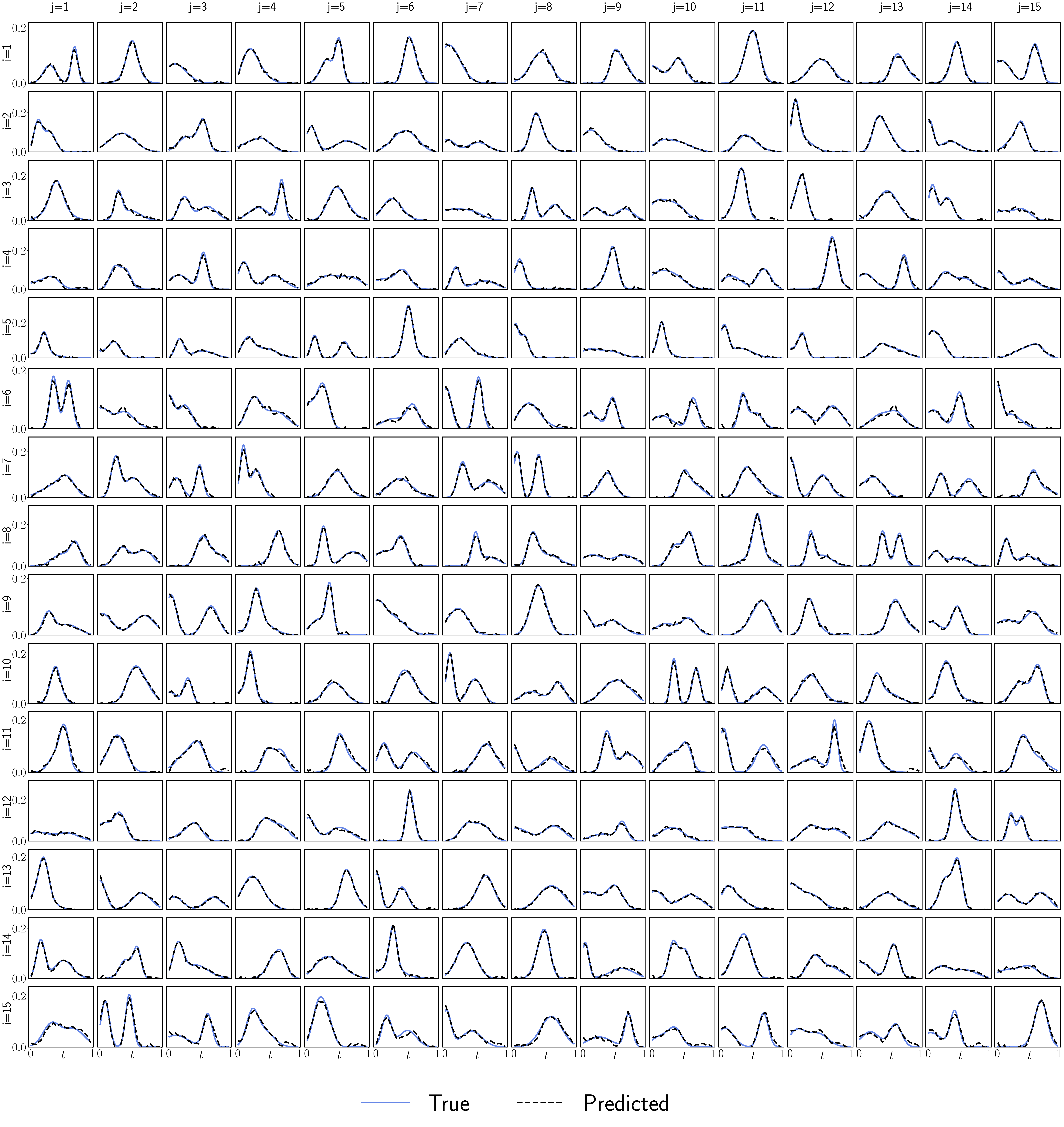}
    \caption{\textit{Bimodal Gaussian kernel} --- Comparison of the predicted and exact kernel matrices.}
    \label{fig:gauss_15d_1m_time_kernels}
\end{figure}

The numerical values of the kernels parameters are given in Appendix in Tables \ref{table:config_gauss_high_dimension_part1} and \ref{table:config_gauss_high_dimension_part2}.
The fits of kernels are displayed in Figure \ref{fig:gauss_15d_1m_time_kernels}. The model is able to capture both unimodal and bimodal shapes of the kernels.

\paragraph{Non-multiplicative bimodal Gaussian kernel.} As an ultimate test, we validate the ability of our estimation method to capture non-multiplicative relationships between the time component and the mark component of the kernel. We set $D=2$, $M=10$ and simulate $5.10^6$ events of a linear marked Hawkes process with non-multiplicative kernel functions that we define, for $1\leq i,j\leq D, \hspace{0.1cm}t\geq 0, \hspace{0.1cm}x\in\mathcal{X}$, as
\begin{equation}\label{eq:non_multiplicative_gauss_kernel}
    \phi^{ij}(t,x):=\frac{\alpha_{ij}}{2\sqrt{2\pi}}\left(\frac{1}{\underbar{$\sigma$}_{ij}}\exp\left(-\frac{1}{2}\left(\frac{t-\underbar{$\mu$}_{ij}}{\underbar{$\sigma$}_{ij}}\right)^2\right)+\frac{1}{\overline{\sigma}_{ij}}\exp\left(-\frac{1}{2}\left(\frac{t-m_{ij}(x)}{\overline{\sigma}_{ij}}\right)^2\right)\right),
\end{equation}

where $m_{ij}(x):=\frac{x-1}{M}\underbar{$\mu$}_{ij}+\frac{M-x+1}{M}\overline{\mu}_{ij}$.

Thus, this kernel is non-multiplicative in the sense that it cannot be decomposed as a product of the form $\phi(t,x)=\varphi(t)f(x)$. The parameters are given in Table \ref{table:config_non_multiplicative_kernel}. The second order statistics are estimated using a linear grid $\{\frac{k}{n}\}_{1\leq k\leq n}$ over $[0,1]$, with $n=75$. The baseline intensity is $\mu=[0.05, 0.05]'$.

\begin{table}[!htb]
    \caption{\textit{Non-multiplicative bimodal Gaussian kernel} --- Parameter configuration of the simulation.}
    \begin{subtable}{.2\linewidth}
      \centering
        \caption{$\alpha$}
        \begin{tabular}{c|cc}
           \toprule
           $\alpha_{ij}$ & 1 & 2 \\ 
           \midrule
            1 & 0.5 & 0.1 \\
            2 & 0.1 & 0.2 \\
           \bottomrule
       \end{tabular}
    \end{subtable}%
    \begin{subtable}{.2\linewidth}
      \centering
        \caption{$\underbar{$\mu$}$}
        \begin{tabular}{c|cc}
           \toprule
           $\underbar{$\mu$}_{ij}$ & 1 & 2 \\ 
           \midrule
            1 & 0.05 & 0.15 \\
            2 & 0.25 & 0.15 \\
           \bottomrule
       \end{tabular}
    \end{subtable}%
    \begin{subtable}{.2\linewidth}
      \centering
        \caption{$\underbar{$\sigma$}$}
        \begin{tabular}{c|cc}
           \toprule
           $\underbar{$\sigma$}_{ij}$ & 1 & 2 \\ 
           \midrule
            1 & 0.1 & 0.05 \\
            2 & 0.2 & 0.1 \\
           \bottomrule
       \end{tabular}
    \end{subtable}%
    \begin{subtable}{.2\linewidth}
      \centering
        \caption{$\overline{\mu}$}
        \begin{tabular}{c|cc}
           \toprule
           $\overline{\mu}_{ij}$ & 1 & 2 \\ 
           \midrule
            1 & 0.5 & 0.7 \\
            2 & 0.6 & 0.8 \\
           \bottomrule
       \end{tabular}
    \end{subtable}%
    \begin{subtable}{.2\linewidth}
      \centering
        \caption{$\overline{\sigma}$}
        \begin{tabular}{c|cc}
           \toprule
           $\overline{\sigma}_{ij}$ & 1 & 2 \\ 
           \midrule
            1 & 0.1 & 0.2 \\
            2 & 0.075 & 0.1 \\
           \bottomrule
       \end{tabular}
    \end{subtable}%
    \label{table:config_non_multiplicative_kernel}
\end{table}

\begin{figure}[!ht]
    \centering
    \includegraphics[width=1.\linewidth]{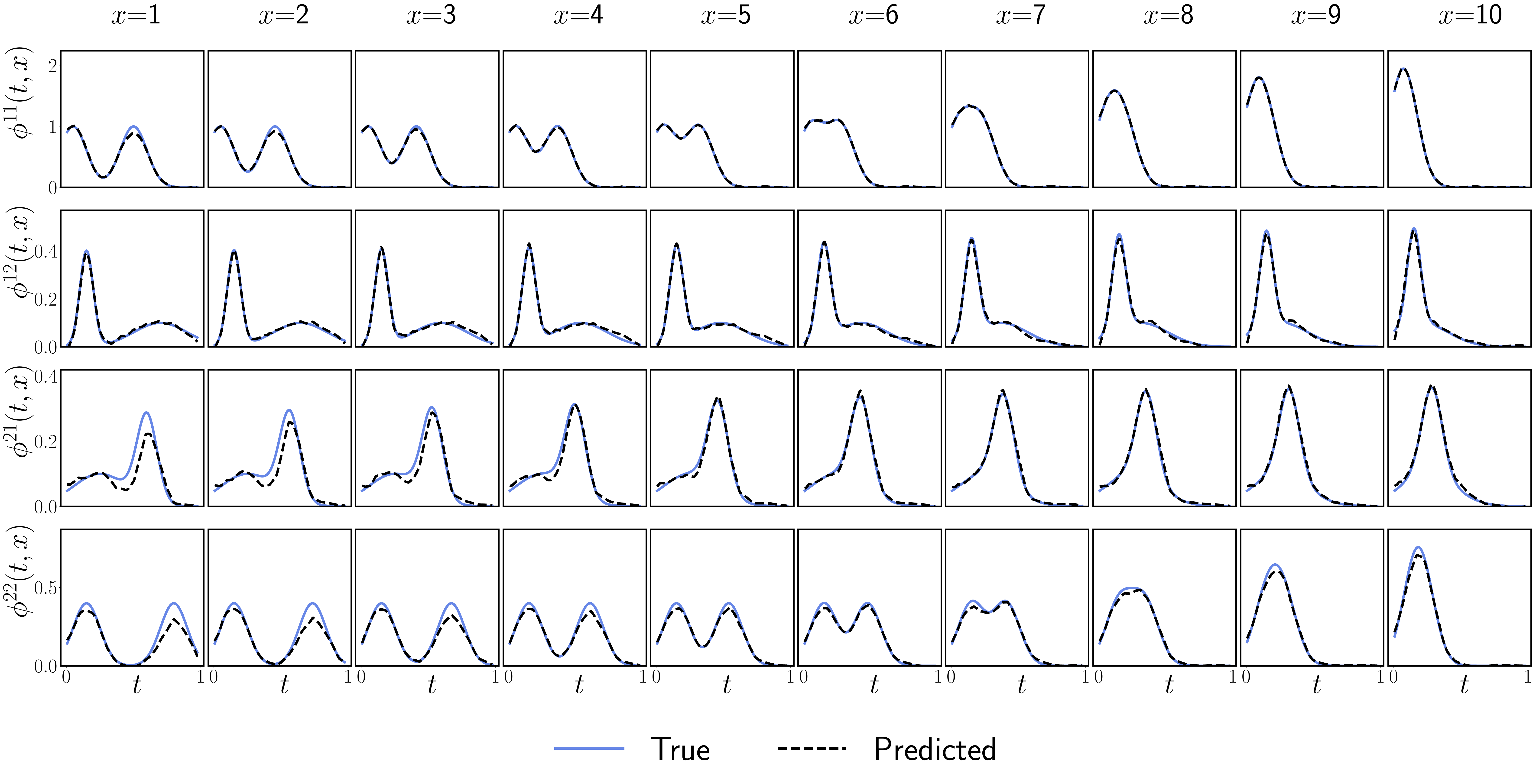}
    \caption{\textit{Non-multiplicative bimodal Gaussian kernel} --- Comparison of the predicted and exact kernel matrices in each mark configuration.}
    \label{fig:non_multiplicative_gauss_2d_10m_kernels}
\end{figure}

The fits of aggregated time kernels and mark kernels are displayed in Figure \ref{fig:non_multiplicative_gauss_2d_10m_kernels}. We observe that the model is able to reproduce the non-multiplicative property of the kernel as the results are satisfactory for all $x\in\mathcal{X}$.

\subsection{Convergence rate with respect to the number of observations}
\label{subsec:CvgceRate}

We measure the rate of decay of the estimation error of the moment-based non-parametric neural Hawkes method for smooth kernels with respect to the number of observations. We set $D=2$, $M=2$ and simulate a linear marked Hawkes process with exponential and power-law kernels given by Equations \eqref{eq:exponentially_decreasing_kernel} and \eqref{eq:slowly_decreasing_kernel}. The numerical values of the parameters are given in Appendix in Table \ref{table:config_power_cvgce_nevents}.

For each case, we simulate a sample of $N$ events and apply the moment-based non-parametric neural Hawkes estimation method. We then compute for the estimated kernels $\hat\phi_N^{i,j}$ a root mean squared error (using a regular subdivision of $[0,T]$ of size $K$):
\begin{equation}
    \Delta_2(N) :=\sqrt{\frac{1}{D^2M(K+1)}\sum_{i,j=1}^D \sum_{m=1}^M \sum_{k=0}^K \left(\hat\phi_N^{i,j}\left(\frac{kT}{K},m\right)-\phi^{i,j}\left(\frac{kT}{K},m\right)\right)^2}.
\end{equation}
For ease of comparison, we also define a normalized version $\overline{\Delta_2}(N)$ of this error by
\begin{equation}
    \overline{\Delta_2}(N) := \frac{\Delta_2(N)}{\sup_{i,j,m} \|\phi^{ij}(\cdot,m)\|_{\infty}}.
\end{equation}
Similarly, we define a $L^\infty$-error
\begin{equation}
    \Delta_\infty(N) := \sup_{i,j,m,k} \left|\hat\phi_N^{i,j}(kT/K,m)-\phi^{i,j}(kT/K,m)\right|,
\end{equation}
and its corresponding normalized error
\begin{equation}
    \overline{\Delta_\infty}(N) := \frac{\Delta_\infty(N)}{\sup_{i,j,m} \|\phi^{ij}(\cdot,m)\|_{\infty}}.
\end{equation}

\begin{figure}
    \centering
    \includegraphics[width=0.5\linewidth]{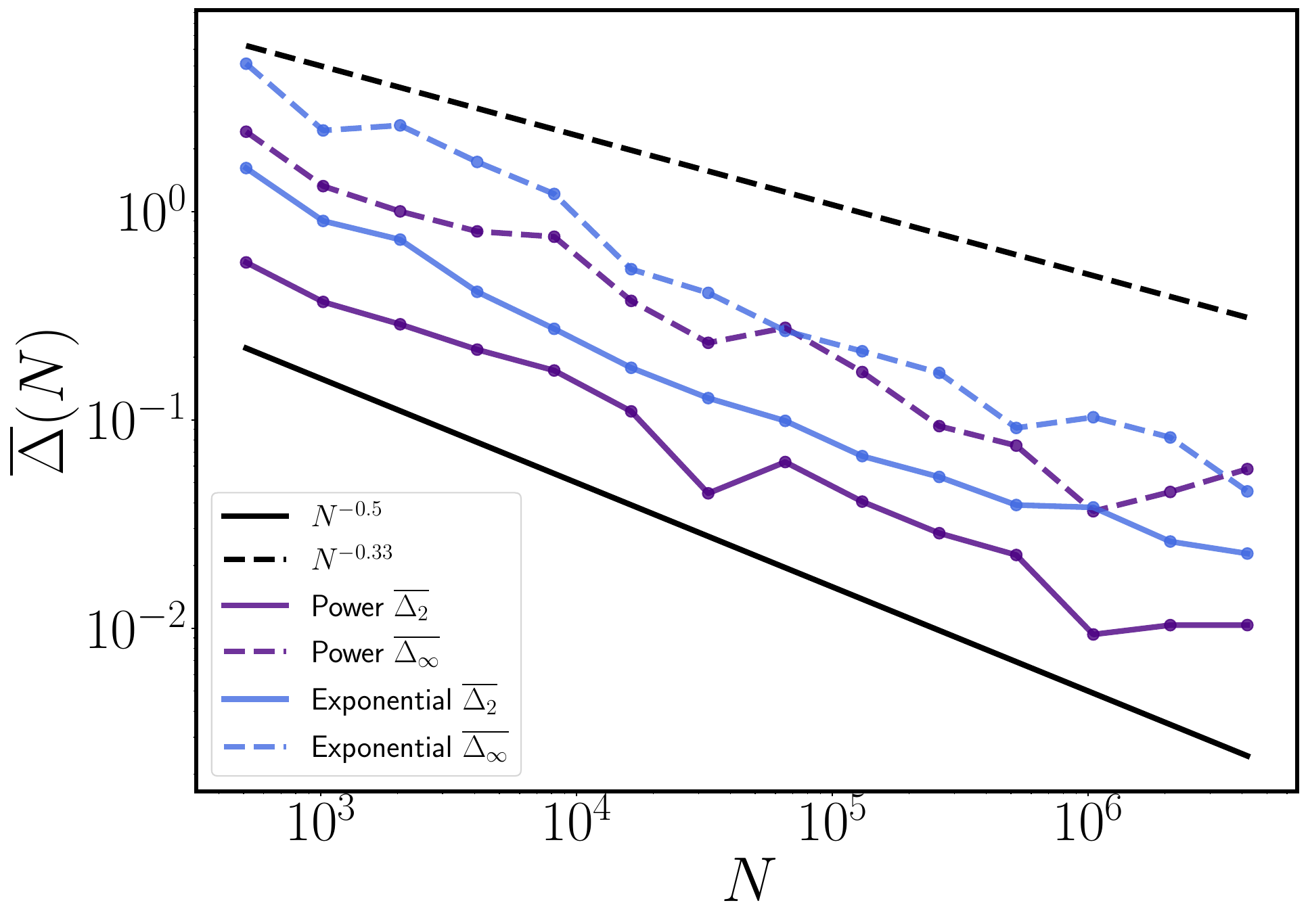}
    \caption{\textit{Robustness experiment} --- Decays of the normalized errors as functions of the number of events $N$ for both power and exponential experiments. We observe a decay rate $\sim N^{-0.5}$ over the 4 decades.}
    \label{fig:robustness_error_vs_nevents}
\end{figure}

Figure \ref{fig:robustness_error_vs_nevents} plots the normalized errors $\overline{\Delta_2}$ and $\overline{\Delta_\infty}$ with respect to the number of events $N$ for both kernels. We empirically observe a decay rate $\sim N^{-0.5}$ over several decades. Note that these results are obtained with the standard configuration of Table \ref{table:hyperparameters_config} without any fine tuning. It was shown in \cite{bacry2016first} that if one assumes that kernels are bounded Lipshitz continuous functions, then the rate of convergence of the $L^\infty$ error on the kernel by the Wiener-Hopf method is rather $\sim N^{-0.33}$ (empirically validated in the same paper on an exponential kernel). 

\subsection{Robustness with respect to the choice of hyperparameters}

We now test the robustness of the moment-based neural Hawkes estimation method  with respect to the choice of various hyperparameters. We simulate the linear marked Hawkes process with power-law kernels defined at Equation \eqref{eq:slowly_decreasing_kernel} and Table \ref{table:config_power_cvgce_nevents} with $N=2.10^6$ events and apply our estimation method with varying number of DGM cells (first experiment), varying number of neurons in the DGM cell (second experiment), varying size of the training set (third experiment) and varying batch size (fourth experiment). Note that when we make each hyperparameter vary, the rest of the hyperparameters is left unchanged with respect to the generic configuration given in Table \ref{table:hyperparameters_config}.

\begin{figure}
    \centering
    \subfloat[Number of DGM cells]{%
        \includegraphics[width=0.25\linewidth]{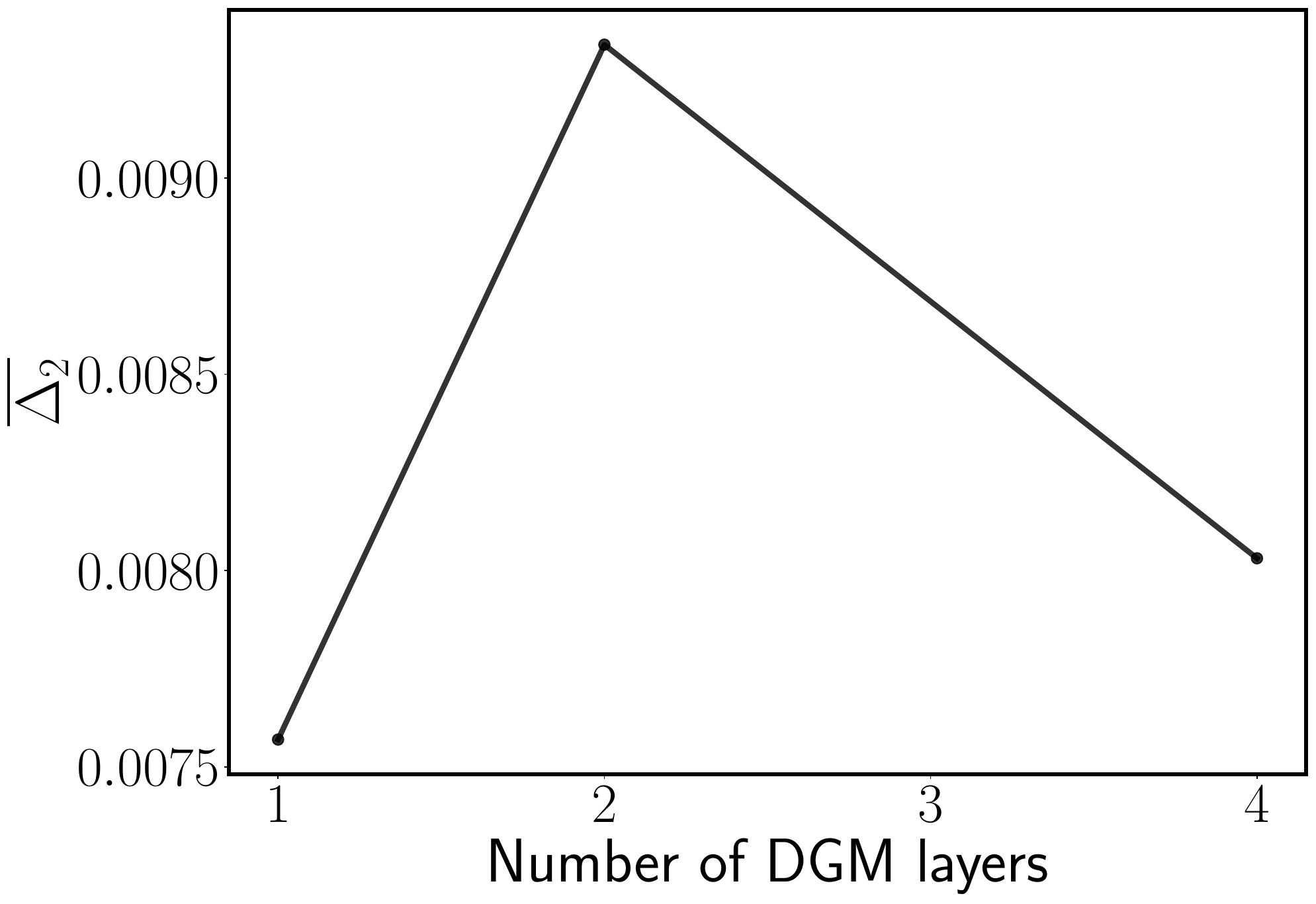}%
    }
    \subfloat[Number of neurons]{%
        \includegraphics[width=0.25\linewidth]{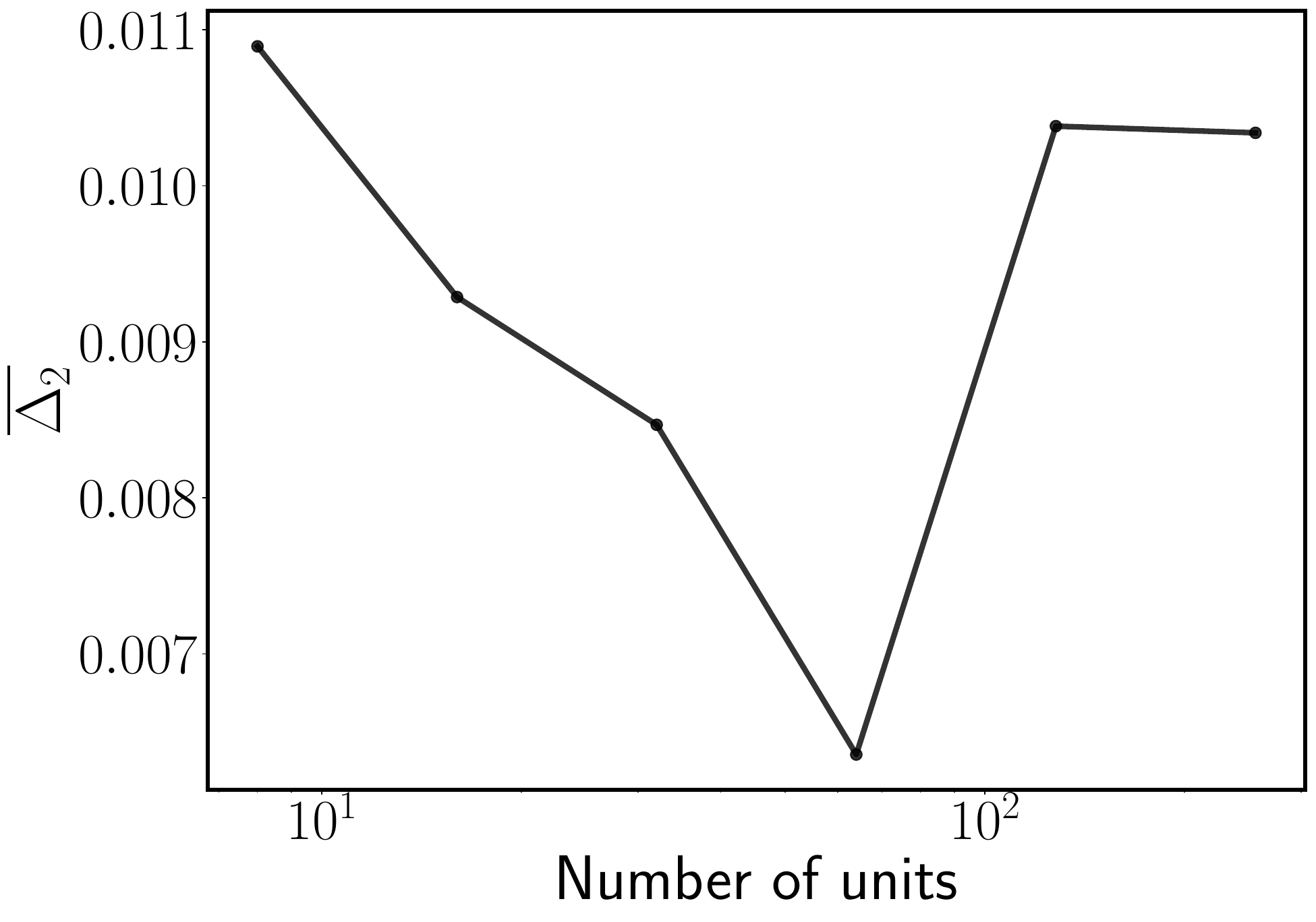}%
    }
    \subfloat[Size of the training set]{%
        \includegraphics[width=0.25\linewidth]{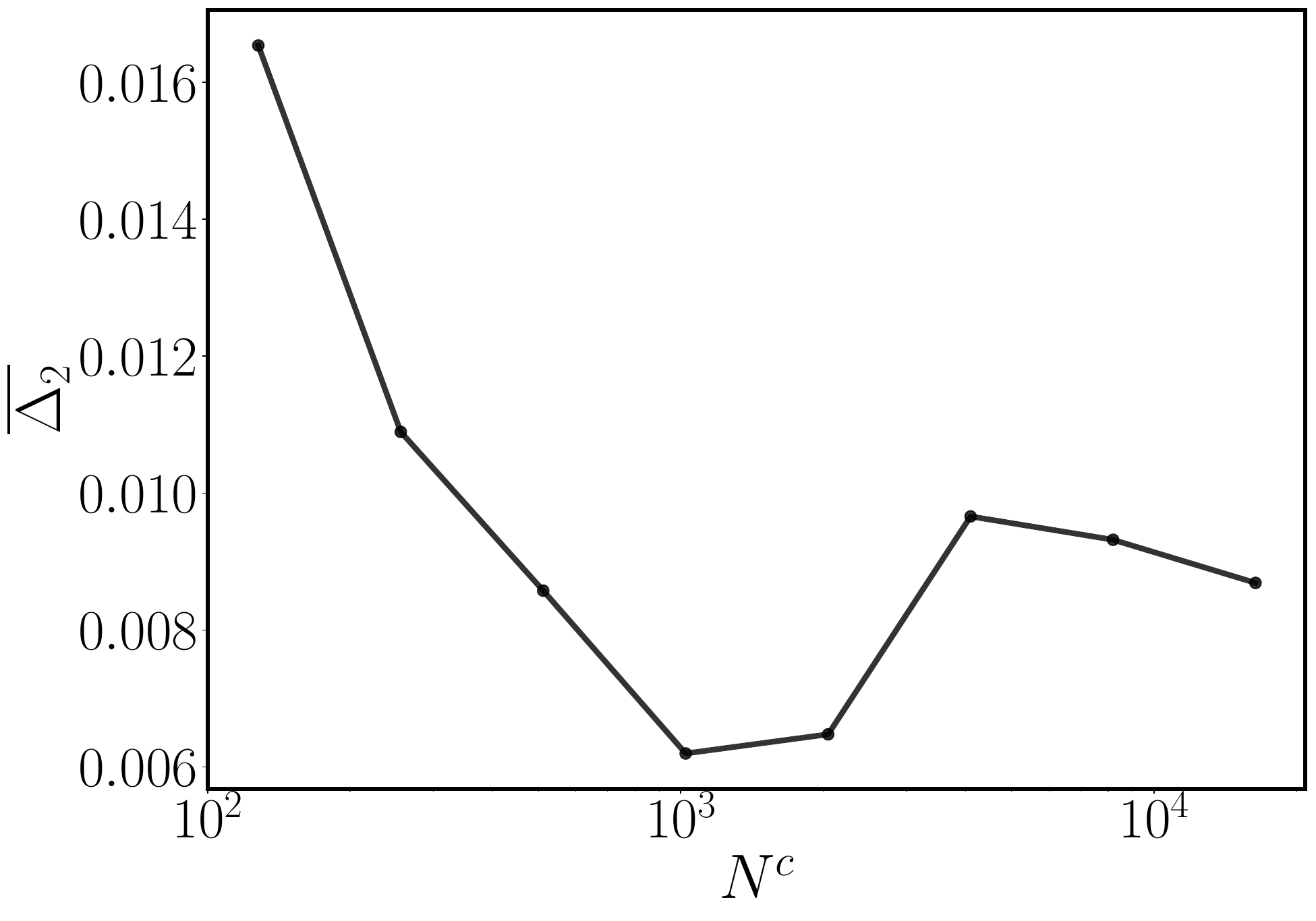}%
    }
    \subfloat[Batch size]{%
        \includegraphics[width=0.25\linewidth]{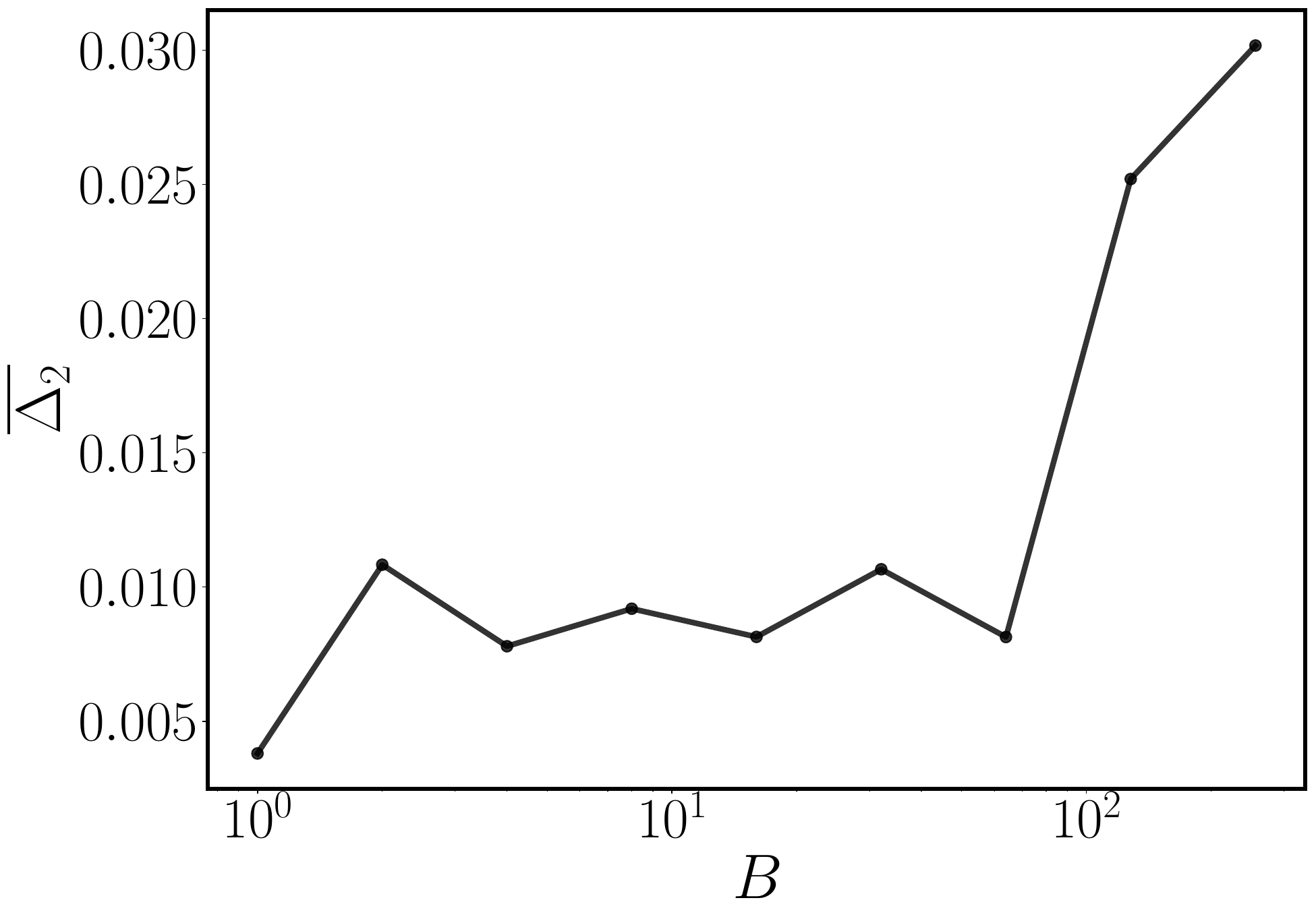}%
    }
    \caption{\textit{Robustness experiment} --- Error $\overline{\Delta_2}$ as a function of the hyperparameters in the four experiments.}
    \label{fig:robustness_l2error}
\end{figure}

\begin{figure}[!ht]
    \centering
    \subfloat[Number of DGM cells]{%
        \includegraphics[width=0.25\linewidth]{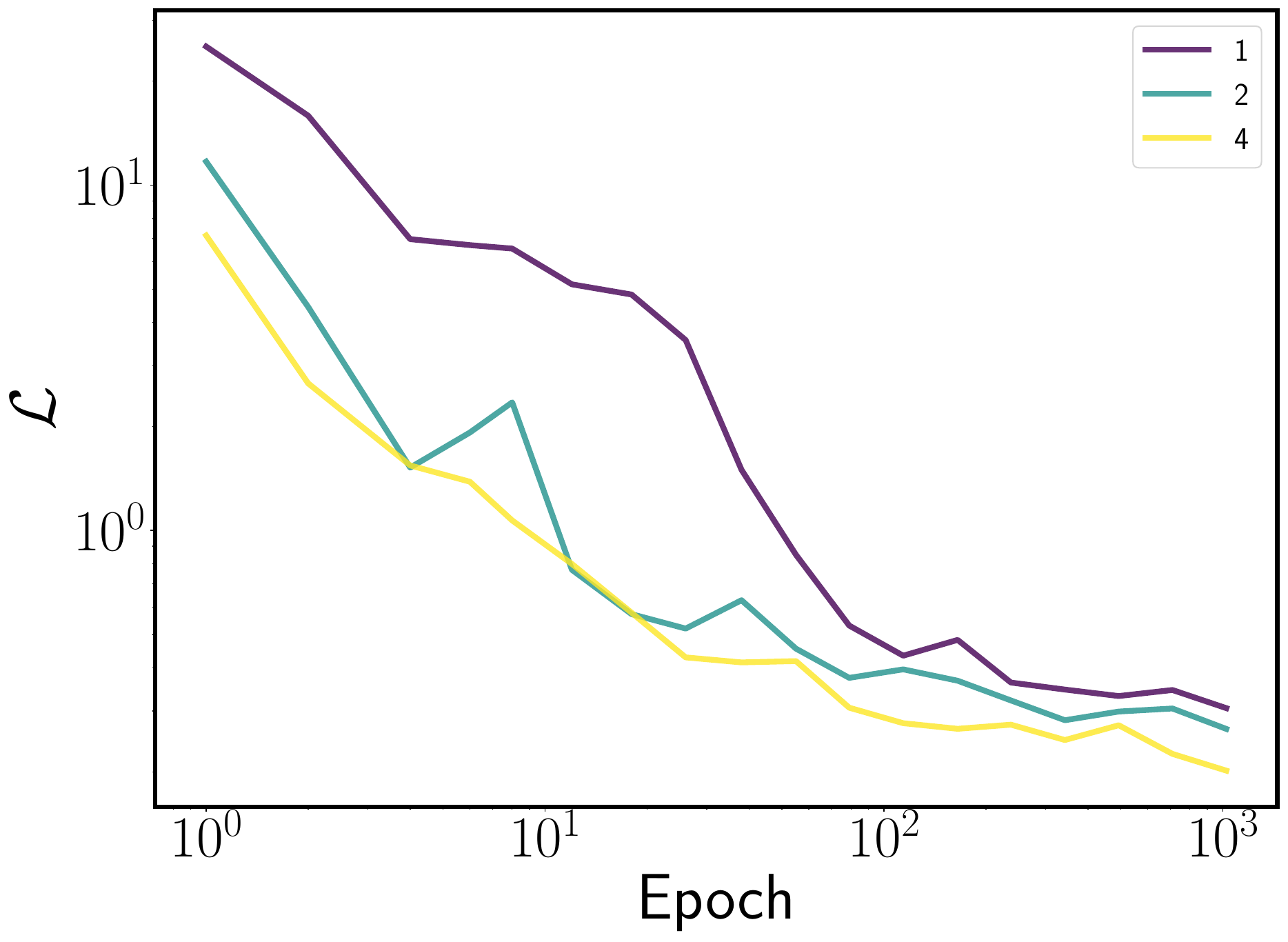}%
    }
    \subfloat[Number of neurons]{%
        \includegraphics[width=0.25\linewidth]{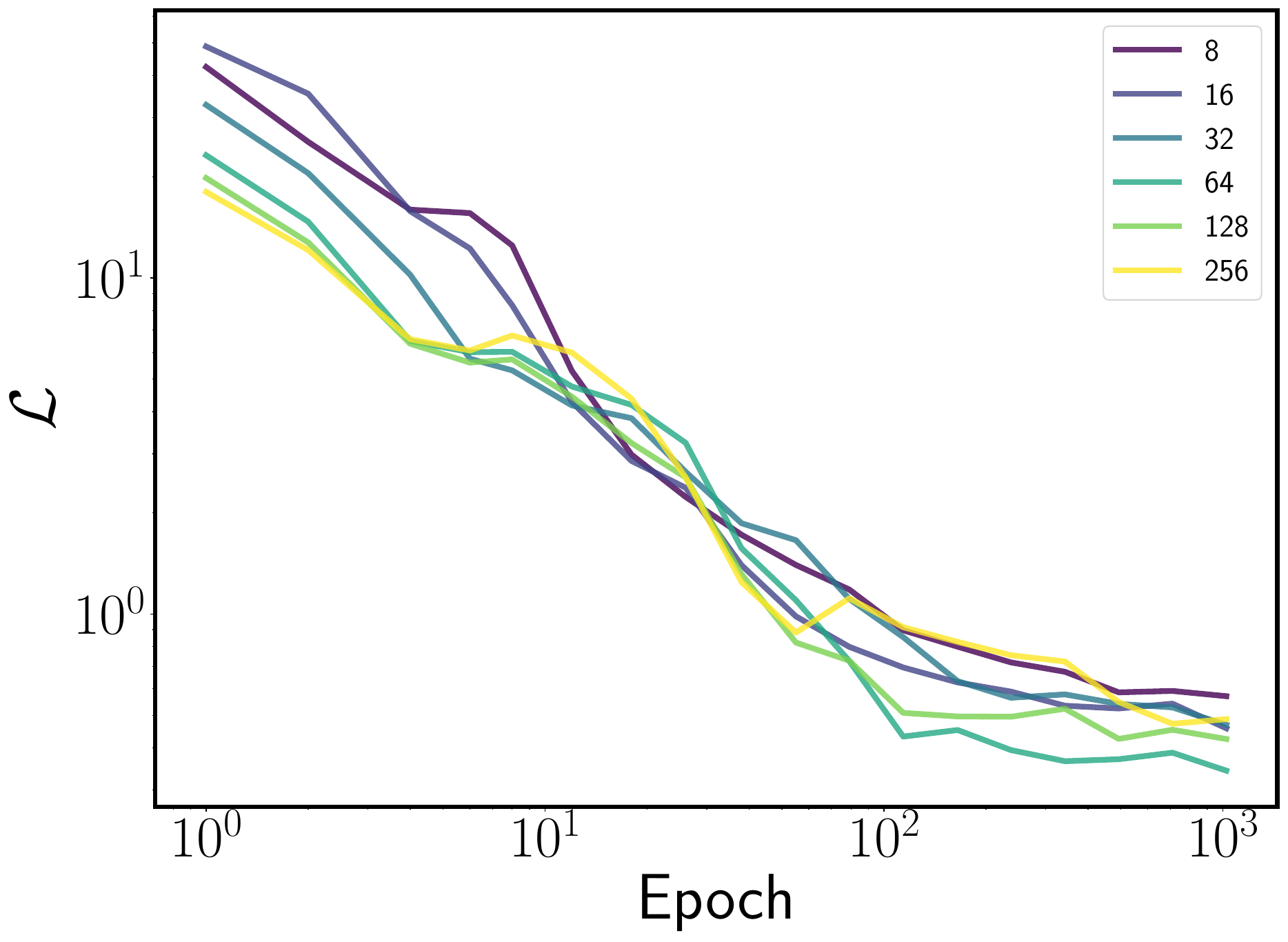}%
    }
    \subfloat[Size of the training set]{%
        \includegraphics[width=0.25\linewidth]{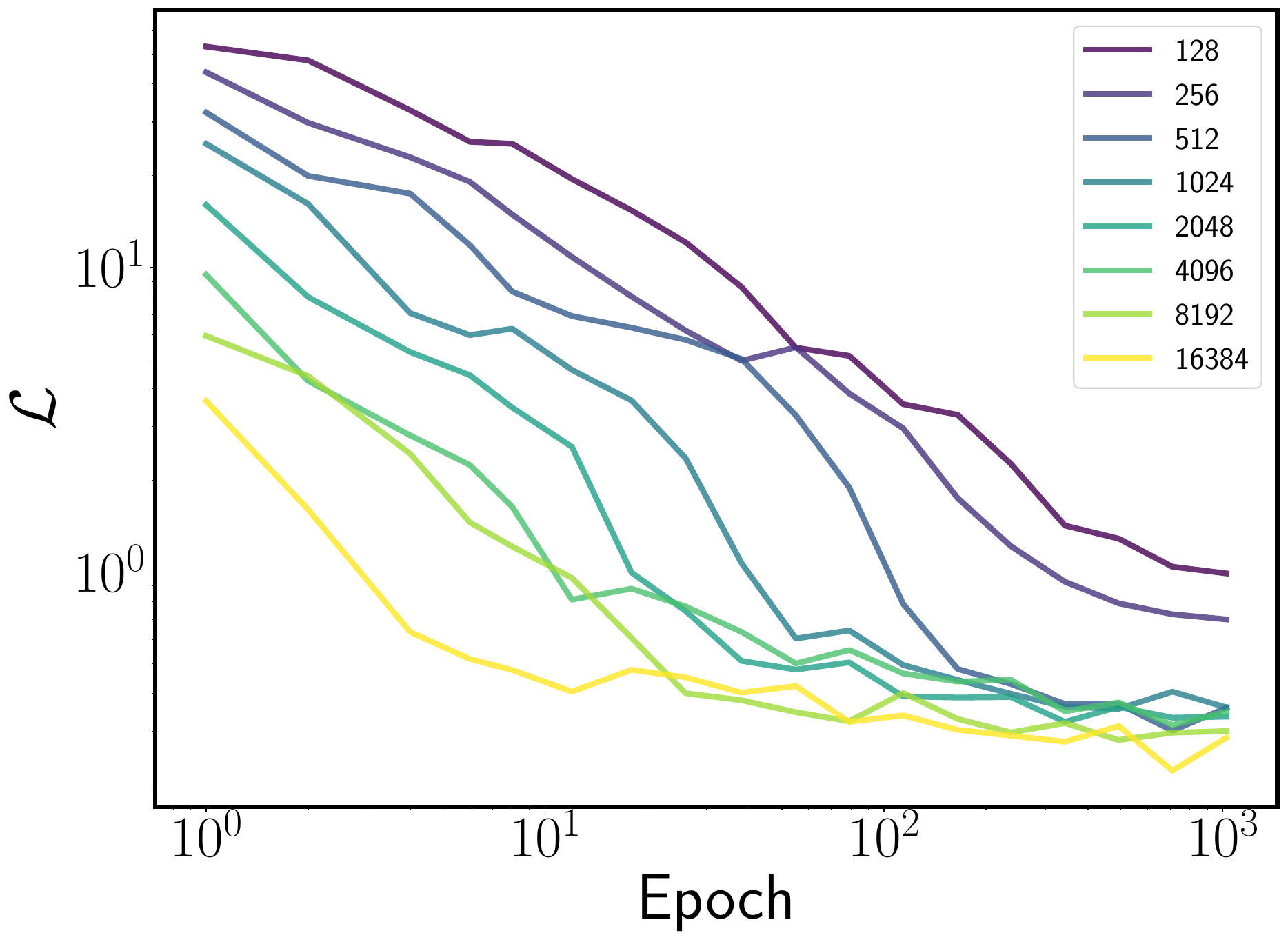}%
    }
    \subfloat[Batch size]{%
        \includegraphics[width=0.25\linewidth]{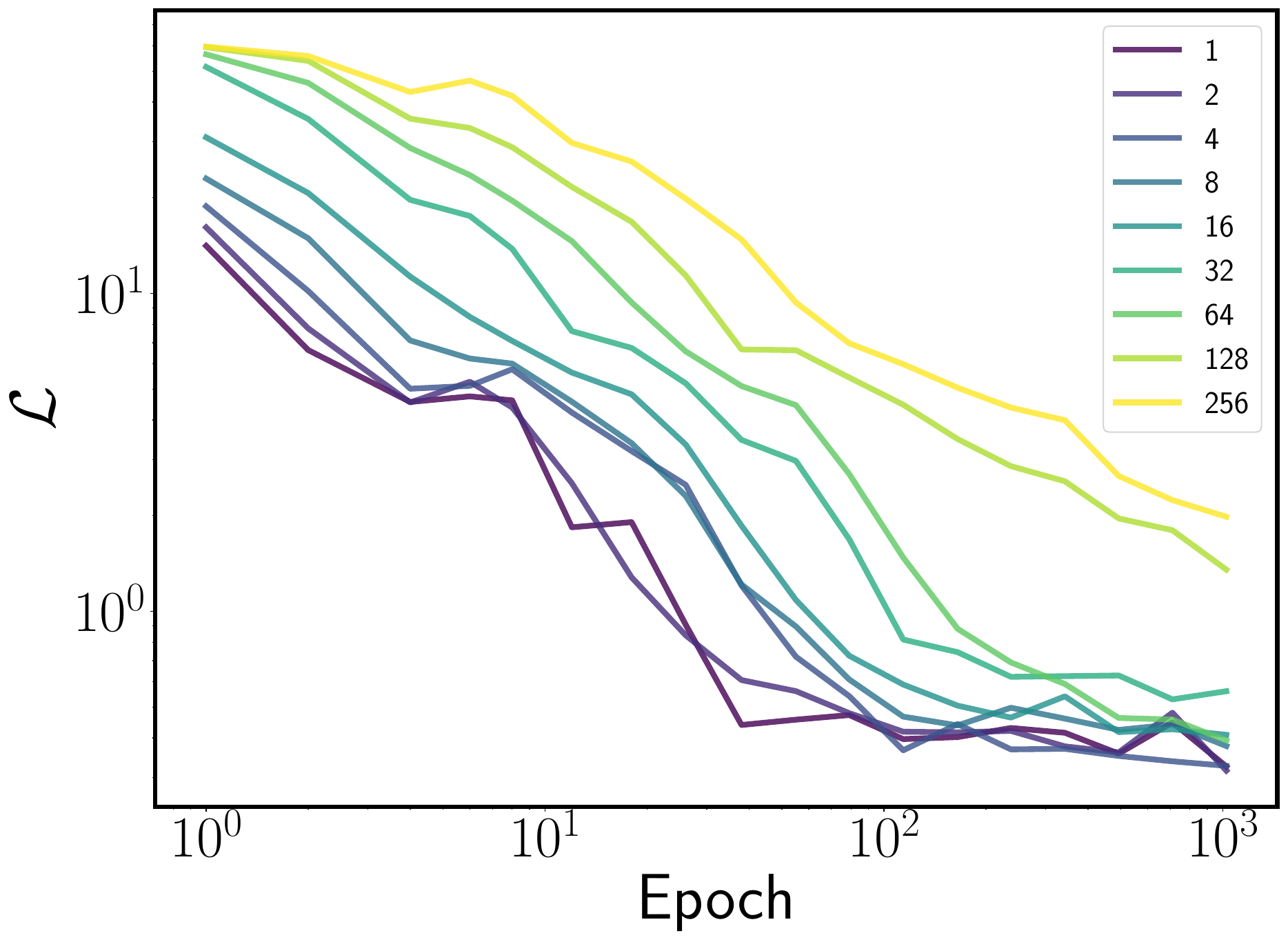}%
    }
    \caption{\textit{Robustness experiment} --- Validation loss as a function of the number of epochs over the four experiments.}
    \label{fig:robustness_losses}
\end{figure}

Figures \ref{fig:robustness_l2error} and \ref{fig:robustness_losses} plot for each experiment the normalized errors $\overline{\Delta_2}$ and the validation losses $\mathcal{L}$ of the trained model for the power law case. Additional material regarding the normalized errors $\overline{\Delta_\infty}$ is displayed in appendix and exhibits similar shapes as $\overline{\Delta_2}$ --- see Figure \ref{fig:robustness_linferror}. The aforementioned losses are the average of the validation losses of the two models $(\hat{\phi}^{1j})_{1\leq j \leq 2}$ and $(\hat{\phi}^{2j})_{1\leq j \leq 2}$, \textit{i.e.} $\mathcal{L}=\frac{1}{2}(\mathcal{L}_1+\mathcal{L}_2)$, across the epochs. We observe that neither the number of cells nor the number of units seem to have a significant impact on the quality of the model. Tuning the number of layers or the number of units does not necessary lead to a solid improvement of the overall performance, but a number of neurons in the range of $[50,100]$ seems a good choice hence justifying our general hyperparameter configuration. Concerning the training set we can formulate a similar conclusion and set a good value in $[1000, 2000]$. By observing losses decays, we can expect that smaller values of the training set should be coupled with a larger number of epochs to ensure that the model explores a sufficient number of collocation points. Thus, the first three experiments do not suggest a high sensitivity of the model to these hyperparameters. Finally, we observe that large values of batch size lowers the learning speed. This result contrasts with the findings of \cite{sankaran2022impact} and may be specific to our problem as the authors explored the case of PDEs and not IEs. We also found that a very small batch size, \textit{i.e.} in the range [1,8], generally improves the performance. All in all, it is very satisfying to observe that the moment-based neural Hawkes estimation method appears robust with respect to the choice of hyperparameters.

\subsection{Stability comparison with the Wiener-Hopf benchmark}

\paragraph{An example with kernels with different orders of magnitude.} We now design a simple experiment to show our non-parametric estimation method can lead to much more stable solutions than the Wiener-Hopf method. We simulate $N=2.10^5$ events of a linear Hawkes process with an exponential kernel in $D=2$ dimensions. We set $M=1$ and the kernel parameters are given in Table \ref{table:config_stability_exponential_kernel}. The baseline intensity is $\mu=[0.05, 0.05]'$. The second order statistics are estimated using the grid of Equation \eqref{eq:g_grid} with parameters $h=0.1$, $n_{\text{lin}}=10$, $n_{\text{log}}=50$.
\begin{table}[!htb]
    \caption{\textit{Benchmark} --- Parameter configuration of the simulation.}
    \begin{subtable}{.5\linewidth}
      \centering
        \caption{$\alpha$}
        \begin{tabular}{c|cc}
           \toprule
           $\alpha_{ij}$ & 1 & 2 \\ 
           \midrule
            1 & 10 & 0.2 \\
            4 & 0.5 & 30 \\ 
           \bottomrule
       \end{tabular}
    \end{subtable}%
    \begin{subtable}{.5\linewidth}
      \centering
        \caption{$\beta$}
        \begin{tabular}{c|cc}
           \toprule
           $\beta_{ij}$ & 1 & 2 \\ 
           \midrule
            1 & 20 & 5 \\
            3 & 2.5 & 40 \\ 
           \bottomrule
       \end{tabular}
    \end{subtable}%
    \label{table:config_stability_exponential_kernel}
\end{table}

\begin{figure}[!ht]
    \centering
    \includegraphics[width=0.75\linewidth]{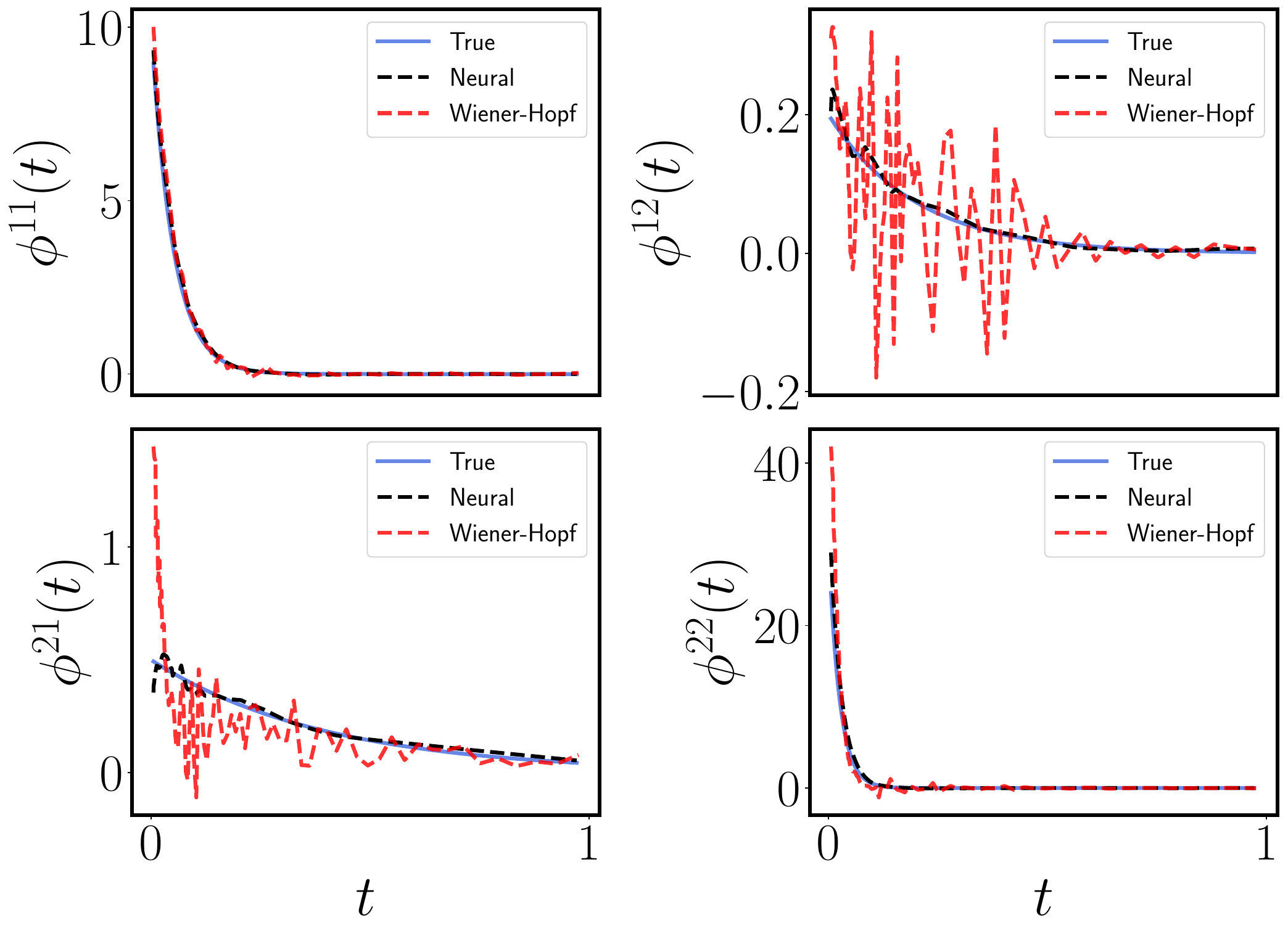}
    \caption{\textit{Benchmark} --- Comparison of the Wiener-Hopf method with the moment-based neural Hawkes.}
    \label{fig:stability_comparison_neural_vs_wh}
\end{figure}

A moment-based neural Hawkes is trained with hyperparameters given in the general configuration setting of Table \ref{table:hyperparameters_config}. 
As a benchmark, we estimate the kernel with the Wiener-Hopf method developed in \cite{bacry2016first}, using our own implementation of the method. We set $Q=200$ a reasonably high number of quadrature points $(t_q)_{q=1,\ldots,Q}$, such that $t_0=0$, $t_Q=T$ and $\delta:=t_{q+1}-t_q=T/(Q-1)$ for any $q$. We then solve the Wiener-Hopf system at each $t_q$, $q=1,\ldots,Q$ to obtain the estimated values $(\phi_q^{ij})_{1\leq q\leq Q}$ of the kernel at time $t_q$. We finally compute the estimated kernel $\widehat{\phi}^{ij}$ at any point $t$ using the equation
\begin{equation}
    \widehat{\phi}^{ij}(t)=G^{ij}(t)-\delta\sum_{k=1}^D\sum_{q=1}^{Q}\phi_q^{ik}K^{kj}(t-t_q),
\end{equation}
with
\begin{equation}\label{eq:K_function}
    K^{kj}(t):=G^{kj}(t)\mathds{1}_{\{t>0\}}+\frac{\Lambda^k}{\Lambda^j}G^{jk}(-t)\mathds{1}_{\{t<0\}}.
\end{equation}
The results are displayed in Figure \ref{fig:stability_comparison_neural_vs_wh}. 

The moment-based neural Hawkes estimation method greatly improves the stability of the results. We observe the Wiener-Hopf method gives satisfactory results for large kernels, \textit{i.e.} $\phi^{11}$ and $\phi^{22}$, but leads to unstable solutions for the other kernels. The resulting oscillatory behaviour was also reported by \cite{cartea2021gradient}.

\begin{figure}[!ht]
    \centering
    \subfloat[Second order statistics $(G^{ij}(t))_{1\leq i,j \leq D}$]{%
        \includegraphics[width=0.5\linewidth]{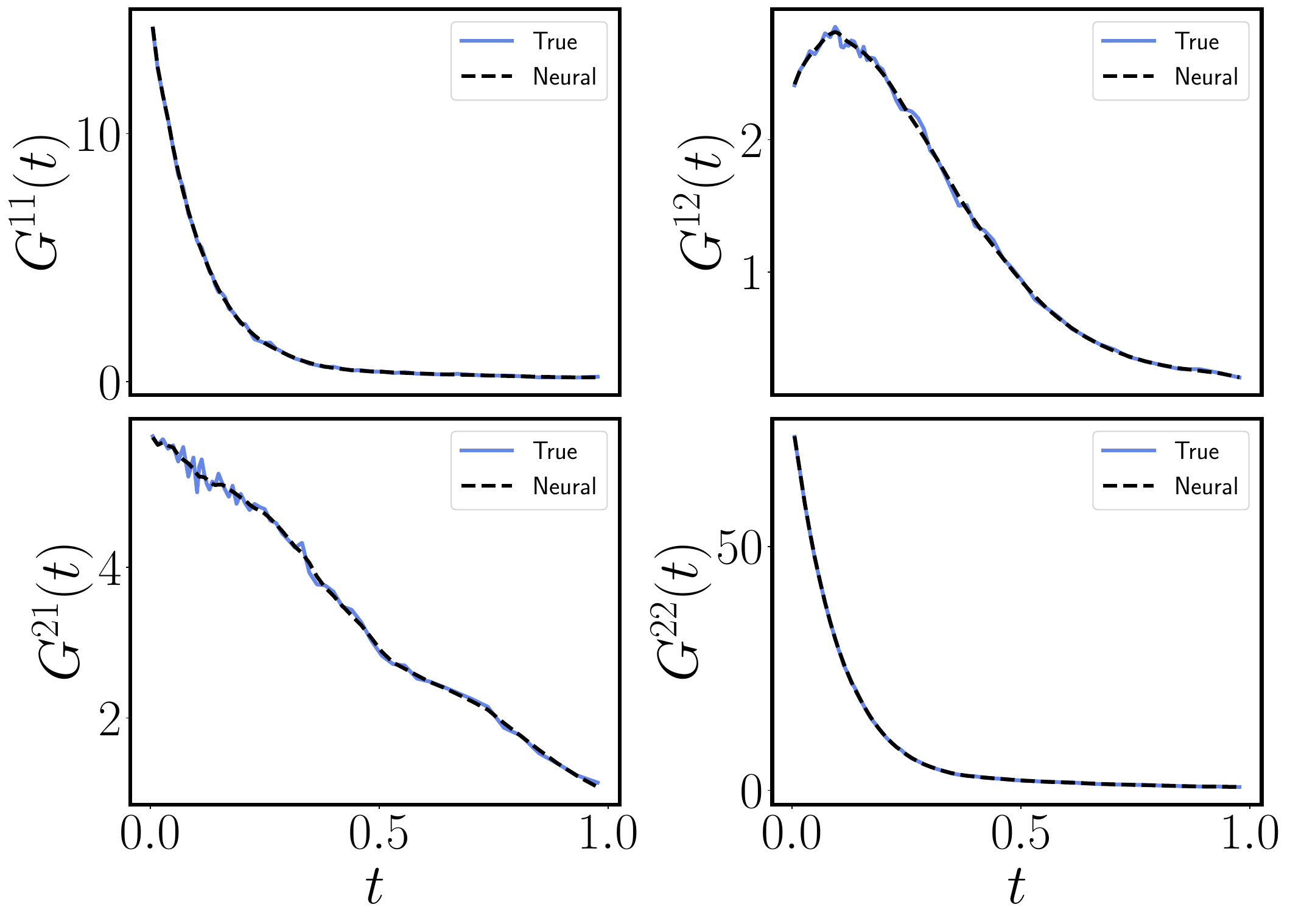}%
    }
    \subfloat[Kernel $(\phi^{ij}(t))_{1\leq i,j \leq D}$]{%
        \includegraphics[width=0.5\linewidth]{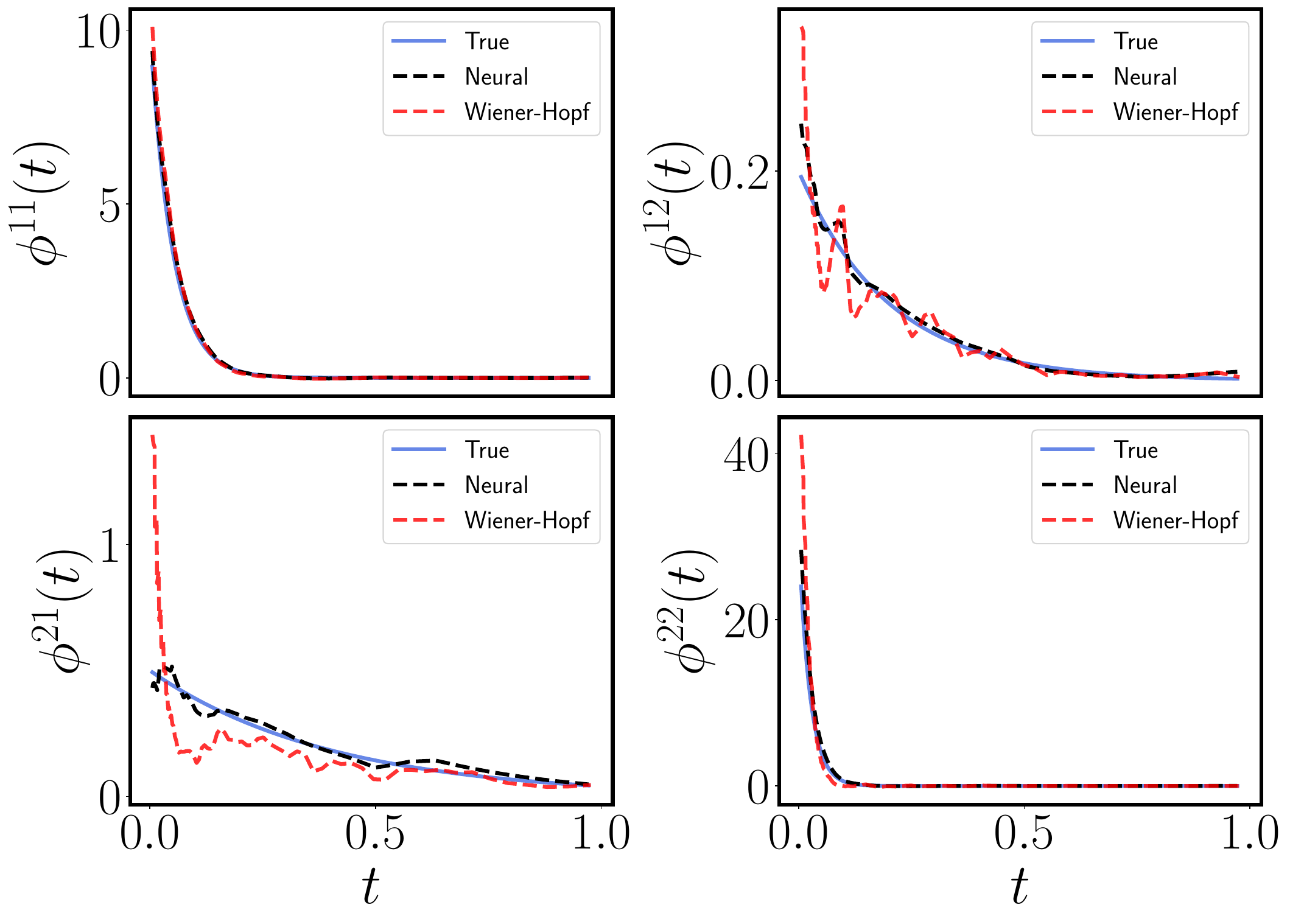}%
    }
    \caption{\textit{Benchmark} --- Fit of the second order statistics with the moment-based neural Hawkes and estimated estimated kernels.}
    \label{fig:stability_comparison_smoothed_neural_vs_wh}
\end{figure}

We now proceed to the following numerical test. We use the estimated moment-based neural Hawkes kernel $(\widetilde{\phi}^{ij})_{1\leq i,j \leq D}$ to compute the fit of the second order statistics 
\begin{equation}
    \widetilde{G}^{ij}(t)=\widetilde{\phi}^{ij}(t)+\sum_{k=1}^D\int_0^T\widetilde{\phi}^{ik}(s)K^{kj}(t-s)\mathrm{d}s, \hspace{0.3cm} 1\leq i,j \leq D, \hspace{0.1cm} 0 \leq t \leq T.
\end{equation}
The computation of functions $K^{ij}$ is done using Equation \eqref{eq:K_function} and the second order statistics $G^{ij}$ that were previously estimated on the data. We then proceed to the estimation of the kernel using the Wiener-Hopf method over the fitted second order statistics $\widetilde{G}^{ij}$ instead of the empirical values of the statistics. The results are displayed in Figure \ref{fig:stability_comparison_smoothed_neural_vs_wh}. The neural Hawkes result remains the best fit, but strikingly the Wiener-Hopf kernel is much less noisy than the previous estimation carried on the raw second order statistics. This numerical experiment suggests that stability issues with Wiener-Hopf may occur depending on the regularity of second order statistics and thus on the binning scheme. Our method appears to depend much less on the irregularities of the estimated statistics that are inherent to the choice of the grid parameters $h$, $n_{\text{lin}}$, $n_{\text{log}}$.

\paragraph{Robustness with respect to the choice of the second order statistics hyperparameters.} In order to assess the sensitivity of our estimation method to the choice of the grid parameters for the computation of the second order statistics, we design the following experiment. We simulate $10^5$ events of the Hawkes process with the non-multiplicative bimodal Gaussian kernel of Equation \eqref{eq:non_multiplicative_gauss_kernel} with parameters given in Table \ref{table:config_non_multiplicative_kernel}. We then proceed to the estimation of the kernels using both the Wiener-Hopf and the neural Hawkes estimation methods, for different second order statistic grid configurations: we use a fully linear grid over the domain $[0,1]$ with varying size $n_{lin}$ ranging from 25 to several hundreds of points. As $n_{lin}$ increases, the number of events per bin decreases and one cannot obtain a smooth estimate of the second order statistic over the whole domain. 

\begin{figure}[!ht]
    \centering
    \includegraphics[width=1.\linewidth]{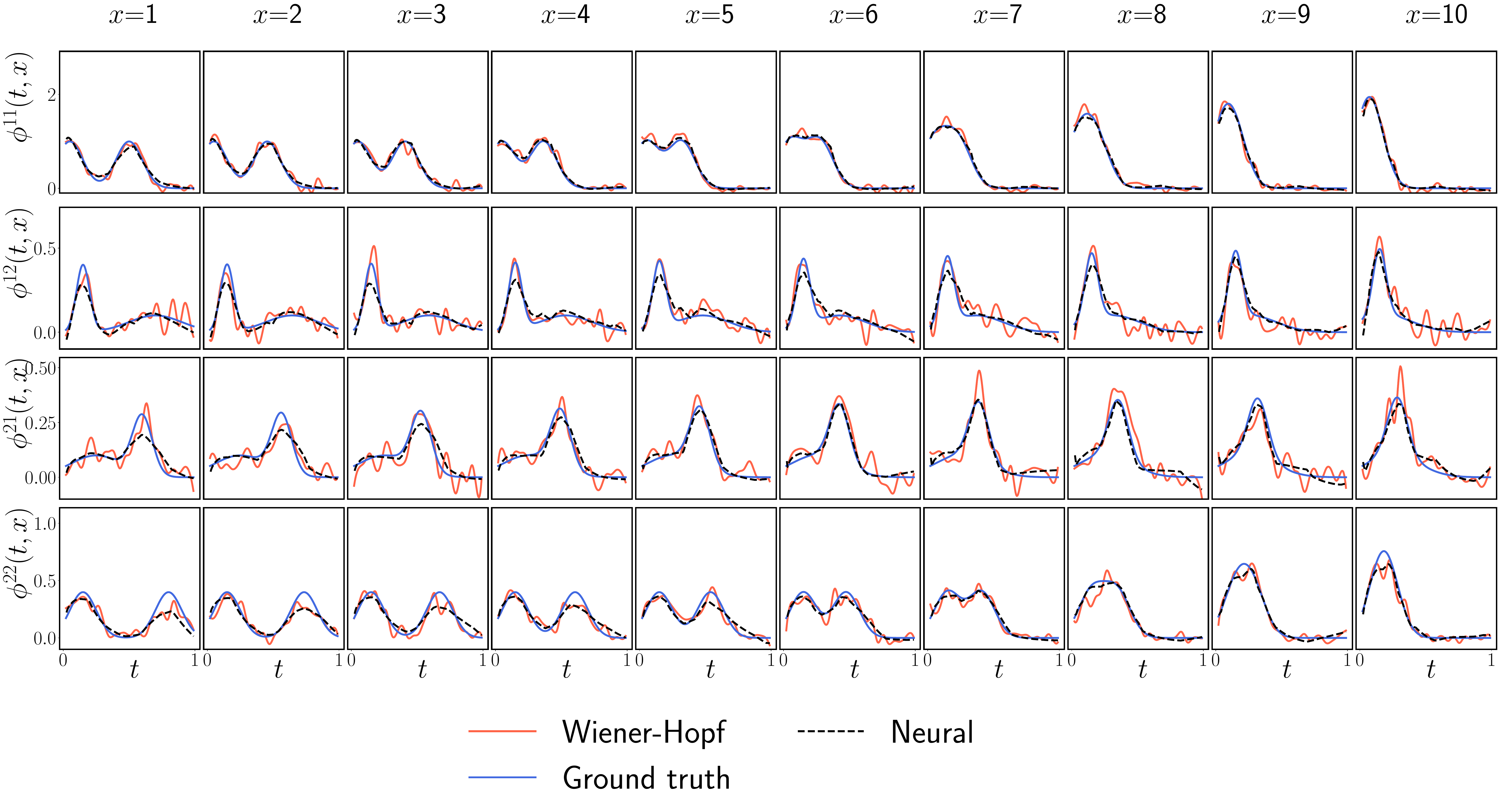}
    \caption{\textit{Robustness to the choice of the second order statistics grid} --- Comparison of the predicted and exact kernel matrices in each mark configuration for the moment-based neural Hawkes estimation method and the Wiener-Hopf methodology. The second order statistics is estimated on a linear grid of 25 points.}
    \label{fig:benchmarking_robustness_statistics_25}
\end{figure}

\begin{figure}[!ht]
    \centering
    \includegraphics[width=1.\linewidth]{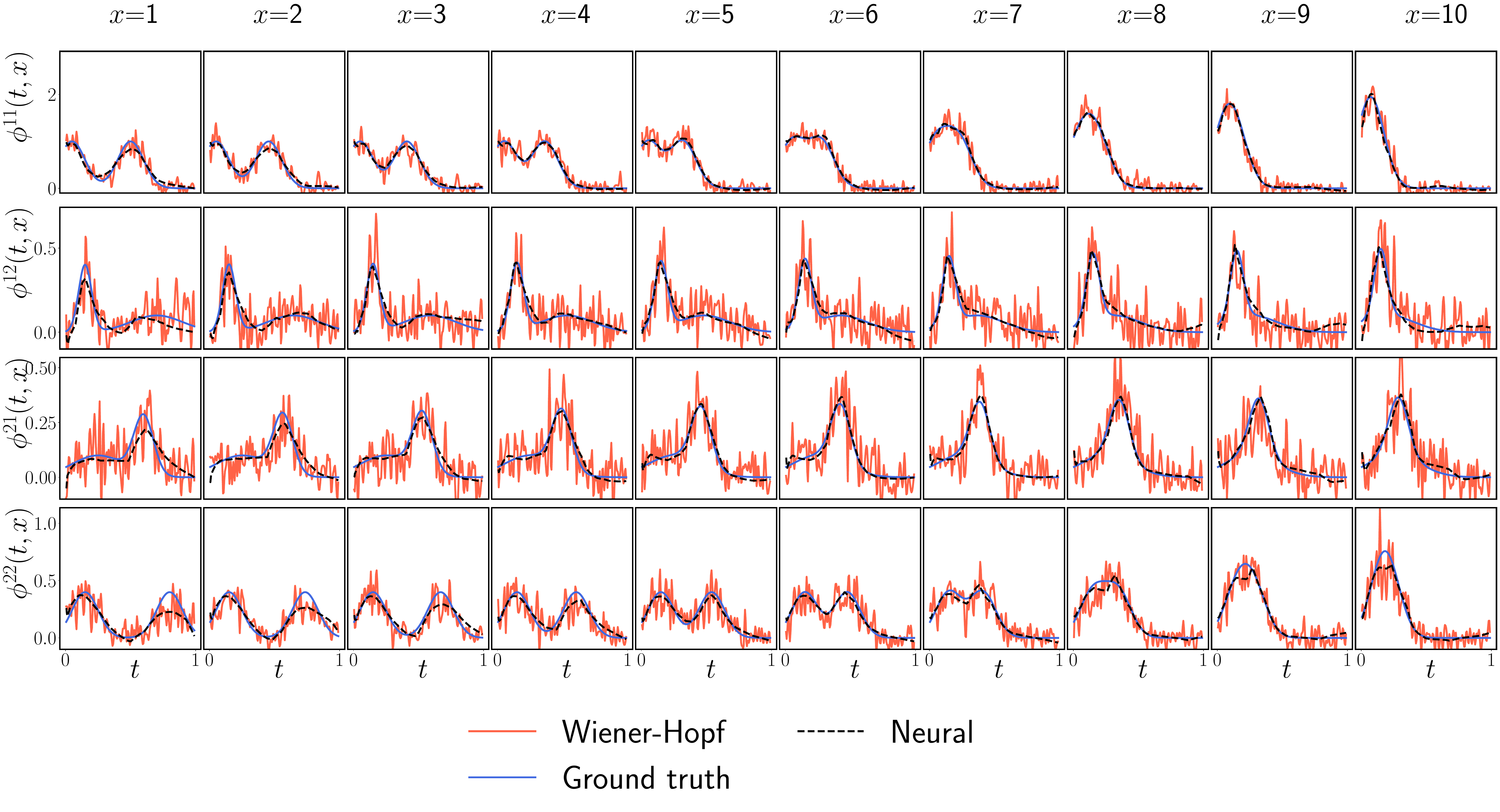}
    \caption{\textit{Robustness to the choice of the second order statistics grid} --- Comparison of the predicted and exact kernel matrices in each mark configuration for the moment-based neural Hawkes estimation method and the Wiener-Hopf methodology. The second order statistics is estimated on a linear grid of 100 points.}
    \label{fig:benchmarking_robustness_statistics_100}
\end{figure}

We display the results in Figure \ref{fig:benchmarking_robustness_statistics_25} for the grid of size $n_{lin}=25$, and in Figure \ref{fig:benchmarking_robustness_statistics_100} for the grid of size $n_{lin}=100$. As expected, we observe that the Wiener-Hopf method is very sensitive to the size of the grid, leading to noisy estimations when the hyperparameter of the second order statistic is poorly chosen. Moreover, for a small size of the grid like in Figure \ref{fig:benchmarking_robustness_statistics_25}, although the fit is quite satisfactory, the Wiener-Hopf method can still generate kernels with oscillatory behaviours which could lead to false discoveries in practical cases, \textit{e.g.} when one seeks to analyze excitation delays and multimodal kernels. Concerning our method, we see that the estimated kernels are much more stable from one grid configuration to another. Even when the second order statistic estimate is noisy, we see that the moment-based neural Hawkes is able to provide very satisfactory results with similar kernel shapes as for smaller grids configurations. We have tested for larger grids up to size 200 and the observation holds: the Wiener-Hopf estimation gets even noisier, whereas our methodology provides similar results. We conclude that the neural Hawkes method is more robust to changes in the second order statistics hyperparameters than Wiener-Hopf, implying that the application of our method to practical cases requires less tuning of the second order statistics grid than in the Wiener-Hopf case.

\section{Application to high-frequency cryptocurrency data}
\label{application_section}

\subsection{Data}

In this section we apply our estimation method to cryptocurrency trade data provided by SUN ZU Lab. The dataset is composed of trades executed between December 1st, 2023 and December 30th, 2023 on Coinbase for 15 cryptocurrency pairs. With every trade is given a timestamp that corresponds to the time at which the matching engine of the market processed the transaction. Timestamps are given with microsecond precision. For each pair, trades that share the same timestamp are aggregated and identified as one single event that was initiated either as a market order or as a marketable limit order. No feed handler disconnection was reported during the time period, ensuring a coherent estimation of the first and second order statistics. Descriptive statistics are displayed in Table \ref{table:coinbase_spillover_descriptive_statistics}.

\begin{table}[!htb]
   \small
   \centering
   \caption{Descriptive statistics of the dataset over the 1 month period.}
   \begin{tabular}{cccc}
   \toprule\toprule
   Pair & Number of events & $\Lambda^i$ (events$.s^{-1}$) & Traded volume (millions USD)\\ 
   \midrule
   BTC-USD & 5,196,281 & 2.00 & 15,803.85\\
   ETH-USD & 3,579,423 & 1.38 & 8,058.49\\
   SOL-USD & 2,734,306 & 1.05 & 5,655.90\\
   ICP-USD & 569,019 & 0.22 & 464.61\\
   SEI-USD & 730,907 & 0.28 & 567.25\\
   AVAX-USD & 1,668,967 & 0.64 & 2,654,.66\\
   XRP-USD & 1,190,267 & 0.46 & 1,091.47\\
   LINK-USD & 1,237,588 & 0.48 & 1,275.48\\
   XLM-USD & 580,121 & 0.22 & 288.27\\
   ADA-USD & 1,436,622 & 0.55 & 1,031.34\\
   MATIC-USD & 896,950 & 0.35 & 753.90\\
   DOT-USD & 810,835 & 0.31 & 415.30\\
   AAVE-USD & 658,330 & 0.25 & 290.75\\
   DOGE-USD & 1,195,724 & 0.46 & 1,174.46\\
   LTC-USD & 967,231 & 0.37 & 405.78\\
   \bottomrule
   \end{tabular}
   \label{table:coinbase_spillover_descriptive_statistics}
\end{table}

\subsection{The impact of volume on the arrival rate}

To illustrate the quality of our moment-based neural Hawkes estimation method in the presence of a marked kernel, we analyze the influence of the traded size on the trade arrival rate. Note that a similar analysis was made in \cite{rambaldi2017role} on traditional financial markets using the Wiener-Hopf estimation method. We focus on the BTC-USD pair, set $D=1$ and $M=15$ and the mark random variable $\xi$ encodes the volume of each trade. The hyperparameter configuration we use is the generic one of Table \ref{table:hyperparameters_config}, we set $T=1$ s and estimate the second order statistics with $h=5.10^{-4}=500$ µs, $n_{\text{lin}}=50$ and $n_{\text{log}}=100$. The discretization of the mark domain and the associated probability mass function $p$ of $\xi$ are displayed in Table \ref{table:coinbase_volume_mark_distribution}. The adopted binning scheme uses a logarithmic grid from $100$ USD to $100,000$ USD. The analysis differs from the one of \citet{rambaldi2017role} as we focus on the shape of the aggregated mark kernel $f$ and the asset classes, and hence the market microstructures, are different. However, our results are consistent in the sense that the shape of the kernel depends on the traded volume. 
\begin{table}
   \small
   \centering
   \caption{\textit{The impact of volume} - Discretization of the mark domain and probability mass function of $\xi$.}
   \begin{tabular}{ccc}
   \toprule\toprule
   m & Volume interval (USD) & $p(m)$\\ 
   \midrule
   1 & (0 - 100] & 0.46\\
   2 & (100 - 170] & 0.046\\
   3 & (170 - 290] & 0.053\\
   4 & (290 - 490] & 0.046\\
   5 & (490 - 840] & 0.043\\
   6 & (840 - 1,425] & 0.053\\
   7 & (1,425 - 2,425] & 0.051\\
   8 & (2,425 - 4,125] & 0.057\\
   9 & (4,125 - 7,000] & 0.078\\
   10 & (7,000 - 12,000] & 0.055\\
   11 & (12,000 - 20,300] & 0.028\\
   12 & (20,300 - 34,500] & 0.015\\
   13 & (34,500 - 58,750] & 0.008\\
   14 & (58,750 - 100,000] & 0.003\\
   15 & (100,000 - $+\infty$] & 0.002\\
   \bottomrule
   \end{tabular}
   \label{table:coinbase_volume_mark_distribution}
\end{table}

\begin{figure}
    \centering
    \subfloat[Kernels $(\phi(t,x))_{x\in\mathcal{X}}$ --- x-axis in logarithmic scale]{%
        \includegraphics[width=0.5\linewidth]{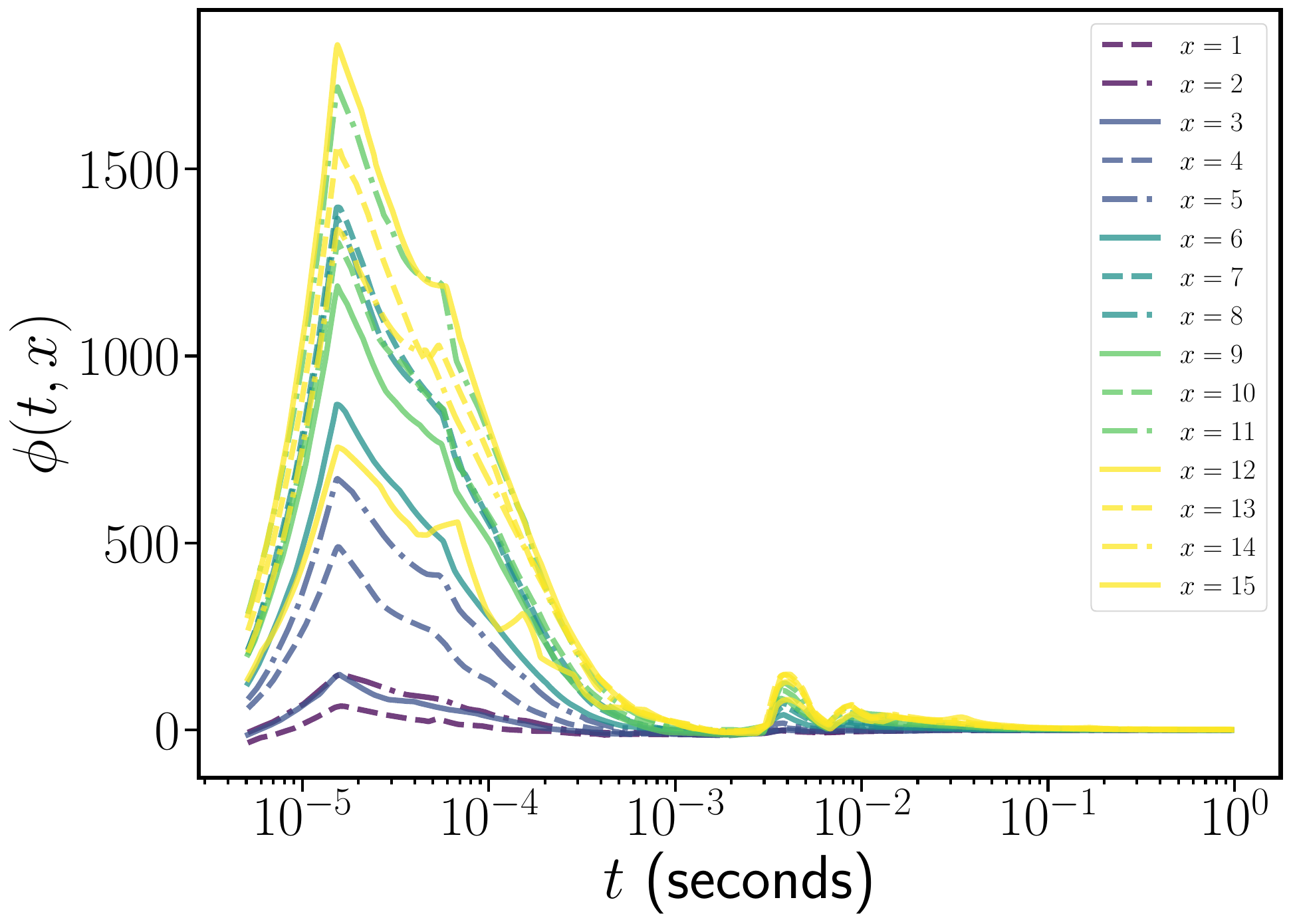}%
    }
    \subfloat[Aggregated mark kernel $f(x)$]{%
        \includegraphics[width=0.5\linewidth]{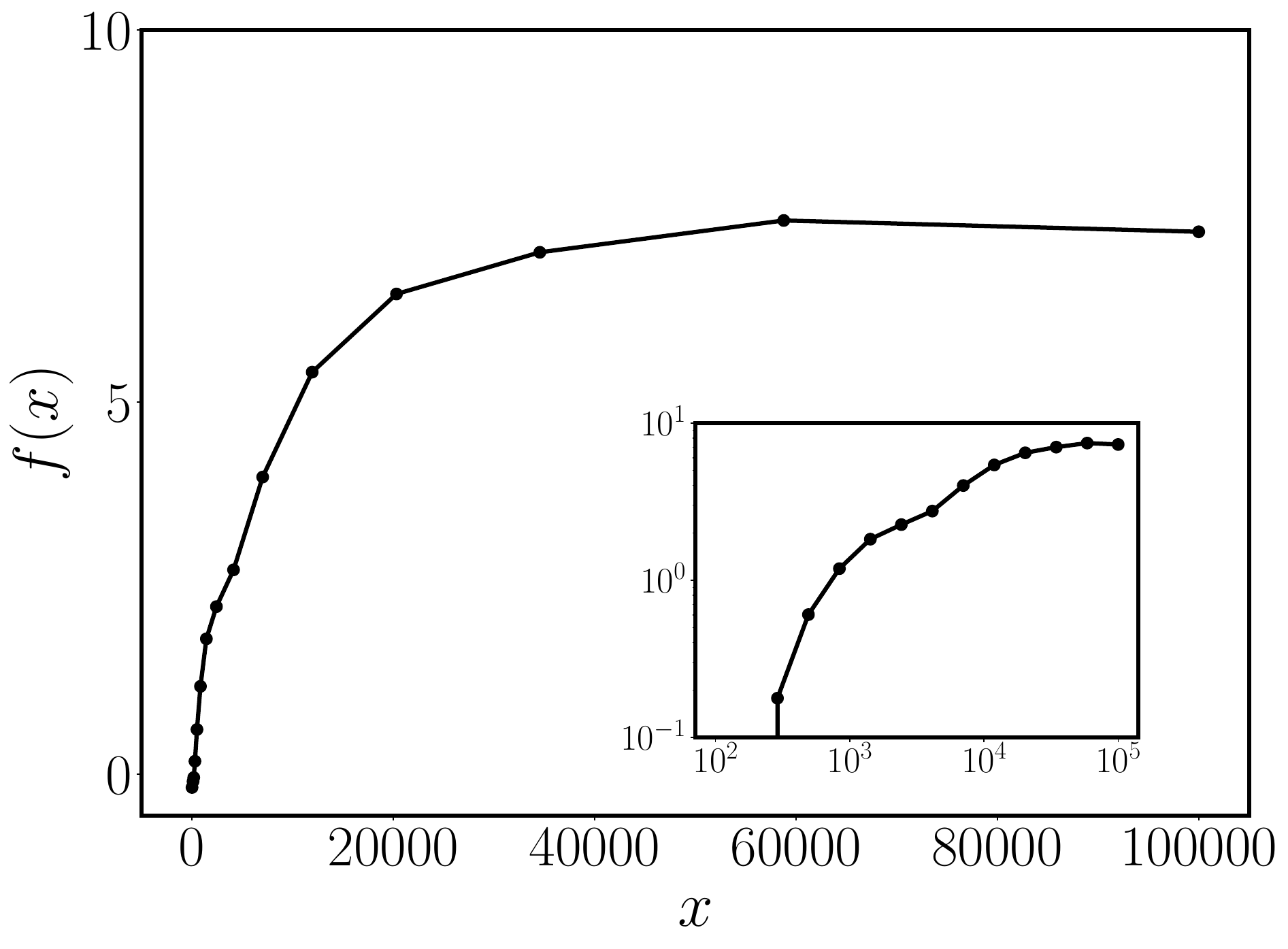}%
    }
    \caption{\textit{The impact of volume} --- Time kernels as functions of time and aggregated mark kernel $f$ as a function of the traded volume (USD). The mark kernel seems to reach a plateau for $x>50,000$.}
    \label{fig:coinbase_volume_kernels}
\end{figure}

Figure \ref{fig:coinbase_volume_kernels} shows the kernels $(\phi(.,m))_{m\in\mathcal{X}}$ and the aggregated mark kernel $f$ computed using Equation \eqref{eq:aggregated_mark_kernel}. We observe that the aggregated kernel is a concave non-decreasing function of the volume and reaches a plateau for sizes that are larger than 50,000 USD. The monotonicity of the function is expected as the larger the size, the greater the amount of information interpreted by the market. A possible explanation of the concavity of the function and its plateau is that the agents consider there is less and less additional information when the size increases and once it reaches a specific value, \textit{i.e.} here 50,000 USD, larger sizes do not bring any additional information.

This finding also provides insights to practitioners regarding the choice of parametric kernels in order flow models. Concerning the kernel functions $(\phi(., x))_{x\in\mathcal{X}}$ they exhibit different latency peaks depending on the traded volume. We see that they all globally share three latency peaks at around 10 µs, 4 to 5 ms and a smaller one near 10 ms. However, for large values of $x$, we observe other peaks: one near 70 µs and one near 200 µs. 

\subsection{Cryptocurrency spillover: a causality analysis of transactions}
\label{subsec:CryptoSpillover}

We analyze the causality relationships between arrival times of transactions amongst the 15 cryptocurrency pairs of Table \ref{table:coinbase_spillover_descriptive_statistics}. To this extent, we proceed to a moment-based non-parametric neural Hawkes estimation in $D=15$ dimensions and set $M=1$. The hyperparameter configuration we use is the generic one of Table \ref{table:hyperparameters_config}, we set $T=1$ s and estimate the second order statistics with $h=5.10^{-4}=500$ µs, $n_{\text{lin}}=50$ and $n_{\text{log}}=100$. Goodness-of-fit is assessed via the comparison of the first and second moments of the fitted process and of the empirical data. Numerical values and the full matrix of the empirical and fitted second order statistics are given in Appendix (Figure \ref{fig:coinbase_spillover_fit_time_kernels}).

\begin{figure}
    \centering
    \subfloat[Branching ratio matrix \\\centering$(\|\phi^{ij}\|)_{1\leq i,j\leq D}$]{%
        \includegraphics[width=0.33\linewidth]{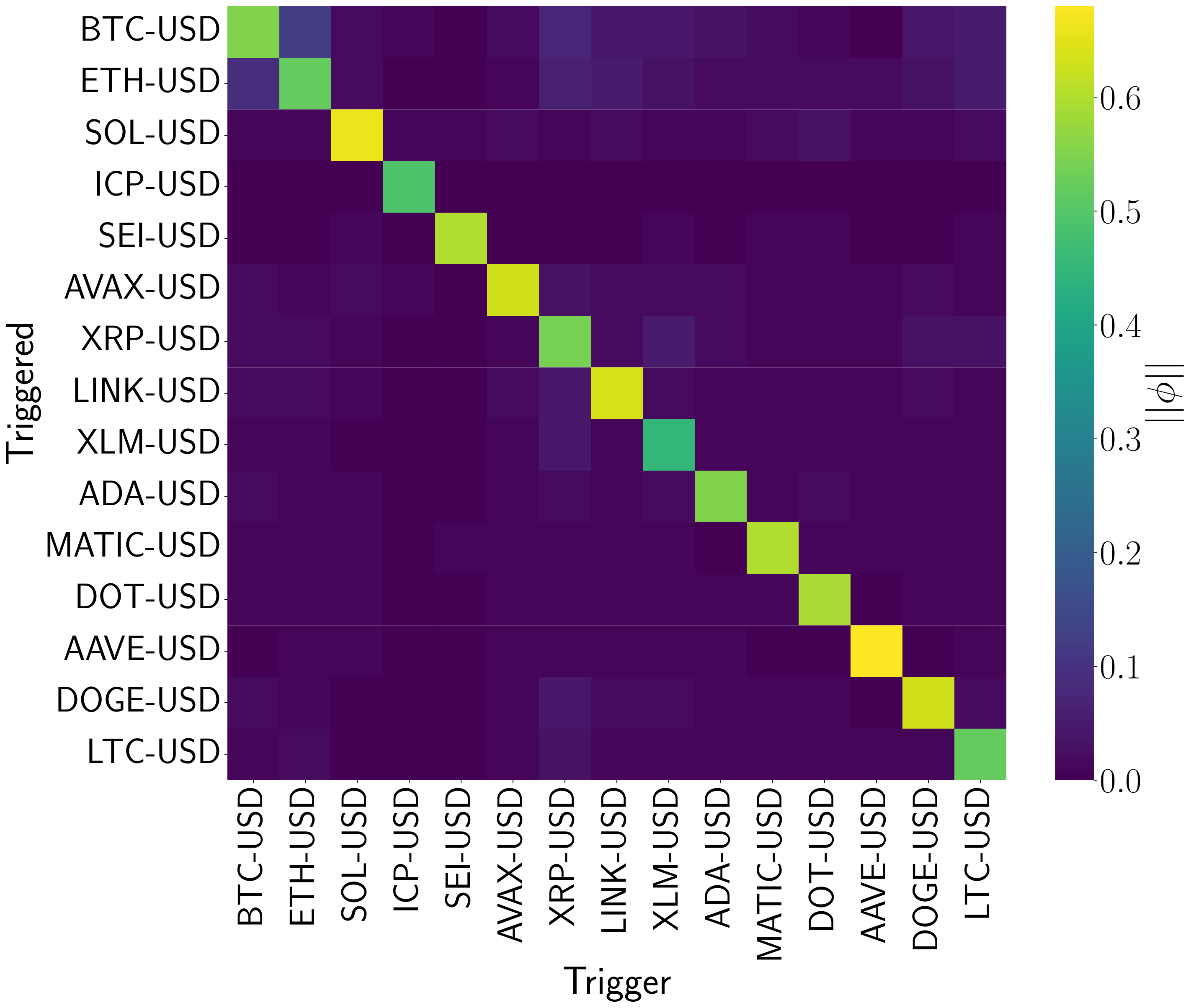}%
    }
    \subfloat[Branching ratio matrix \\\centering$(\|\phi^{ij}\|)_{1\leq i,j\leq D}$ --- diagonal removed]{%
        \includegraphics[width=0.33\linewidth]{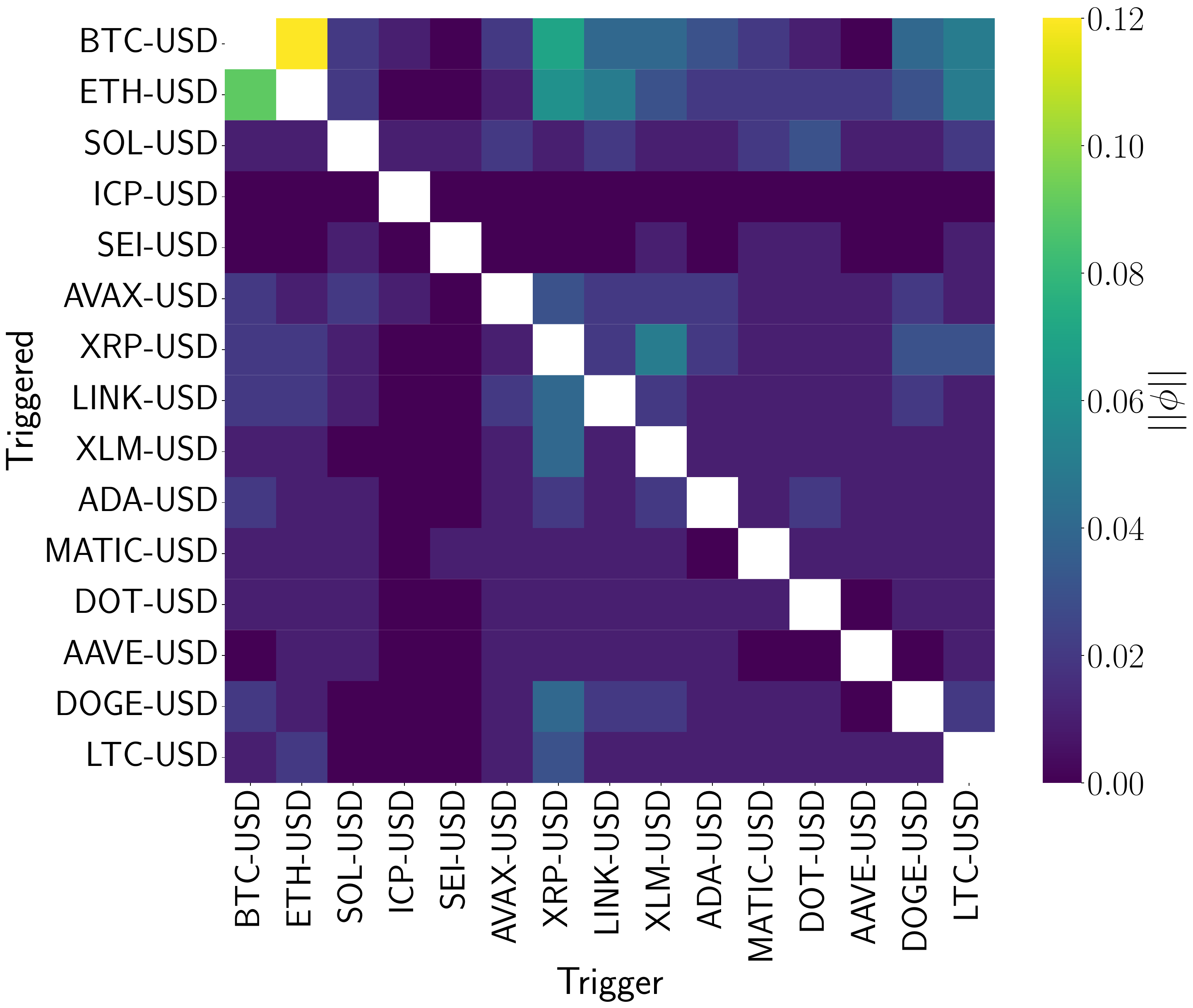}%
    }\\
    \subfloat[Kernels $(\phi^{ii}(t))_{1\leq i \leq D}$\\ --- x-axis in logarithmic scale]{%
        \includegraphics[width=0.33\linewidth]{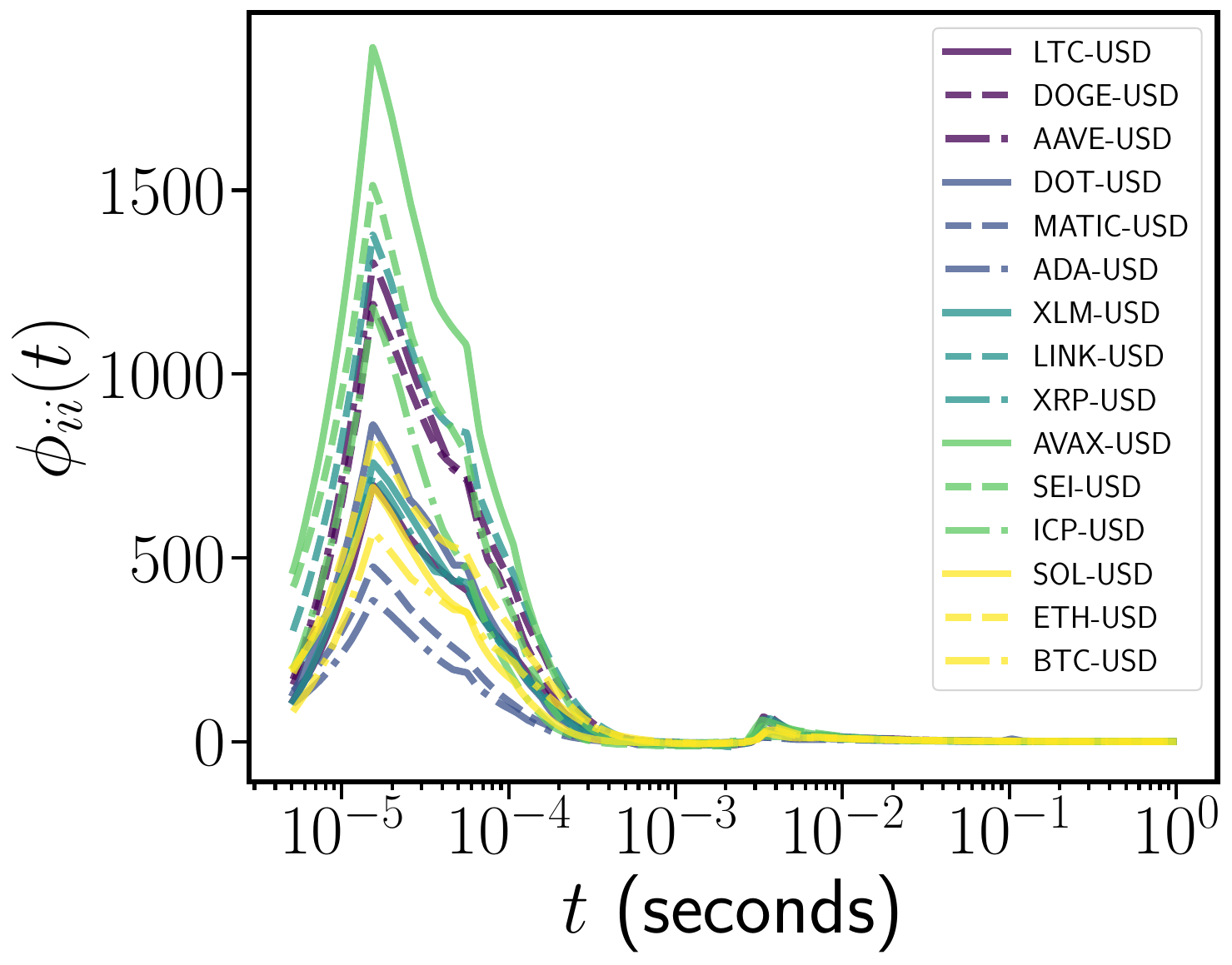}%
    }
    \subfloat[Kernels $(\phi^{ii}(t))_{1\leq i \leq D}$\\ --- Both axes in logarithmic scale]{%
        \includegraphics[width=0.33\linewidth]{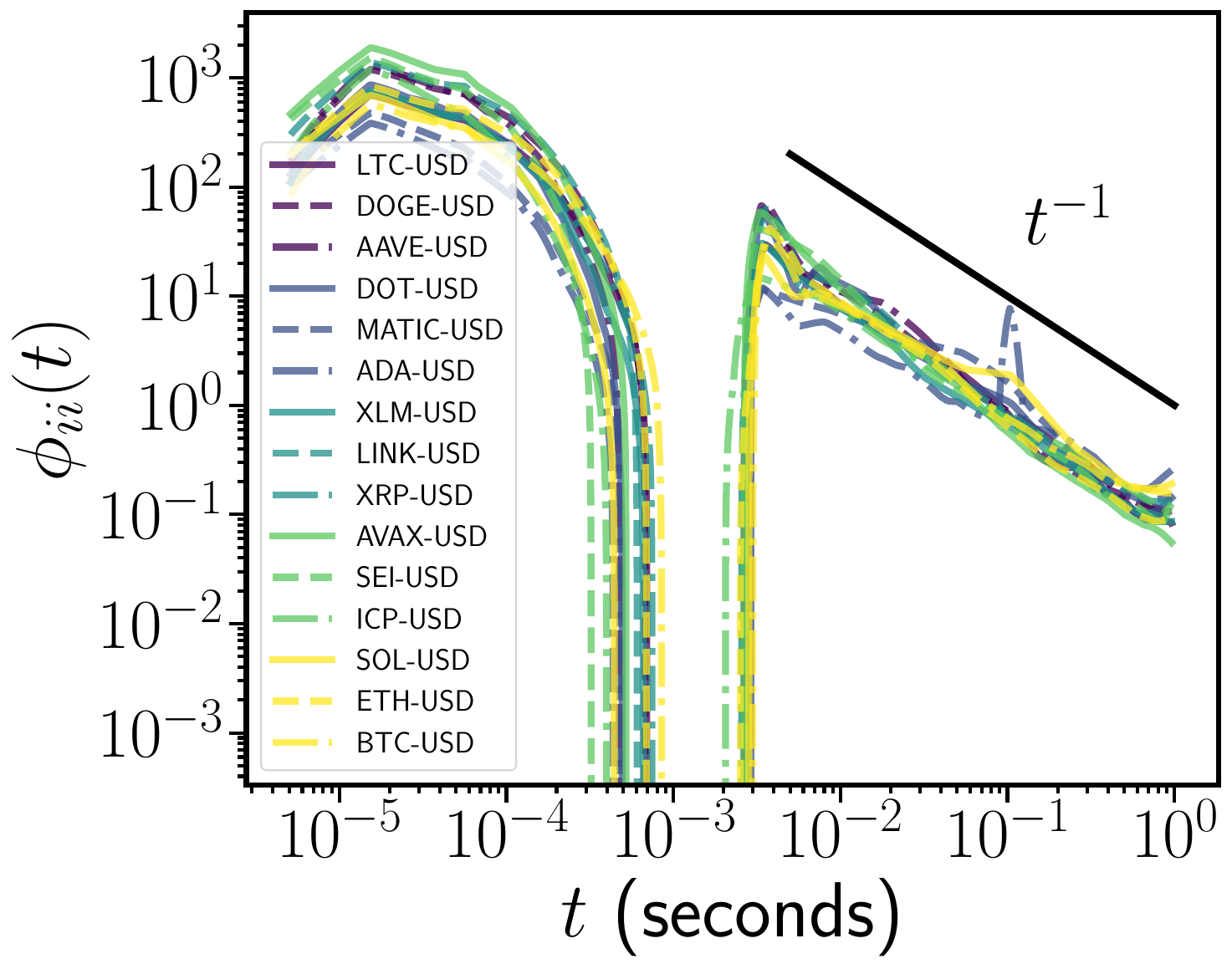}%
    }
    \caption{\textit{Cryptocurrency spillover} --- Branching ratios and diagonal of the kernel matrix. Note the near-critical power law above 3 milliseconds with exponent close to 1.}
    \label{fig:coinbase_spillover_diag_time_kernels}
\end{figure}

The kernel $\phi^{ij}$ characterizes the impact of the arrival of a transaction in the $j$-th pair on the intensity of the arrival rate of transactions in the $i$-th pair. The branching ratio matrix $(\|\phi^{ij}\|)_{1\leq i,j \leq D}$ and the diagonal components $(\phi^{ii})_{1\leq i \leq D}$ of the kernel are displayed in Figure \ref{fig:coinbase_spillover_diag_time_kernels}. The branching ratio $\mathcal{R}$ of the system is 0.80. We observe that Bitcoin (BTC) and Ethereum (ETH) are not the most endogeneous components of the system as other less traded coins show a much larger branching ratio for their self-excitation component, e.g., AAVE, DOGE, SOL or LINK. Furthermore, as observed for the BTC-USD pair in the previous analysis, the kernels exhibit mainly three peaks at around 10 µs, 2 to 5 ms. The self-excitation component of ADA shows a significant peak at 100 ms. We recover the near-critical power law behavior $\sim t^{-\gamma}$ with $\gamma\simeq1$ of the kernel for $t>3$ ms, consistent with the microstructure literature --- see e.g., \cite{bacry2012non, fosset2022non}. All kernels share an inhibitory behaviour between 500 µs and 3 ms. Note that we checked the duration distribution and observed a gap in this specific time frame as well. We provide the entire estimated kernel in Figure \ref{fig:coinbase_spillover_time_kernels} in appendix. The results highlight the stability of the method as no fine tuning was made for this experiment. 

If one denotes by $N^{i\leftarrow j}$ the counting process of events of type $i$ that are caused by the arrival of events of type $j$, we have from \cite{bacry2015hawkes}
\begin{equation}
    \mathbb{E}\left(\mathrm{d}N^{i\leftarrow j}_t\right)=\Lambda^j\|\phi^{ij}\|\mathrm{d}t.
\end{equation}
Therefore the ratio of the arrival rate of events of type $i$ with a direct ancestor of type $j$ to the overall arrival rate of events of type $i$ is a measure of the causal relationship from $j$ to $i$. We thus define the spillover ratio
\begin{equation}\label{eq:spillover_ratio}
    S^{ij}:=\frac{\Lambda^j}{\Lambda^i}\|\phi^{ij}\|\mathds{1}_{\{i\neq j\}}, \hspace{0.3cm}1\leq i,j \leq D.
\end{equation}

Questions that naturally arise when analyzing the microstructure of a highly correlated multi-asset framework are the following. Which asset does lead the others and which asset does not? Can we sort the assets according to their causal strength amongst the selected trading universe and is the traded volume a good indicator of such property?

To answer these questions and gain insights about the direction of causal relationships, we propose to look at two aggregated versions of the spillover ratio of Equation \eqref{eq:spillover_ratio}. The first measure is what we call the leader ratio
\begin{equation}
    L^j:=\frac{\Lambda^j}{\sum_{i\neq j}\Lambda^i}\sum_{i\neq j}\|\phi^{ij}\|, \hspace{0.3cm} 1\leq j\leq D.
\end{equation}
This measure is the ratio of the arrival rate of all the events --- except those of type $j$ --- with a direct ancestor of type $j$ to the overall arrival rate of all the events that are not of type $j$. The larger the ratio, the greater the causal strength of type $j$ over the trading universe. In other words, a large ratio $L^j$ indicates that a large proportion of the observed events that are not of type $j$ is directly caused by an event of type $j$, suggesting that the $j$-th pair leads the other pairs. On the contrary, a small ratio $L^j\rightarrow 0$ indicates that very few events that are not of type $j$ are directly caused by an event of type $j$, suggesting that the $j$-th pair does not lead the other pairs. The second measure we define is the receiver ratio
\begin{equation}
    R^i:=\frac{\sum_{j\neq i}\Lambda^j\|\phi^{ij}\|}{\Lambda^i}, \hspace{0.3cm} 1\leq i\leq D.
\end{equation}
It is the ratio of the arrival rate of the events of type $i$ that are directly caused by events of other types than $i$ to the overall arrival rate of events of type $i$. When $R^i\rightarrow 1$, the causal strength of other types of events over events of type $i$ is large. Hence, a large ratio $R^i$ indicates that a large proportion of the observed events of type $i$ is directly caused by an event of another type, suggesting that the $i$-th pair is a receiver amongst the pairs. On the contrary, a small ratio $R^i\rightarrow 0$ indicates that very few events of type $i$ are directly caused by an event of another type, suggesting that the $i$-th pair is not a receiver.

Usually, practitioners refer to the participation rate of the assets to gain insights about possible leader/lager relationships. For example, it is well-known  --- \textit{e.g.} see \citet{jia2023seesaw} --- that the most traded cryptocurrencies, \textit{i.e.} Bitcoin and Ethereum, lead the overall market. But the question can be trickier for less traded coins such as Cardano (ADA) and Ripple (XRP). We define the participation rate
\begin{equation}
    \nu^j:=\frac{V^j}{\sum_{i=1}^DV^i},
\end{equation}
where $V^j$ is the traded volume of pair $j$ and we compare the ranking with respect to this measure with the ranking with respect to the aggregated spillover ratio.

\begin{figure}
    \centering
    \subfloat[Spillover ratio matrix \\\centering$(S^{ij})_{1\leq i,j\leq D}$]{%
        \includegraphics[width=0.25\linewidth]{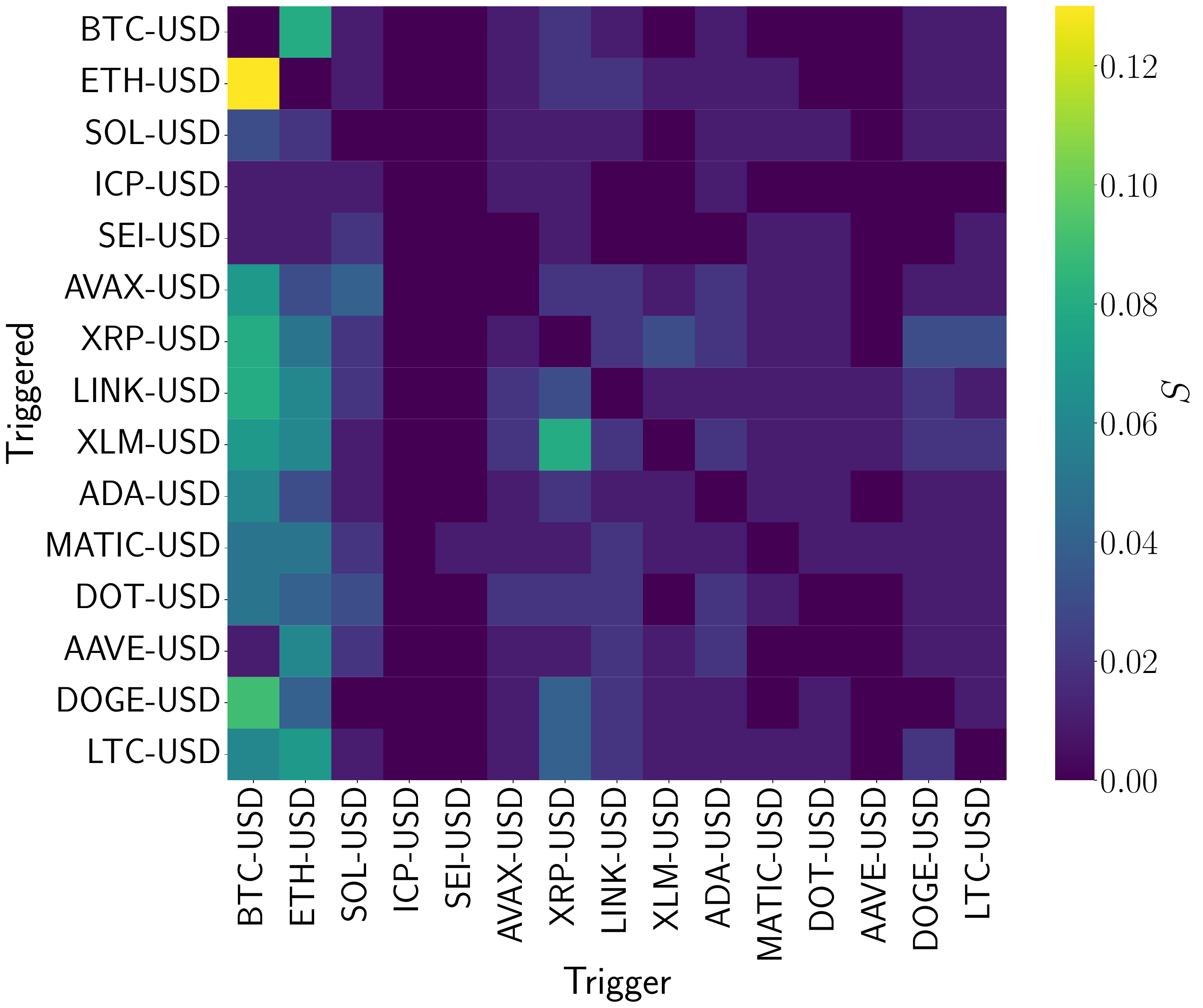}%
    }
    \subfloat[Leader ratios \\\centering$(L^{j})_{1\leq j\leq D}$]{%
        \includegraphics[width=0.25\linewidth]{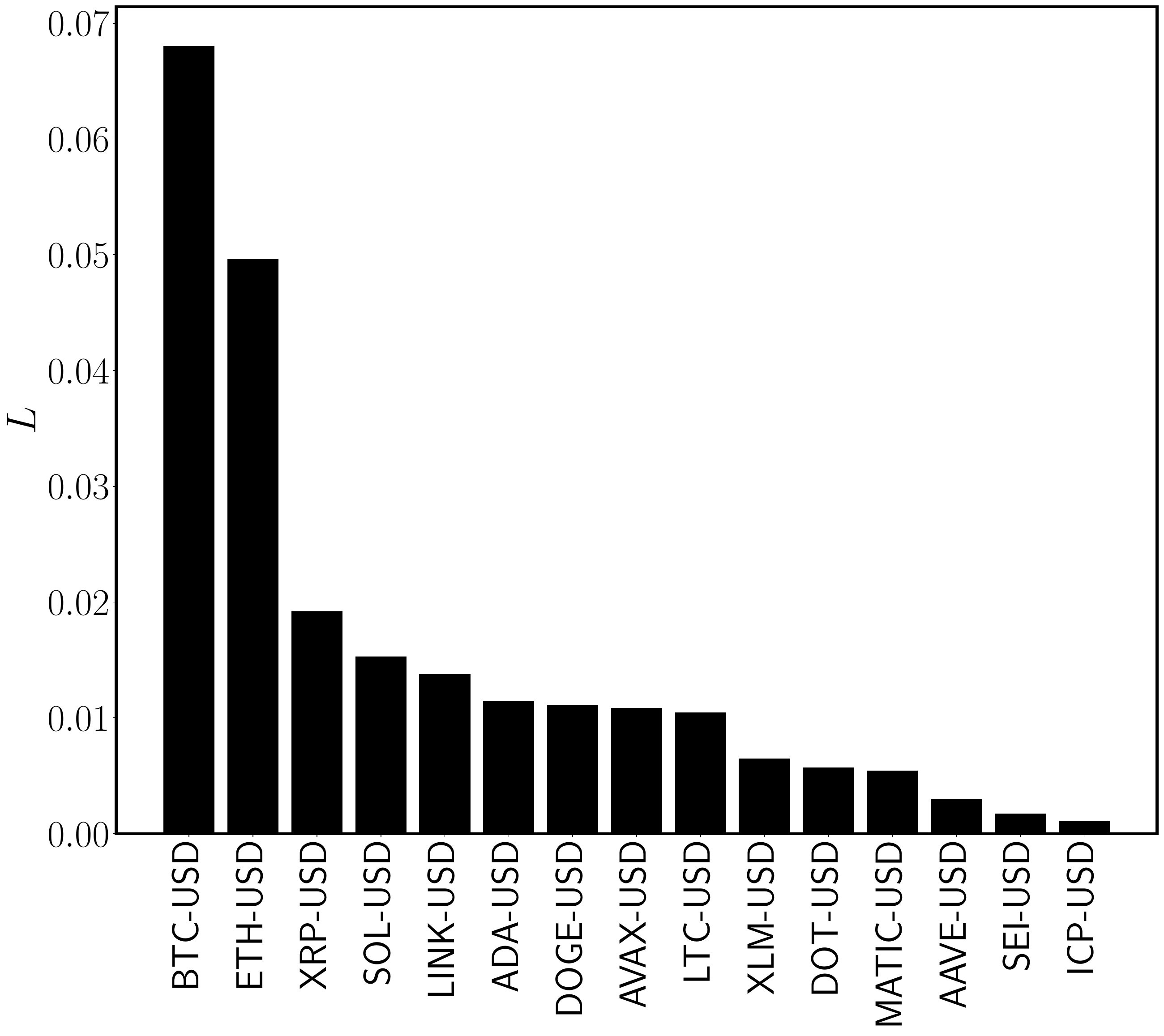}%
    }
    \subfloat[Receiver ratios \\\centering$(R^{i})_{1\leq i\leq D}$]{%
        \includegraphics[width=0.25\linewidth]{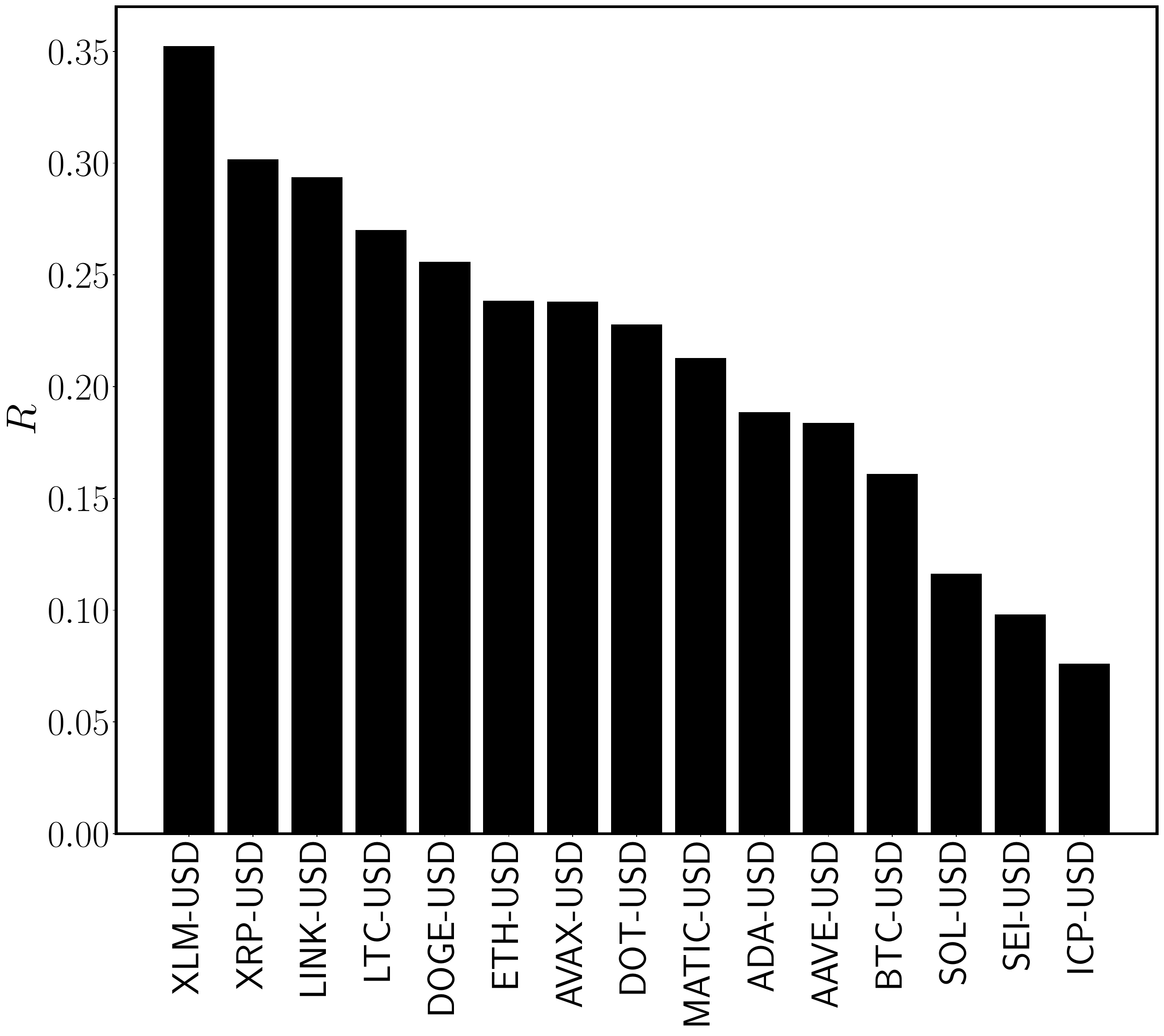}%
    }
    \subfloat[Participation rates $(\nu^{j})_{1\leq j\leq D}$]{%
        \includegraphics[width=0.25\linewidth]{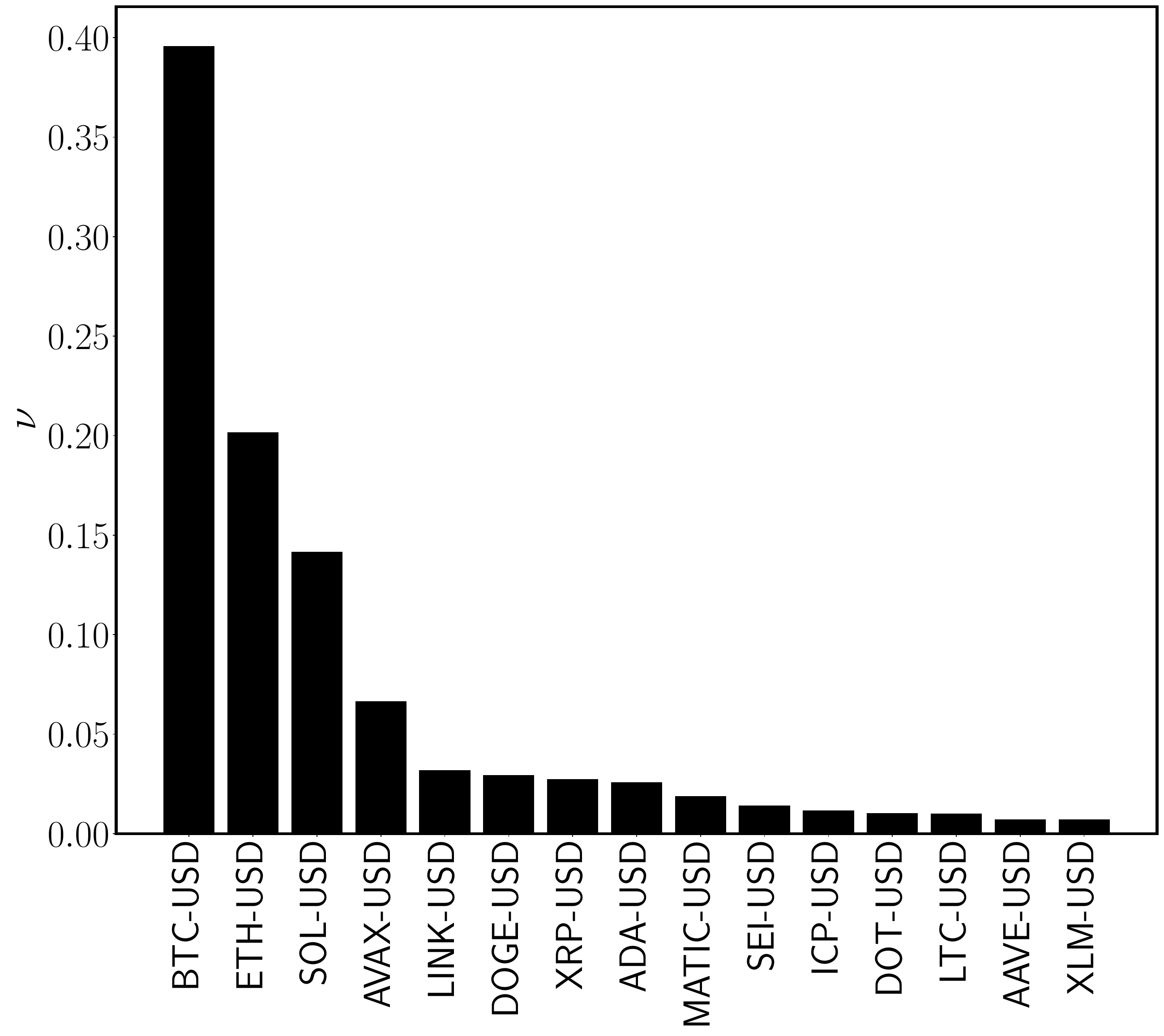}%
    }
    \caption{\textit{Cryptocurrency spillover} --- Spillover ratios and ranking of the leader, receiver measures and participation rates amongst the 15 cryptocurrency pairs.}
    \label{fig:coinbase_spillover_leadlag_measures}
\end{figure}

The results are displayed in Figure \ref{fig:coinbase_spillover_leadlag_measures}. First of all we clearly see discrepancies between the branching ratio matrix of Figure \ref{fig:coinbase_spillover_diag_time_kernels} and the spillover ratio matrix which is due to the inherent differences in market activity amongst the coins. We observe a strong causality from BTC and ETH to other coins as expected. There is a strong bi-directional causality between BTC and ETH which is consistent with the conclusions of recent empirical studies --- see for example \citet{kumar2019volatility, moratis2021quantifying}. Some of the least traded coins such as ICP, SEI and AAVE are strongly self-exciting but exhibit weak mutual excitation with other coins. Interestingly, we notice a bi-directional causality relationship between XRP and XLM, the strongest direction being from XRP to XLM. This finding gives credit to the widespread hypothesis that Ripple and Stellar (XLM) are two highly correlated cryptocurrencies\footnote{Although we did not find any empirical analysis of the XRP/XLM relationship in the academic literature, we refer the reader to a blog discussion \url{https://u.today/intriguing-xrp-and-stellar-xlm-correlation-deciphered-heres-explanation} (Accessed on Jan. 9th, 2024).}. In fact, the ideas behind these protocols are similar and the causal relationship between both prices is still a debate in the decentralized finance community. Furthermore, we observe that the top three leaders are BTC, ETH and XRP and the top three receivers are XLM, XRP and LINK. It shows that XRP is both a strong transmitter and a strong receiver which is consistent with the studies of \citet{kim2021causal, moratis2021quantifying}. Comparing the ratios with the participation rates, we see that high market activity does not necessarily mean high shock transmission. For example, the traded volume on AVAX is much greater than the one on XRP on this time period but XRP shows a significantly higher leader ratio. We also observe that least traded coins are not necessarily strong receivers, \textit{e.g.} ICP. Finally, it is noteworthy that we trained several models with different hyperparameters configurations and all the results were mostly identical.

\subsection{Intraday seasonality and robustness of empirical results}
\label{subsec:datastationarity}
The linear marked Hawkes process model and the second order characterization of Theorem \ref{th:characterization_theorem} rely on the assumption of stationarity in the data generating process. However, in practice, financial data often exhibits non-stationary features. It is well-known that equity trading data exhibits significant intraday seasonality, meaning that the intensity of trades can vary widely throughout the day. Applying our method without accounting for this seasonality could lead to inaccurate estimations of the kernel, and could inflate the branching ratios \citep{filimonov2015apparent, wehrli2021scale}. We thus conduct robustness checks for the empirical results of the previous sections. 

We start by computing the trading intensity at time $t$ on 5-minutes bins each day. Then, we compute the average of these intensities per bin over the month, with a 95\% confidence interval. The results for the 3 most traded coins in the sample are displayed in Appendix in Figure \ref{fig:coinbase_intraday_intensity}. We observe an intraday seasonality, with a high activity and variance from 16:00 UTC to 20:00 UTC. The intensity appears to be reasonably stable from 07:00 to 12:00 UTC, exhibiting smaller variations and no strong trend. We therefore restrict our sample by focusing on the time interval 7:00-12:00 UTC every day and implement two alternative estimation strategies.

In the first alternative estimation strategy, we apply our estimation procedure of the Hawkes process over the whole month of December 2023, using only the selected time window each day. The covariance estimator is computed once on this subsample. We then compute the fitted Hawkes kernels with the neural Hawkes method, as well as the matrix of branching ratios and our leader-receiver metrics. We find that the estimated branching ratio of the system is 78\%, which is close to the one obtained when the estimation is performed over the entire period of time (80\%). We display in Appendix the estimated branching ratio matrix in Figure \ref{fig:coinbase_spillover_branching_ratio_seasonality}, and the spillover ratios with the leader-receiver rankings in Figure \ref{fig:coinbase_spillover_leadlag_measures_seasonality}. We find that the differences between the branching ratios estimated on the full time period and the ones estimated on the restricted time period are small. The relative difference between the two is about a few percent for the largest branching ratios, and about 15 \% to 20\% for the smallest ones, which are also the most prone to estimation errors. Note that these results are in line with those of \citet{rambaldi2017role}, who performed a similar experiment to test for the robustness of the Wiener-Hopf method to seasonality effects. Concerning the spillover ratios and the leader-receiver measures, we also find small discrepancies, but they do not change our conclusions. We conclude that the main findings of the spillover analysis are robust to intraday seasonality effects.

For a complete view, we propose a second alternative estimation strategy, in which we use our estimation procedure daily from 7:00 to 12:00 UTC, i.e. we obtain 30 fits of our model. We then compute the matrix of median branching ratios and boxplots of our leader-receiver metrics. Results are given in Appendix on Figure \ref{fig:coinbase_spillover_leadlag_measures_seasonality_boxplot}. We observe that the median spillover ratios $(S^{ij})_{1\leq i,j \leq D}$ are close to the ones we obtained in the two previous settings and exhibit similar features, such as the causality relationship from XRP to XLM. Furthermore, the box plots of leader-receiver rankings offer a similar yet interestingly complementary view. Boxplots of the leader ratios confirm that BTC and ETH are clear leaders, ICP and SEI close the ranking. For the receiver ratios, median ranking is also very close to the previous ones, although the larger variance observed for some coins may alter the ranking on a daily basis. All in all, it is very interesting to observe that, even though daily fits of the model are more prone to estimation errors since much less data is used for each fit than when a single model is estimated on the entire period of time, (daily) median metrics are in line with the previous (full sample) results.

\section{Conclusion}

In this paper we proposed a new non-parametric estimation method for linear multivariate Hawkes processes that relies on two main tools:
\begin{itemize}
    \item The second order statistics characterization theorem of multivariate linear Hawkes processes originally proven in \cite{bacry2016first};
    \item The physics-informed neural networks that are able to solve non-linear PDEs in high dimension, provided that they are properly trained -- see \cite{wang2023expert}.
\end{itemize}
The Fredhlom equation that links second order statistics to the kernel function is solved using a neural network that is trained using the latest advances in PINNs literature. The benefits of our approach is threefold.

\begin{itemize}
    \item The application of the Wiener-Hopf method requires the inversion of a matrix that can be arbitrarily large in high dimension, leading to potential computation issues and approximation errors. The neural network model is able to work around this problem as it is well suited to high dimension inference;
    \item The performance of the model can be evaluated by tracking the estimated second order statistics at any point, allowing us to evaluate its ability to fit the data;
    \item Once the model is trained, it is able to compute the value of the kernel at any given point. This contrasts with the Wiener-Hopf method for which the computation of the kernel between two quadrature points is done either by linear interpolation or by inversion of the characterization equation.
\end{itemize}

We provided a general hyperparameter configuration that can be used to fit a first non-parametric model before further tuning, as it showed robust performance across the numerical experiments. The model shows a great performance on simulated data with general kernels. We explored the influence of some hyperparameters on the final performance and found the model to be essentially sensitive to the batch size, the learning rate and the number of hidden layers. In general, a small batch size, a learning rate of $0.001$ and 1 to 3 DGM cells provide very satisfactory results.

As an example of practical applications we conducted a non-parametric estimation of the kernel matrix of transactions on a centralized exchange in two settings. The first one is univariate marked Hawkes estimated on BTC-USD trades, mapping the traded volume to the mark process. The second one is a multivariate Hawkes estimated on a universe of 15 cryptocurrency pairs. We analyzed the shape of the mark component of the kernel and found it appears to behave as a concave function of the volume. We introduced a spillover measure and leader/receiver ratios that can be used for identifying the directions of the causal relationships of shocks amongst a correlated asset universe. The estimated metrics indicate that the traded volume is not necessarily representative of the leader-lager relationships. The use of our method can provide reliable insights about the causality structure of large complex systems.

\bibliography{main}
\bibliographystyle{apalike}

\addtocontents{toc}{\protect\setcounter{tocdepth}{1}}

\section*{Appendix}
This appendix provides additional figures and tables that do not appear in the main text in order to enhance readability.

\subsection{Numerical validation -- Additional material}
\begin{figure}[!ht]
    \centering
    \subfloat[Number of DGM cells]{%
        \includegraphics[width=0.25\linewidth]{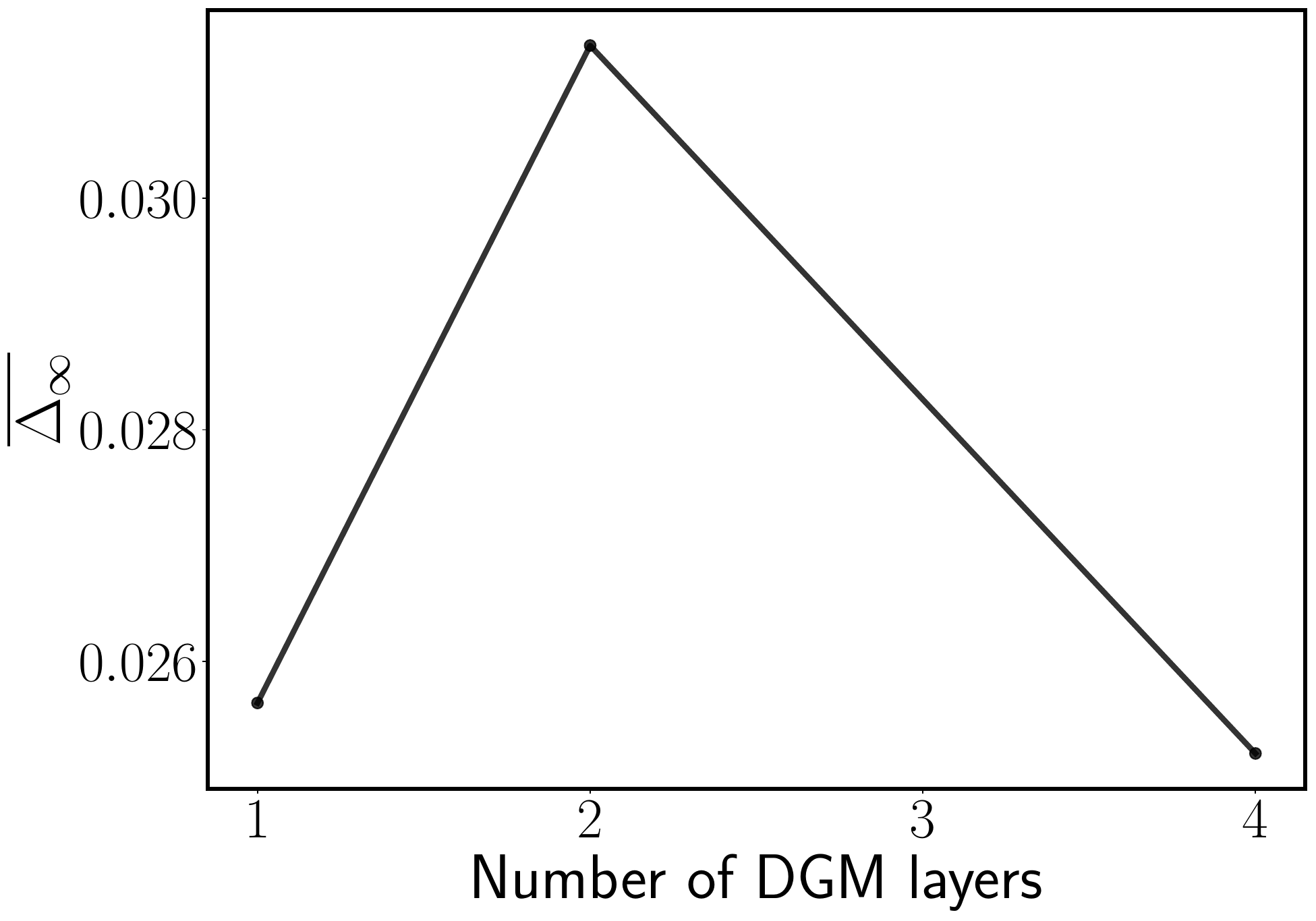}%
    }
    \subfloat[Number of neurons]{%
        \includegraphics[width=0.25\linewidth]{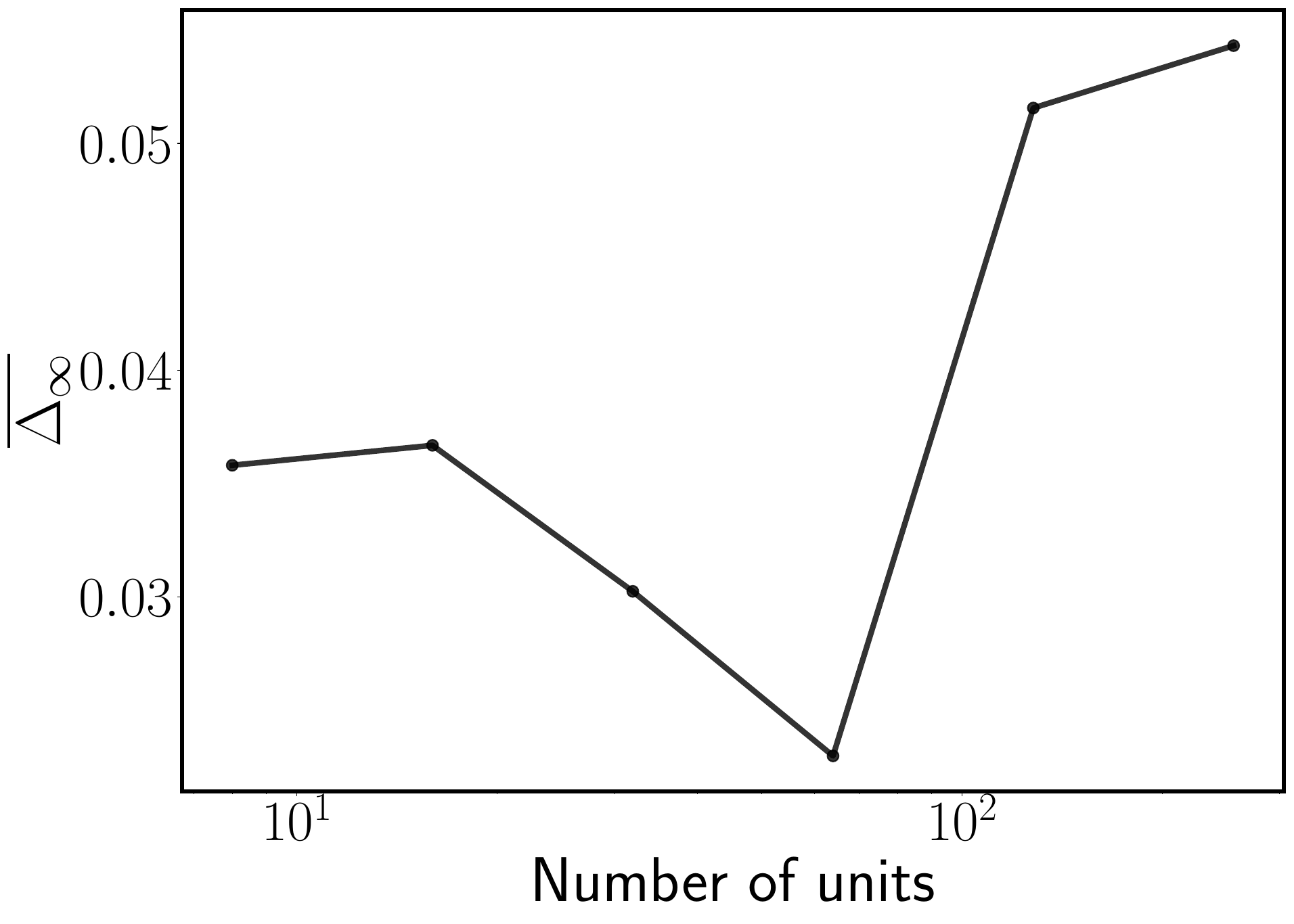}%
    }
    \subfloat[Size of the training set]{%
        \includegraphics[width=0.25\linewidth]{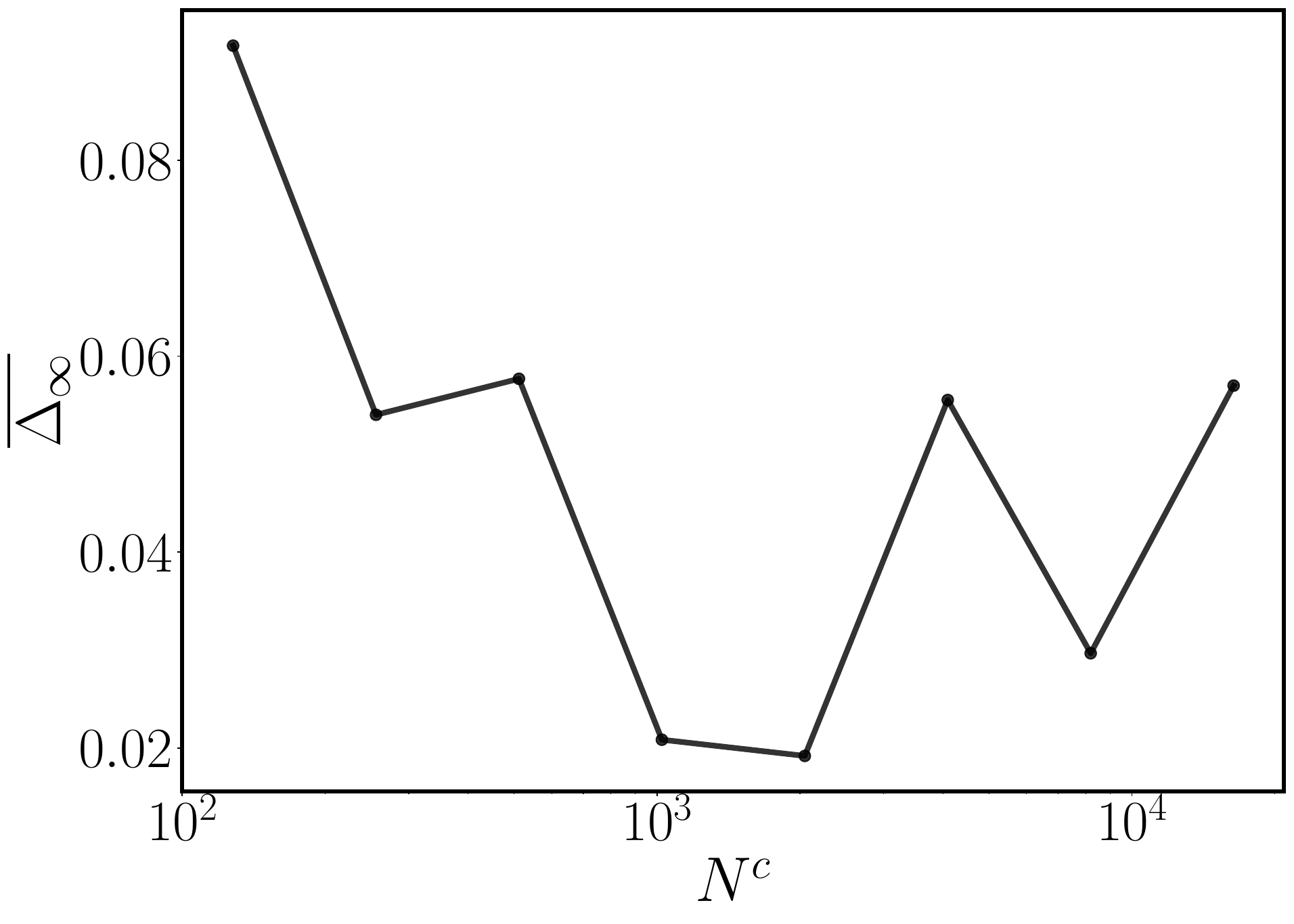}%
    }
    \subfloat[Batch size]{%
        \includegraphics[width=0.25\linewidth]{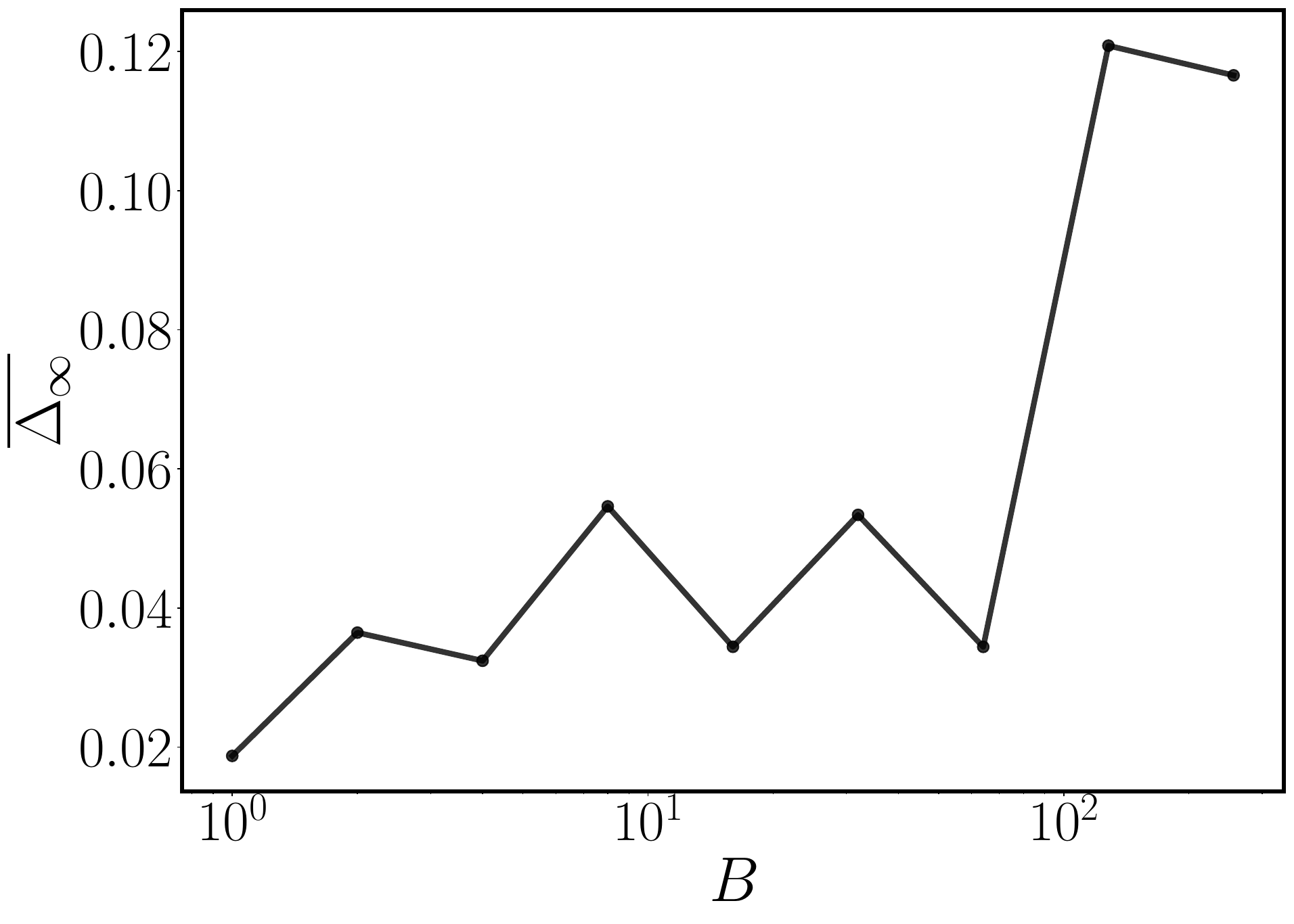}%
    }
    \caption{\textit{Robustness experiment} --- Error $\overline{\Delta_\infty}$ as a function of the hyperparameters in the four experiments.}
    \label{fig:robustness_linferror}
\end{figure}

We give the true values of the parameters of the high-dimensional bimodal Gaussian kernel of Section \ref{subsec:VariousKernels}. The baseline intensity is 
\begin{equation}
    \mu=[0.07, 0.20, 0.09, 0.16, 0.12, 0.16, 0.12, 0.10, 0.15, 0.09, 0.14, 0.11, 0.09, 0.14, 0.16]'.
\end{equation}
and the five $15\times 15$ matrices defining the kernel parameters are given in Tables \ref{table:config_gauss_high_dimension_part1} and \ref{table:config_gauss_high_dimension_part2}.
We also give the values of the parameters used in Section \ref{subsec:CvgceRate} in Tables \ref{table:config_exp_cvgce_nevents} and \ref{table:config_power_cvgce_nevents}. Finally, Figure \ref{fig:robustness_linferror} displays $L_\infty$ errors of the robustness tests of Section \ref{subsec:CvgceRate}.
\begin{table}[!ht]
    \small
    \centering
    \caption{\textit{Bimodal Gaussian kernel} --- Parameter configuration of the simulation (Part 1).}
    \begin{subtable}{\linewidth}
        \centering
        \caption{$\alpha$}
        \begin{tabular}{c|ccccccccccccccc}
            \toprule
            $\alpha_{ij}$ & 1 & 2 & 3 & 4 & 5 & 6 & 7 & 8 & 9 & 10 & 11 & 12 & 13 & 14 & 15 \\ 
            \midrule
            1 & 0.03 & 0.04 & 0.03 & 0.04 & 0.04 & 0.04 & 0.06 & 0.05 & 0.04 & 0.05 & 0.06 & 0.04 & 0.03 & 0.04 & 0.06 \\
            2 & 0.05 & 0.05 & 0.06 & 0.04 & 0.05 & 0.05 & 0.04 & 0.05 & 0.05 & 0.04 & 0.03 & 0.06 & 0.06 & 0.05 & 0.04 \\
            3 & 0.06 & 0.03 & 0.05 & 0.05 & 0.06 & 0.04 & 0.04 & 0.04 & 0.04 & 0.06 & 0.06 & 0.05 & 0.06 & 0.05 & 0.03 \\
            4 & 0.03 & 0.04 & 0.06 & 0.06 & 0.06 & 0.05 & 0.04 & 0.03 & 0.05 & 0.06 & 0.06 & 0.06 & 0.06 & 0.04 & 0.04 \\
            5 & 0.03 & 0.03 & 0.04 & 0.05 & 0.04 & 0.06 & 0.04 & 0.05 & 0.03 & 0.04 & 0.05 & 0.04 & 0.04 & 0.05 & 0.04 \\
            6 & 0.05 & 0.05 & 0.04 & 0.06 & 0.06 & 0.04 & 0.06 & 0.04 & 0.04 & 0.04 & 0.03 & 0.06 & 0.04 & 0.05 & 0.04 \\
            7 & 0.05 & 0.06 & 0.04 & 0.06 & 0.04 & 0.04 & 0.06 & 0.06 & 0.03 & 0.04 & 0.05 & 0.06 & 0.04 & 0.04 & 0.04 \\
            8 & 0.04 & 0.05 & 0.05 & 0.05 & 0.06 & 0.06 & 0.04 & 0.05 & 0.05 & 0.06 & 0.06 & 0.04 & 0.05 & 0.03 & 0.04 \\
            9 & 0.03 & 0.06 & 0.06 & 0.04 & 0.04 & 0.06 & 0.04 & 0.06 & 0.05 & 0.03 & 0.05 & 0.03 & 0.04 & 0.04 & 0.05 \\
            10 & 0.03 & 0.06 & 0.03 & 0.05 & 0.04 & 0.05 & 0.06 & 0.04 & 0.05 & 0.05 & 0.05 & 0.05 & 0.04 & 0.06 & 0.06 \\
            11 & 0.05 & 0.05 & 0.05 & 0.04 & 0.04 & 0.05 & 0.04 & 0.04 & 0.06 & 0.05 & 0.06 & 0.05 & 0.06 & 0.04 & 0.05 \\
            12 & 0.04 & 0.06 & 0.03 & 0.05 & 0.06 & 0.04 & 0.05 & 0.06 & 0.03 & 0.03 & 0.04 & 0.05 & 0.05 & 0.05 & 0.04 \\
            13 & 0.06 & 0.05 & 0.03 & 0.06 & 0.04 & 0.04 & 0.05 & 0.04 & 0.06 & 0.05 & 0.03 & 0.05 & 0.03 & 0.06 & 0.05 \\
            14 & 0.06 & 0.04 & 0.05 & 0.04 & 0.05 & 0.05 & 0.05 & 0.05 & 0.04 & 0.05 & 0.06 & 0.04 & 0.06 & 0.03 & 0.03 \\
            15 & 0.06 & 0.06 & 0.05 & 0.05 & 0.06 & 0.05 & 0.05 & 0.05 & 0.04 & 0.03 & 0.05 & 0.04 & 0.04 & 0.05 & 0.05 \\
            \bottomrule
        \end{tabular}
    \end{subtable}

    \vspace{0.3cm}
    
    \begin{subtable}{\linewidth}
        \centering
        \caption{$\underbar{$\mu$}$}
        \begin{tabular}{c|ccccccccccccccc}
            \toprule
            $\underbar{$\mu$}_{ij}$ & 1 & 2 & 3 & 4 & 5 & 6 & 7 & 8 & 9 & 10 & 11 & 12 & 13 & 14 & 15 \\ 
            \midrule
            1 & 0.71 & 0.53 & 0.04 & 0.30 & 0.32 & 0.60 & 0.10 & 0.46 & 0.67 & 0.43 & 0.53 & 0.59 & 0.72 & 0.47 & 0.61 \\
            2 & 0.30 & 0.29 & 0.56 & 0.42 & 0.65 & 0.36 & 0.03 & 0.43 & 0.09 & 0.47 & 0.37 & 0.17 & 0.30 & 0.02 & 0.40 \\
            3 & 0.42 & 0.50 & 0.25 & 0.71 & 0.45 & 0.30 & 0.40 & 0.29 & 0.20 & 0.08 & 0.34 & 0.20 & 0.38 & 0.31 & 0.41 \\
            4 & 0.36 & 0.42 & 0.56 & 0.59 & 0.71 & 0.44 & 0.64 & 0.07 & 0.43 & 0.07 & 0.69 & 0.70 & 0.73 & 0.27 & 0.50 \\
            5 & 0.18 & 0.01 & 0.55 & 0.44 & 0.61 & 0.53 & 0.22 & 0.17 & 0.09 & 0.16 & 0.30 & 0.20 & 0.61 & 0.18 & 0.41 \\
            6 & 0.61 & 0.40 & 0.26 & 0.61 & 0.27 & 0.30 & 0.55 & 0.21 & 0.11 & 0.16 & 0.35 & 0.13 & 0.61 & 0.08 & 0.22 \\
            7 & 0.59 & 0.26 & 0.12 & 0.30 & 0.47 & 0.34 & 0.29 & 0.40 & 0.34 & 0.69 & 0.57 & 0.53 & 0.07 & 0.20 & 0.58 \\
            8 & 0.73 & 0.72 & 0.54 & 0.66 & 0.75 & 0.41 & 0.70 & 0.31 & 0.68 & 0.39 & 0.57 & 0.52 & 0.65 & 0.47 & 0.49 \\
            9 & 0.59 & 0.04 & 0.69 & 0.30 & 0.37 & 0.27 & 0.25 & 0.39 & 0.00 & 0.22 & 0.71 & 0.30 & 0.68 & 0.47 & 0.59 \\
            10 & 0.41 & 0.55 & 0.31 & 0.11 & 0.50 & 0.60 & 0.08 & 0.72 & 0.43 & 0.72 & 0.70 & 0.38 & 0.47 & 0.29 & 0.38 \\
            11 & 0.44 & 0.32 & 0.49 & 0.70 & 0.66 & 0.62 & 0.66 & 0.01 & 0.73 & 0.30 & 0.04 & 0.72 & 0.13 & 0.46 & 0.41 \\
            12 & 0.11 & 0.06 & 0.44 & 0.69 & 0.53 & 0.58 & 0.36 & 0.65 & 0.64 & 0.20 & 0.35 & 0.70 & 0.42 & 0.43 & 0.22 \\
            13 & 0.21 & 0.01 & 0.72 & 0.30 & 0.63 & 0.42 & 0.67 & 0.49 & 0.10 & 0.05 & 0.10 & 0.32 & 0.37 & 0.29 & 0.18 \\
            14 & 0.16 & 0.42 & 0.15 & 0.71 & 0.49 & 0.23 & 0.39 & 0.43 & 0.63 & 0.34 & 0.27 & 0.38 & 0.56 & 0.19 & 0.16 \\
            15 & 0.70 & 0.09 & 0.11 & 0.41 & 0.31 & 0.51 & 0.43 & 0.59 & 0.74 & 0.38 & 0.69 & 0.45 & 0.18 & 0.43 & 0.68 \\
            \bottomrule
        \end{tabular}
    \end{subtable}

    \vspace{0.3cm}
    
    \begin{subtable}{\linewidth}
        \centering
        \caption{$\underbar{$\sigma$}$}
        \begin{tabular}{c|ccccccccccccccc}
            \toprule
            $\underbar{$\sigma$}_{ij}$ & 1 & 2 & 3 & 4 & 5 & 6 & 7 & 8 & 9 & 10 & 11 & 12 & 13 & 14 & 15 \\ 
            \midrule
            1  & 0.05 & 0.08 & 0.18 & 0.16 & 0.10 & 0.10 & 0.18 & 0.14 & 0.10 & 0.11 & 0.11 & 0.15 & 0.16 & 0.08 & 0.09 \\
            2  & 0.09 & 0.20 & 0.08 & 0.13 & 0.19 & 0.17 & 0.13 & 0.14 & 0.16 & 0.19 & 0.13 & 0.17 & 0.09 & 0.06 & 0.08 \\
            3  & 0.10 & 0.16 & 0.09 & 0.05 & 0.13 & 0.18 & 0.20 & 0.06 & 0.13 & 0.15 & 0.07 & 0.07 & 0.18 & 0.09 & 0.13 \\
            4  & 0.12 & 0.09 & 0.06 & 0.18 & 0.19 & 0.11 & 0.18 & 0.12 & 0.12 & 0.16 & 0.11 & 0.11 & 0.07 & 0.11 & 0.16 \\
            5  & 0.19 & 0.18 & 0.18 & 0.18 & 0.08 & 0.06 & 0.13 & 0.07 & 0.15 & 0.05 & 0.20 & 0.06 & 0.17 & 0.10 & 0.16 \\
            6  & 0.07 & 0.16 & 0.11 & 0.19 & 0.09 & 0.20 & 0.06 & 0.15 & 0.14 & 0.18 & 0.06 & 0.17 & 0.14 & 0.14 & 0.16 \\
            7  & 0.13 & 0.07 & 0.09 & 0.09 & 0.15 & 0.18 & 0.08 & 0.06 & 0.14 & 0.14 & 0.17 & 0.12 & 0.20 & 0.08 & 0.08 \\
            8  & 0.08 & 0.14 & 0.09 & 0.08 & 0.17 & 0.09 & 0.17 & 0.08 & 0.18 & 0.11 & 0.18 & 0.15 & 0.07 & 0.16 & 0.17 \\
            9  & 0.18 & 0.16 & 0.12 & 0.07 & 0.05 & 0.19 & 0.14 & 0.12 & 0.11 & 0.17 & 0.12 & 0.08 & 0.18 & 0.09 & 0.13 \\
            10 & 0.11 & 0.13 & 0.07 & 0.17 & 0.16 & 0.14 & 0.06 & 0.10 & 0.18 & 0.06 & 0.14 & 0.19 & 0.18 & 0.09 & 0.14 \\
            11 & 0.12 & 0.13 & 0.12 & 0.10 & 0.16 & 0.14 & 0.15 & 0.09 & 0.16 & 0.13 & 0.07 & 0.05 & 0.19 & 0.11 & 0.10 \\
            12 & 0.18 & 0.14 & 0.11 & 0.14 & 0.18 & 0.07 & 0.12 & 0.17 & 0.08 & 0.18 & 0.14 & 0.18 & 0.14 & 0.05 & 0.05 \\
            13 & 0.08 & 0.09 & 0.14 & 0.18 & 0.09 & 0.10 & 0.12 & 0.16 & 0.17 & 0.15 & 0.10 & 0.18 & 0.18 & 0.10 & 0.15 \\
            14 & 0.07 & 0.10 & 0.08 & 0.09 & 0.19 & 0.20 & 0.14 & 0.10 & 0.19 & 0.07 & 0.13 & 0.11 & 0.09 & 0.14 & 0.19 \\
            15 & 0.17 & 0.07 & 0.20 & 0.16 & 0.10 & 0.14 & 0.17 & 0.18 & 0.06 & 0.10 & 0.08 & 0.16 & 0.15 & 0.07 & 0.11 \\
            \bottomrule
        \end{tabular}
    \end{subtable}
    \label{table:config_gauss_high_dimension_part1}
\end{table}

\begin{table}[!ht]
    \small
    \centering
    \caption{\textit{Bimodal Gaussian kernel} --- Parameter configuration of the simulation (Part 2).}
    \begin{subtable}{\linewidth}
        \centering
        \caption{$\overline{\mu}$}
        \begin{tabular}{c|ccccccccccccccc}
            \toprule
            $\overline{\mu}_{ij}$ & 1 & 2 & 3 & 4 & 5 & 6 & 7 & 8 & 9 & 10 & 11 & 12 & 13 & 14 & 15 \\ 
            \midrule
            1 & 0.31 & 0.52 & 0.16 & 0.17 & 0.52 & 0.52 & 0.04 & 0.38 & 0.51 & 0.00 & 0.47 & 0.40 & 0.61 & 0.45 & 0.07 \\
            2 & 0.12 & 0.43 & 0.30 & 0.17 & 0.07 & 0.57 & 0.54 & 0.35 & 0.19 & 0.16 & 0.48 & 0.09 & 0.45 & 0.36 & 0.27 \\
            3 & 0.42 & 0.30 & 0.62 & 0.33 & 0.57 & 0.24 & 0.03 & 0.66 & 0.67 & 0.38 & 0.28 & 0.12 & 0.50 & 0.07 & 0.07 \\
            4 & 0.10 & 0.24 & 0.16 & 0.10 & 0.31 & 0.12 & 0.18 & 0.10 & 0.49 & 0.35 & 0.16 & 0.68 & 0.10 & 0.58 & 0.04 \\
            5 & 0.22 & 0.24 & 0.18 & 0.16 & 0.13 & 0.53 & 0.32 & 0.02 & 0.45 & 0.14 & 0.05 & 0.04 & 0.35 & 0.02 & 0.65 \\
            6 & 0.37 & 0.02 & 0.00 & 0.25 & 0.02 & 0.60 & 0.03 & 0.32 & 0.48 & 0.67 & 0.55 & 0.69 & 0.33 & 0.50 & 0.00 \\
            7 & 0.35 & 0.54 & 0.50 & 0.09 & 0.52 & 0.44 & 0.70 & 0.05 & 0.42 & 0.52 & 0.39 & 0.02 & 0.27 & 0.66 & 0.73 \\
            8 & 0.50 & 0.36 & 0.74 & 0.65 & 0.27 & 0.13 & 0.50 & 0.54 & 0.15 & 0.62 & 0.59 & 0.33 & 0.38 & 0.12 & 0.14 \\
            9 & 0.28 & 0.72 & 0.04 & 0.37 & 0.21 & 0.01 & 0.12 & 0.41 & 0.47 & 0.60 & 0.57 & 0.37 & 0.57 & 0.02 & 0.15 \\
            10 & 0.40 & 0.69 & 0.00 & 0.21 & 0.39 & 0.42 & 0.49 & 0.29 & 0.67 & 0.35 & 0.10 & 0.31 & 0.28 & 0.32 & 0.66 \\
            11 & 0.55 & 0.22 & 0.26 & 0.49 & 0.52 & 0.13 & 0.68 & 0.58 & 0.39 & 0.55 & 0.69 & 0.33 & 0.16 & 0.02 & 0.61 \\
            12 & 0.59 & 0.33 & 0.28 & 0.42 & 0.03 & 0.54 & 0.62 & 0.04 & 0.32 & 0.39 & 0.03 & 0.33 & 0.69 & 0.50 & 0.39 \\
            13 & 0.15 & 0.67 & 0.09 & 0.19 & 0.71 & 0.02 & 0.62 & 0.71 & 0.51 & 0.63 & 0.27 & 0.01 & 0.55 & 0.48 & 0.71 \\
            14 & 0.53 & 0.61 & 0.44 & 0.56 & 0.26 & 0.28 & 0.30 & 0.52 & 0.04 & 0.53 & 0.41 & 0.63 & 0.01 & 0.59 & 0.52 \\
            15 & 0.33 & 0.48 & 0.63 & 0.23 & 0.17 & 0.18 & 0.02 & 0.62 & 0.26 & 0.15 & 0.06 & 0.11 & 0.59 & 0.04 & 0.73 \\
            \bottomrule
        \end{tabular}
    \end{subtable}

    \vspace{0.3cm}
    
    \begin{subtable}{\linewidth}
        \centering
        \caption{$\overline{\sigma}$}
        \begin{tabular}{c|ccccccccccccccc}
            \toprule
            $\overline{\sigma}_{ij}$ & 1 & 2 & 3 & 4 & 5 & 6 & 7 & 8 & 9 & 10 & 11 & 12 & 13 & 14 & 15 \\ 
            \midrule
            1  & 0.10 & 0.14 & 0.18 & 0.11 & 0.06 & 0.08 & 0.17 & 0.18 & 0.08 & 0.16 & 0.12 & 0.15 & 0.10 & 0.12 & 0.15 \\
            2  & 0.07 & 0.19 & 0.15 & 0.16 & 0.08 & 0.16 & 0.16 & 0.08 & 0.17 & 0.16 & 0.17 & 0.06 & 0.12 & 0.17 & 0.14 \\
            3  & 0.20 & 0.06 & 0.15 & 0.16 & 0.17 & 0.13 & 0.19 & 0.12 & 0.14 & 0.16 & 0.12 & 0.12 & 0.15 & 0.06 & 0.15 \\
            4  & 0.18 & 0.08 & 0.16 & 0.09 & 0.18 & 0.19 & 0.07 & 0.06 & 0.07 & 0.17 & 0.17 & 0.07 & 0.15 & 0.16 & 0.09 \\
            5  & 0.06 & 0.09 & 0.07 & 0.10 & 0.06 & 0.11 & 0.19 & 0.05 & 0.18 & 0.14 & 0.07 & 0.13 & 0.12 & 0.09 & 0.12 \\
            6  & 0.06 & 0.13 & 0.08 & 0.12 & 0.15 & 0.10 & 0.08 & 0.20 & 0.07 & 0.08 & 0.12 & 0.15 & 0.15 & 0.07 & 0.06 \\
            7  & 0.20 & 0.14 & 0.06 & 0.05 & 0.15 & 0.19 & 0.17 & 0.06 & 0.10 & 0.08 & 0.11 & 0.07 & 0.14 & 0.10 & 0.19 \\
            8  & 0.15 & 0.12 & 0.17 & 0.15 & 0.06 & 0.16 & 0.05 & 0.20 & 0.18 & 0.08 & 0.07 & 0.05 & 0.06 & 0.09 & 0.06 \\
            9  & 0.09 & 0.18 & 0.09 & 0.19 & 0.14 & 0.13 & 0.19 & 0.14 & 0.18 & 0.12 & 0.15 & 0.13 & 0.10 & 0.18 & 0.20 \\
            10 & 0.07 & 0.17 & 0.14 & 0.06 & 0.16 & 0.11 & 0.12 & 0.18 & 0.17 & 0.05 & 0.07 & 0.14 & 0.08 & 0.19 & 0.08 \\
            11 & 0.07 & 0.16 & 0.16 & 0.09 & 0.08 & 0.09 & 0.16 & 0.15 & 0.07 & 0.09 & 0.12 & 0.19 & 0.08 & 0.09 & 0.18 \\
            12 & 0.19 & 0.10 & 0.16 & 0.11 & 0.11 & 0.05 & 0.13 & 0.19 & 0.19 & 0.16 & 0.14 & 0.14 & 0.20 & 0.13 & 0.06 \\
            13 & 0.18 & 0.16 & 0.14 & 0.16 & 0.13 & 0.06 & 0.19 & 0.15 & 0.13 & 0.18 & 0.17 & 0.13 & 0.19 & 0.07 & 0.14 \\
            14 & 0.15 & 0.06 & 0.20 & 0.11 & 0.16 & 0.06 & 0.15 & 0.08 & 0.05 & 0.09 & 0.11 & 0.18 & 0.19 & 0.19 & 0.15 \\
            15 & 0.14 & 0.06 & 0.07 & 0.09 & 0.09 & 0.08 & 0.07 & 0.17 & 0.20 & 0.15 & 0.14 & 0.13 & 0.10 & 0.18 & 0.09 \\
            \bottomrule
        \end{tabular}
    \end{subtable}
    \label{table:config_gauss_high_dimension_part2}
\end{table}

\begin{table}[!ht]
    \caption{A 2-variate exponentially decreasing kernel.}
    \begin{subtable}{.5\linewidth}
      \centering
        \caption{$\alpha$}
        \begin{tabular}{c|cc}
           \toprule
           $\alpha_{ij}$ & 1 & 2 \\ 
           \midrule
            1 & 1 & 0.25 \\
            2 & 0.5 & 0.75 \\
           \bottomrule
       \end{tabular}
    \end{subtable}%
    \begin{subtable}{.5\linewidth}
      \centering
        \caption{$\beta$}
        \begin{tabular}{c|cc}
           \toprule
           $\beta_{ij}$ & 1 & 2 \\ 
           \midrule
            1 & 2 & 1 \\
            2 & 1 & 1.5 \\
           \bottomrule
       \end{tabular}
    \end{subtable}%
    \label{table:config_exp_cvgce_nevents}
\end{table}
\begin{table}[!htbp]
    \caption{A 2-variate power-law decreasing kernel.}
    \begin{subtable}{.33\linewidth}
      \centering
        \caption{$\alpha$}
        \begin{tabular}{c|cc}
           \toprule
           $\alpha_{ij}$ & 1 & 2 \\ 
           \midrule
            1 & 0.012 & 0.008 \\
            2 & 0.004 & 0.005 \\
           \bottomrule
       \end{tabular}
    \end{subtable}%
    \begin{subtable}{.33\linewidth}
      \centering
        \caption{$\beta$}
        \begin{tabular}{c|cc}
           \toprule
           $\beta_{ij}$ & 1 & 2 \\ 
           \midrule
            1 & 1.3 & 1.3 \\
            2 & 1.3 & 1.3 \\
           \bottomrule
       \end{tabular}
    \end{subtable}%
    \begin{subtable}{.33\linewidth}
      \centering
        \caption{$\gamma$}
        \begin{tabular}{c|cc}
           \toprule
           $\gamma_{ij}$ & 1 & 2 \\ 
           \midrule
            1 & 0.0005 & 0.0005 \\
            2 & 0.0005 & 0.0005 \\
           \bottomrule
       \end{tabular}
    \end{subtable}%
    \label{table:config_power_cvgce_nevents}
\end{table}

\subsection{Cryptocurrency spillover -- Additional material}

\begin{figure}
    \centering
    \includegraphics[width=1.\linewidth]{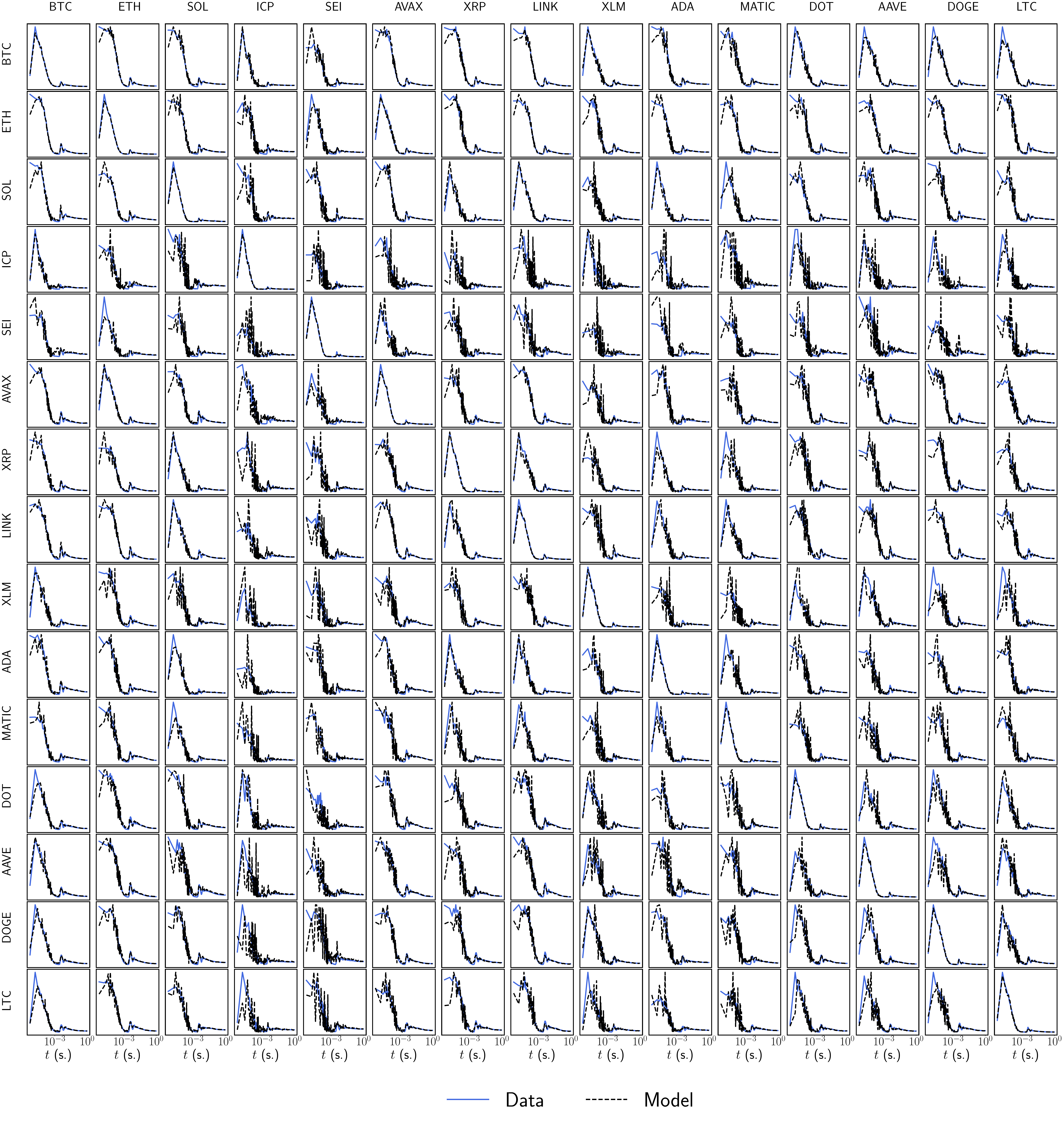}
    \caption{\textit{Cryptocurrency spillover} --- Fit of the second order statistics with the moment-based neural Hawkes kernel $(\phi^{ij})_{1\leq i,j \leq D}$ for a universe of 15 cryptocurrencies using trade data on Coinbase from 2023/12/01 to 2023/12/30. Note the scale of the y-axis might change from one kernel to another as they do not necessary share the same orders of magnitude. We chose to remove the ticks of the y-axis for illustration purposes. The second order statistics of the fitted model were computed over simulations using $10^6$ events.}
    \label{fig:coinbase_spillover_fit_time_kernels}
\end{figure}

\paragraph{Goodness-of-fit.}
We evaluate the goodness-of-fit of the fitted model of Section \ref{subsec:CryptoSpillover} by comparing its first and second moments with the moments of the data. We simulate $10^6$ events of a linear Hawkes process with the fitted kernels using the thinning algorithm, and then estimate the second order statistics on the simulated data with the same grid as the one we used for the real data. For the first moment, we obtain
\begin{equation}
    \Lambda=[2.21, 1.48, 1.11, 0.25, 0.31, 0.71, 0.50 , 0.52, 0.23, 0.61, 0.39, 0.35, 0.28, 0.48, 0.40]'
\end{equation}
which can be compared to the ``true'' vector computed on the data and given in Table \ref{table:coinbase_spillover_descriptive_statistics}. We obtain a satisfactory fit, the mean absolute relative error being $9.3\%$ with respect to the values computed on the data. Figure \ref{fig:coinbase_spillover_fit_time_kernels} shows the second order statistics computed on simulated data and their corresponding fit with the real ones. Again, it is very satisfactory to observe that the model is able to reproduce the multimodal shape of the second order statistics, and produces a good fit concerning the order of magnitude of the excitations. 

\begin{figure}[!ht]
    \centering
    \includegraphics[width=1.\linewidth]{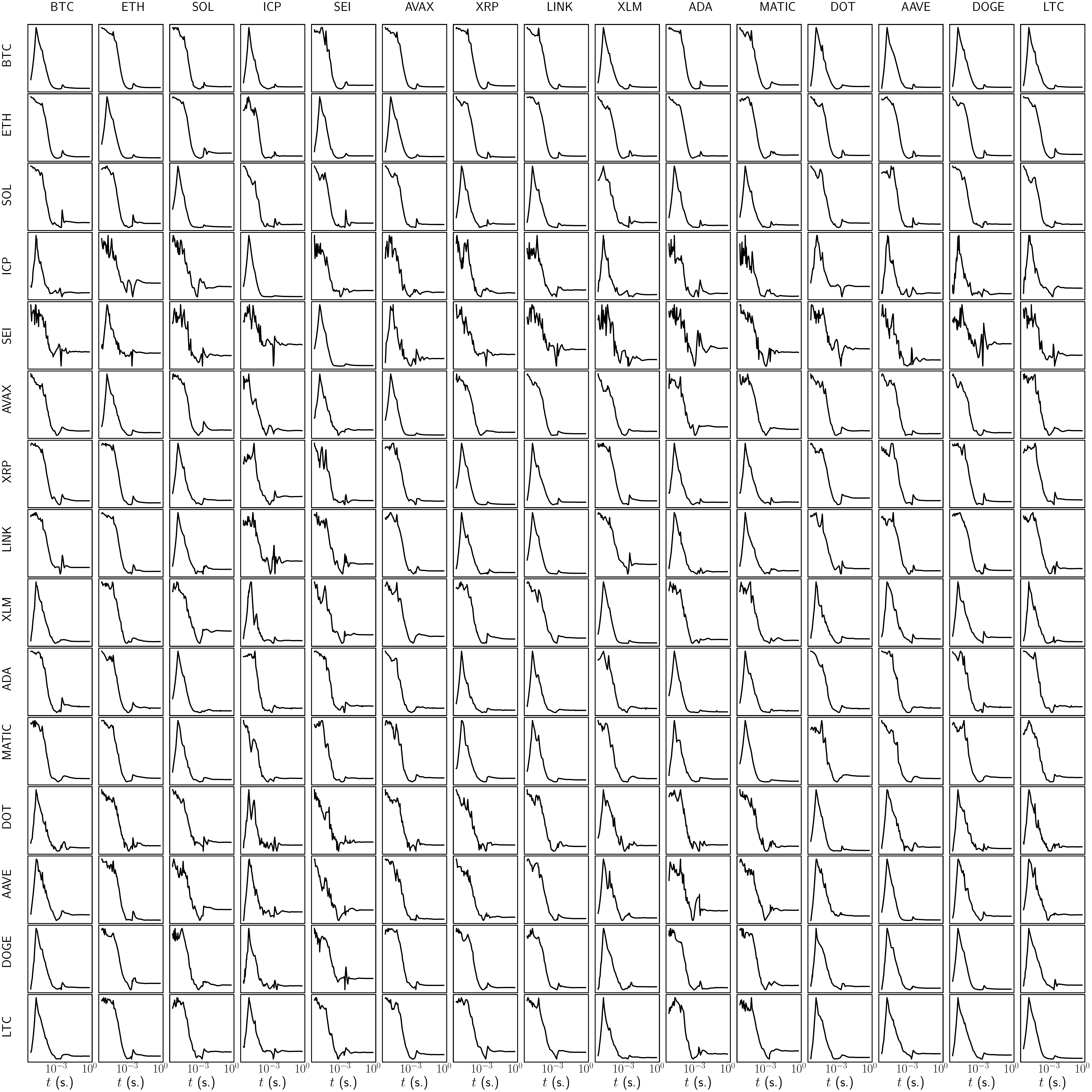}
    \caption{\textit{Cryptocurrency spillover} --- Moment-based neural Hawkes kernel $(\phi^{ij})_{1\leq i,j \leq D}$ as a function of time for a universe of 15 cryptocurrencies using trade data on Coinbase from 2023/12/01 to 2023/12/30. Note the scale of the y-axis might change from one kernel to another as they do not necessary share the same orders of magnitude. We chose to remove the ticks of the y-axis for illustration purposes.}
    \label{fig:coinbase_spillover_time_kernels}
\end{figure}

\paragraph{Fitted kernels.}
Figure \ref{fig:coinbase_spillover_time_kernels} displays the full 15-dimensional kernel estimated in the empirical analysis of Section \ref{subsec:CryptoSpillover}.

\begin{figure}
    \centering
    \subfloat[BTC-USD]{%
        \includegraphics[width=0.33\linewidth]{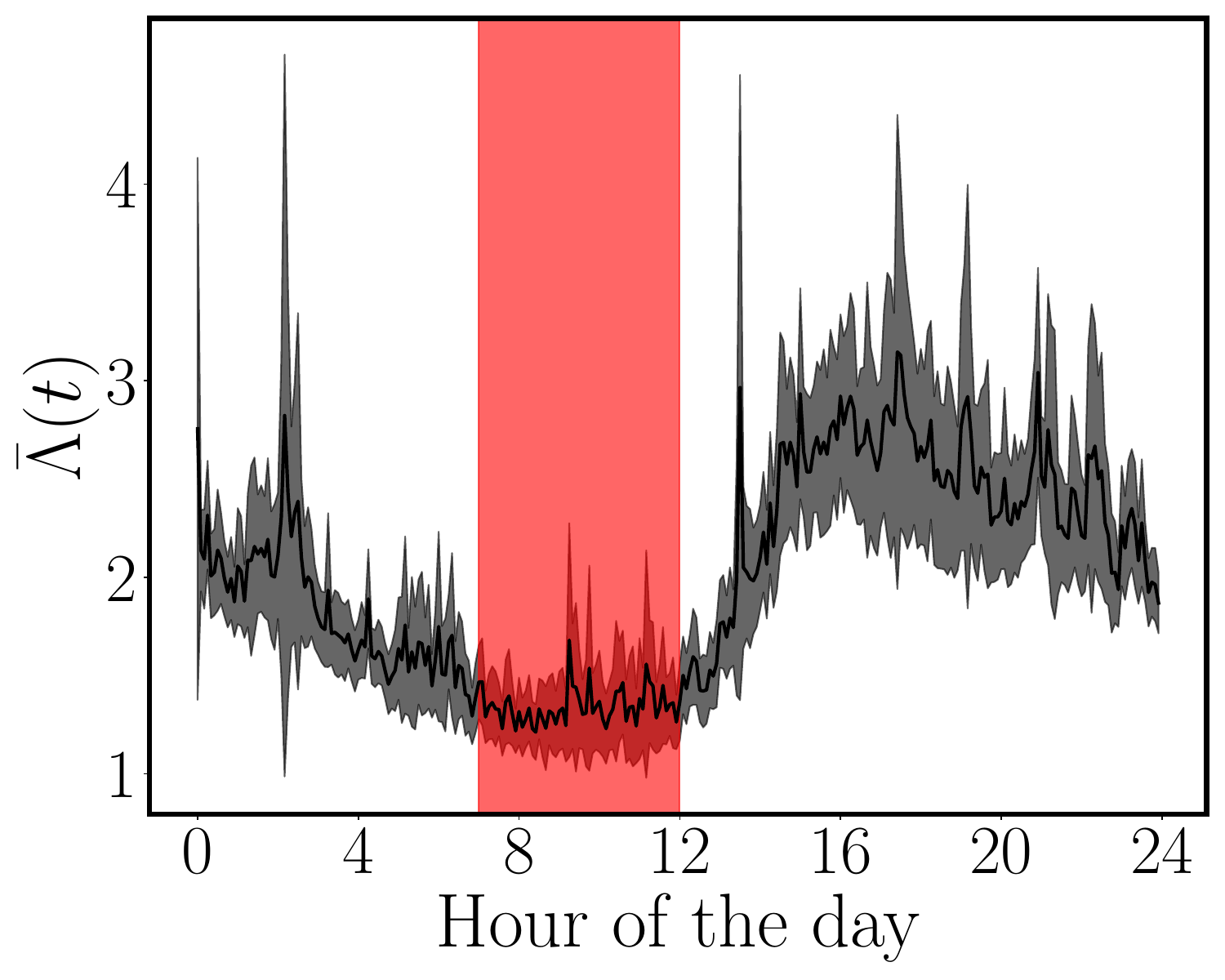}%
    }
    \subfloat[ETH-USD]{%
        \includegraphics[width=0.33\linewidth]{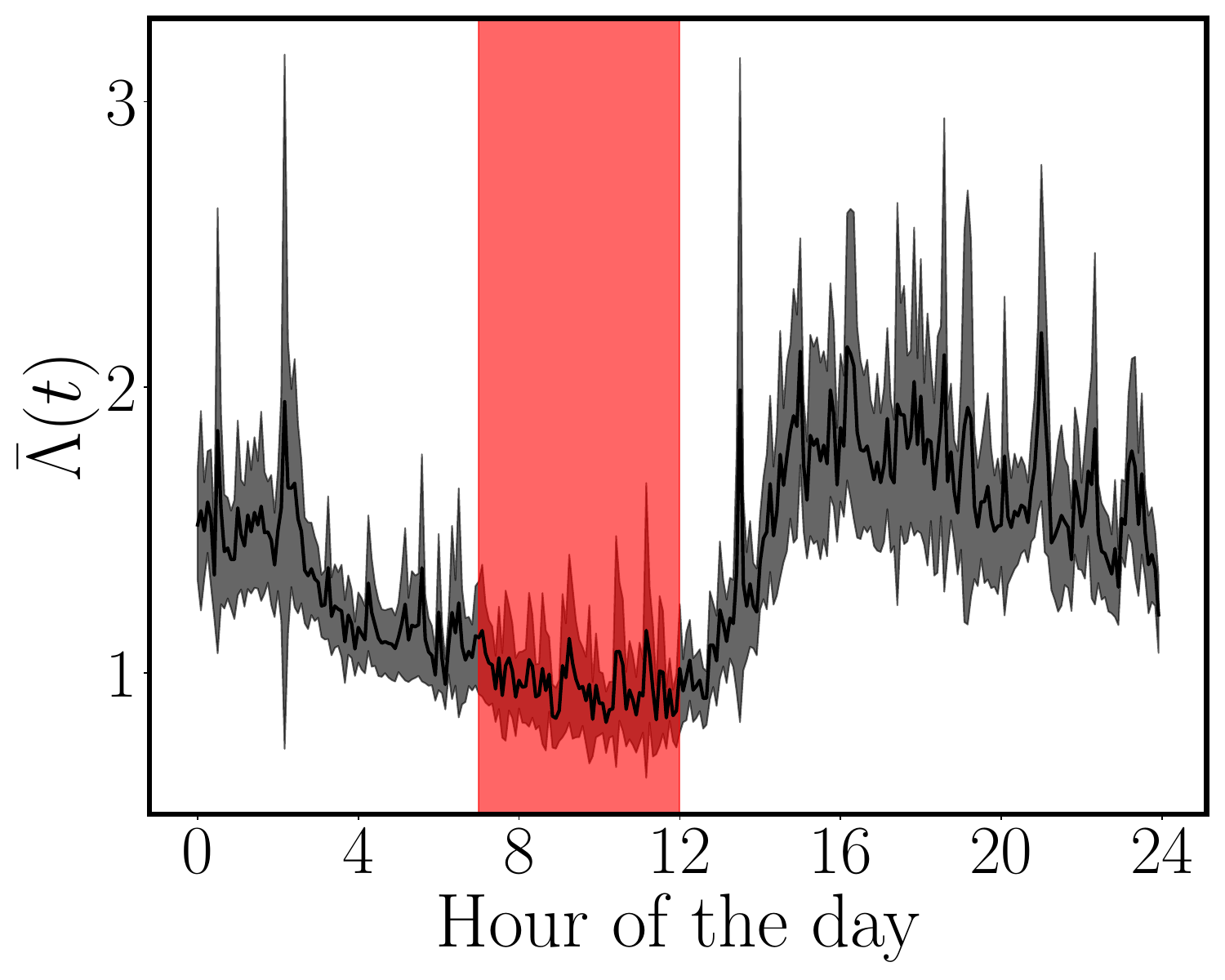}%
    }
    \subfloat[SOL-USD]{%
        \includegraphics[width=0.33\linewidth]{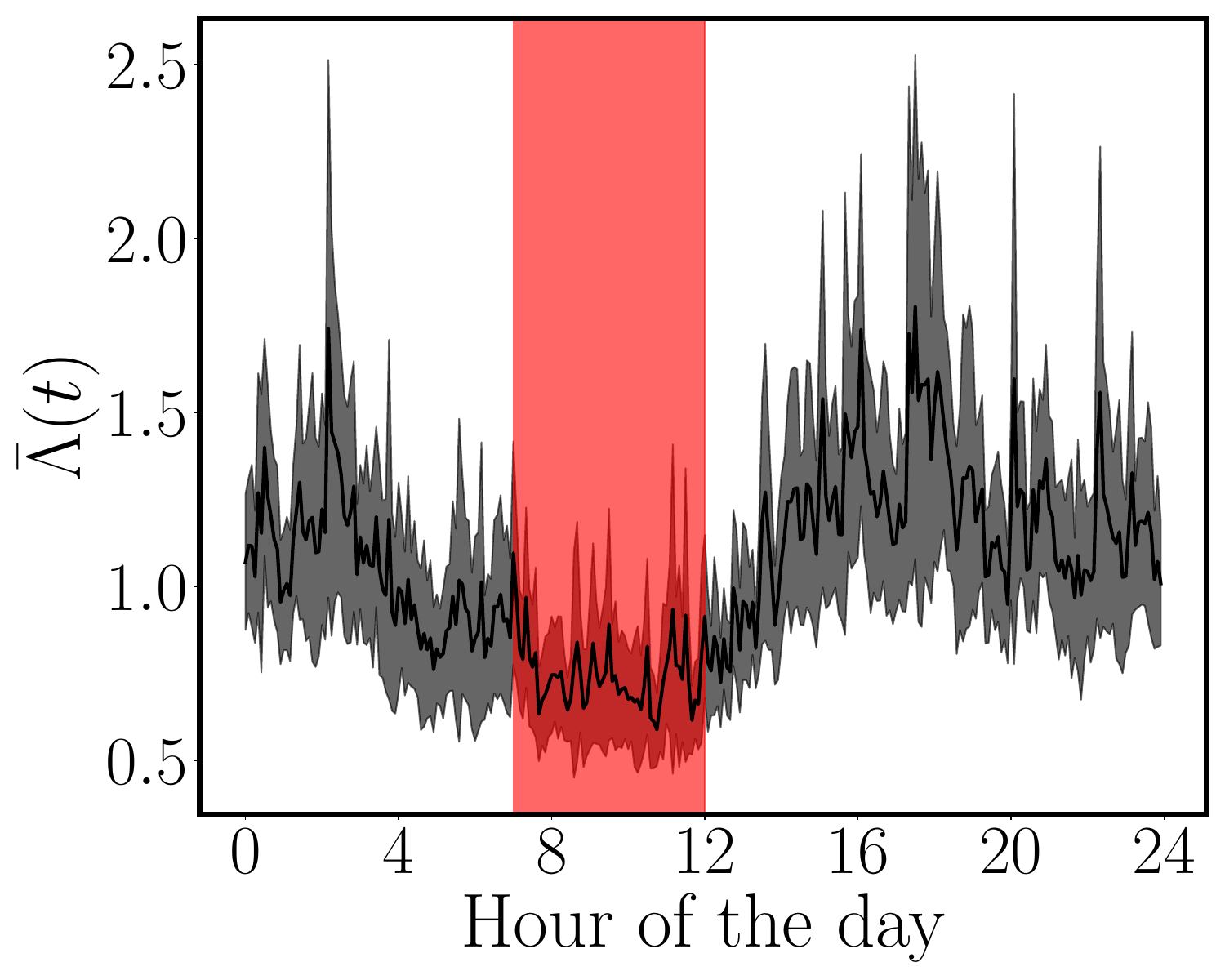}%
    }
    \caption{\textit{Intraday seasonality} --- Intensity of marketable orders expressed in number of events per second, computed over bins of 5 minutes. The black plain line is the average intensity, and the dark shade represents its 95\% confidence interval. Both statistics are computed using the 30 days of data. The red shaded aread represents the time period from 7:00 to 12:00 that is used in the robustness check experiment.}
    \label{fig:coinbase_intraday_intensity}
\end{figure}

\begin{figure}
    \centering
    \subfloat[Branching ratio matrix \\\centering$(\|\phi^{ij}\|)_{1\leq i,j\leq D}$]{%
        \includegraphics[width=0.25\linewidth]{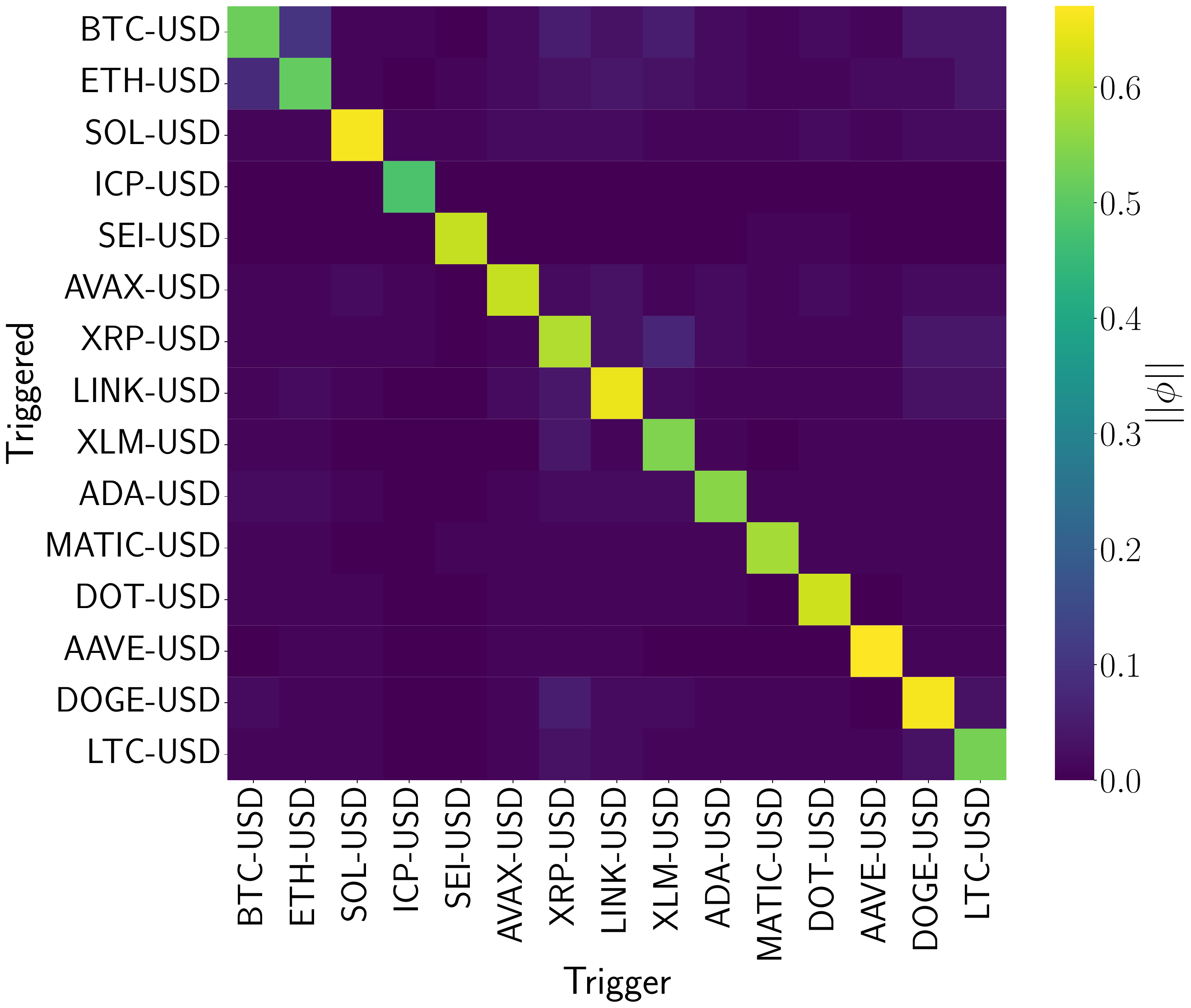}%
    }
    \subfloat[Branching ratio matrix \\\centering$(\|\phi^{ij}\|)_{1\leq i,j\leq D}$ --- diagonal removed]{%
        \includegraphics[width=0.25\linewidth]{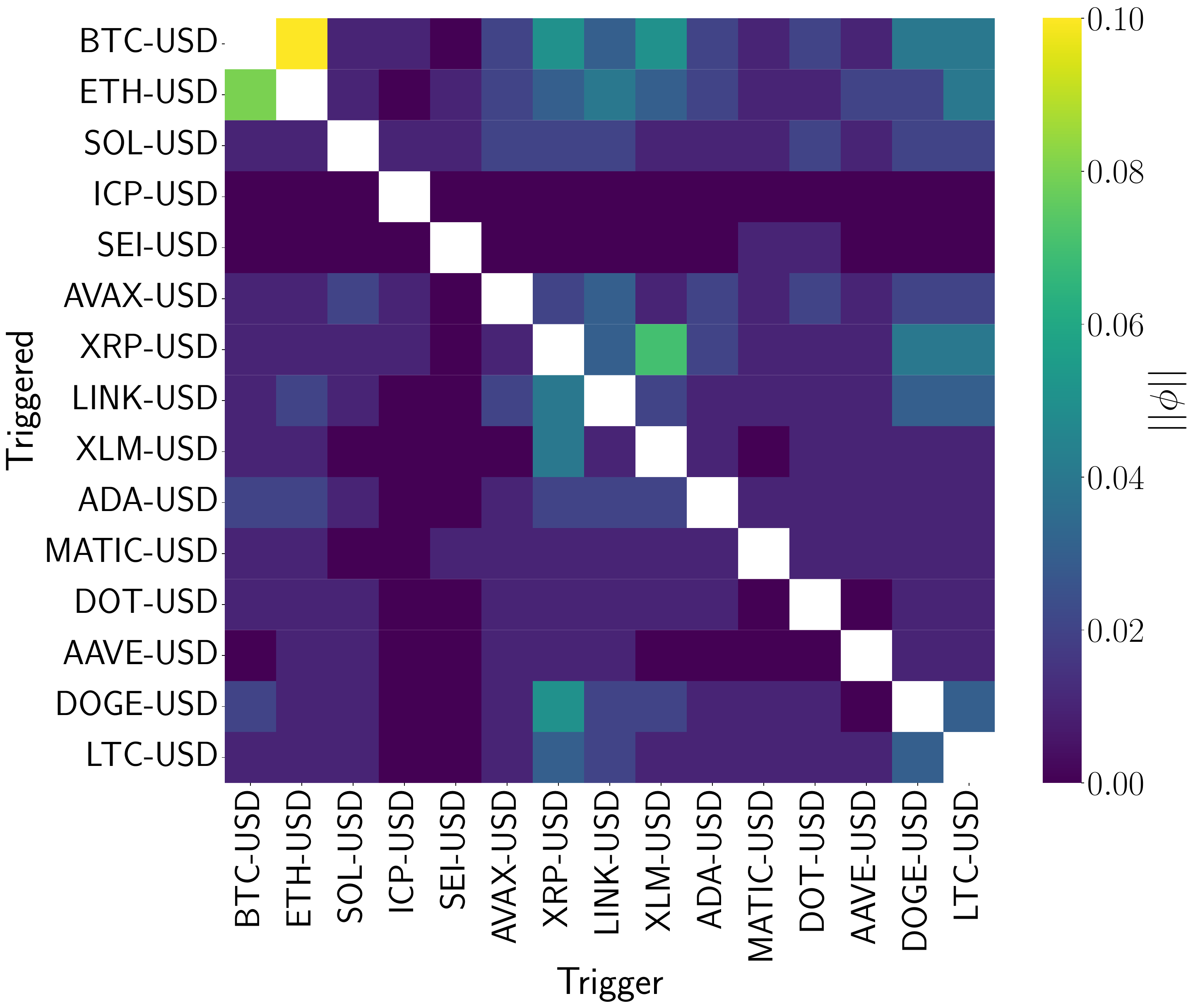}%
    }
    \subfloat[Kernels $(\phi^{ii}(t))_{1\leq i \leq D}$\\ --- x-axis in logarithmic scale]{%
        \includegraphics[width=0.25\linewidth]{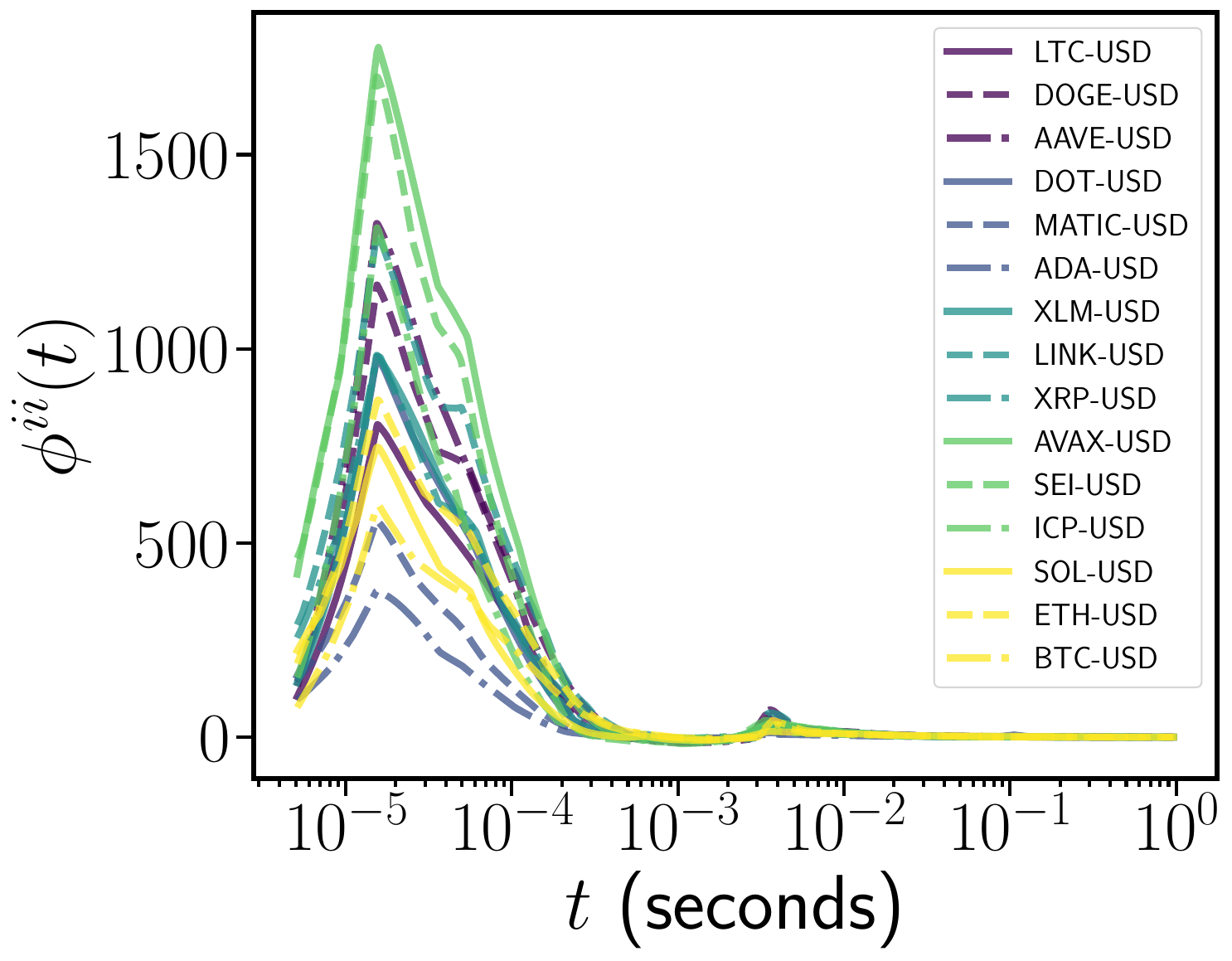}%
    }
    \subfloat[Kernels $(\phi^{ii}(t))_{1\leq i \leq D}$\\ --- Both axes in logarithmic scale]{%
        \includegraphics[width=0.25\linewidth]{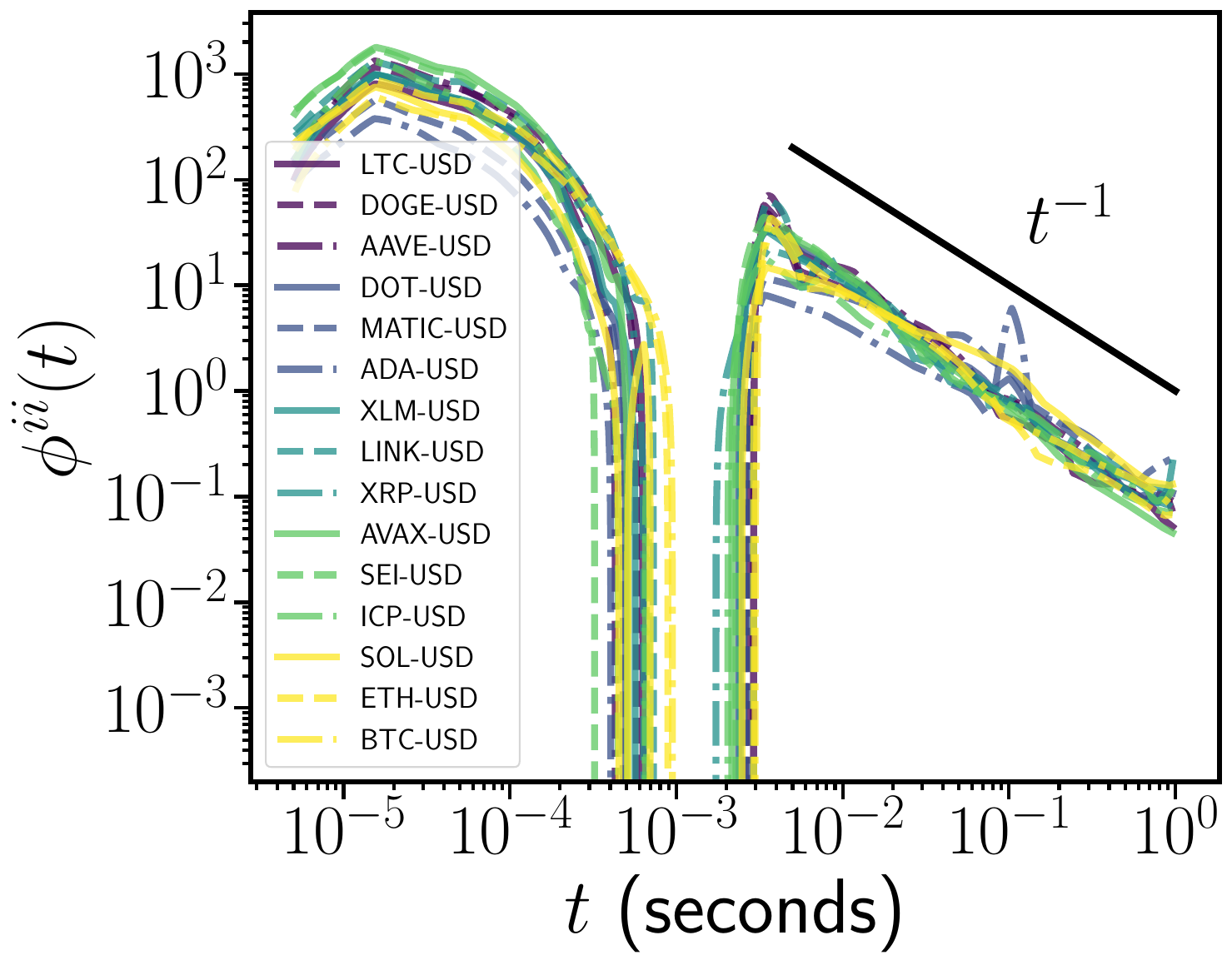}%
    }
    \caption{\textit{Cryptocurrency spillover} --- Branching ratios the kernel matrix. Estimation performed on the time windows spanning from 07:00 to 12:00 UTC.}
    \label{fig:coinbase_spillover_branching_ratio_seasonality}
\end{figure}

\begin{figure}
    \centering
    \subfloat[Spillover ratio matrix \\\centering$(S^{ij})_{1\leq i,j\leq D}$]{%
        \includegraphics[width=0.25\linewidth]{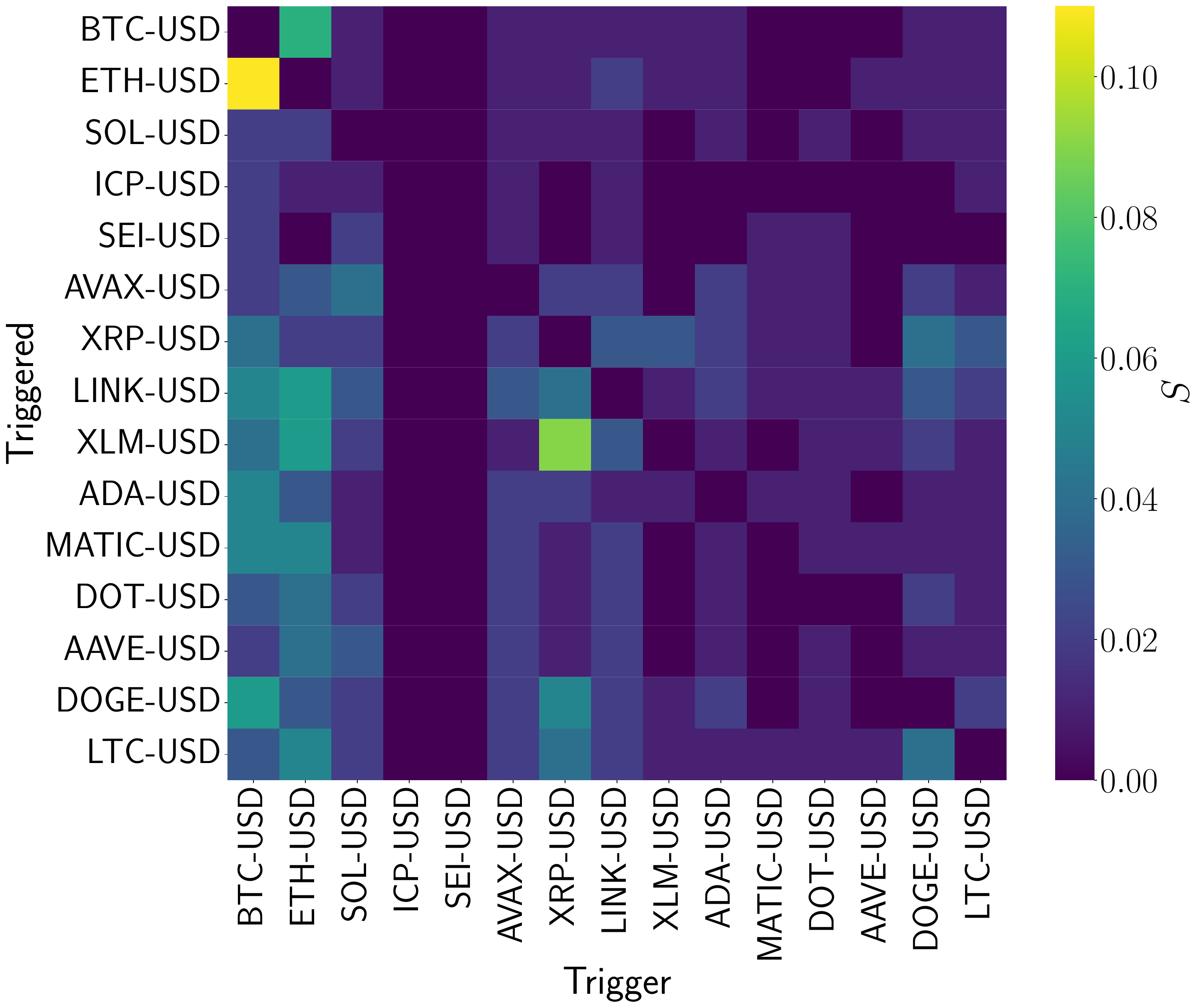}%
    }
    \subfloat[Leader ratios \\\centering$(L^{j})_{1\leq j\leq D}$]{%
        \includegraphics[width=0.25\linewidth]{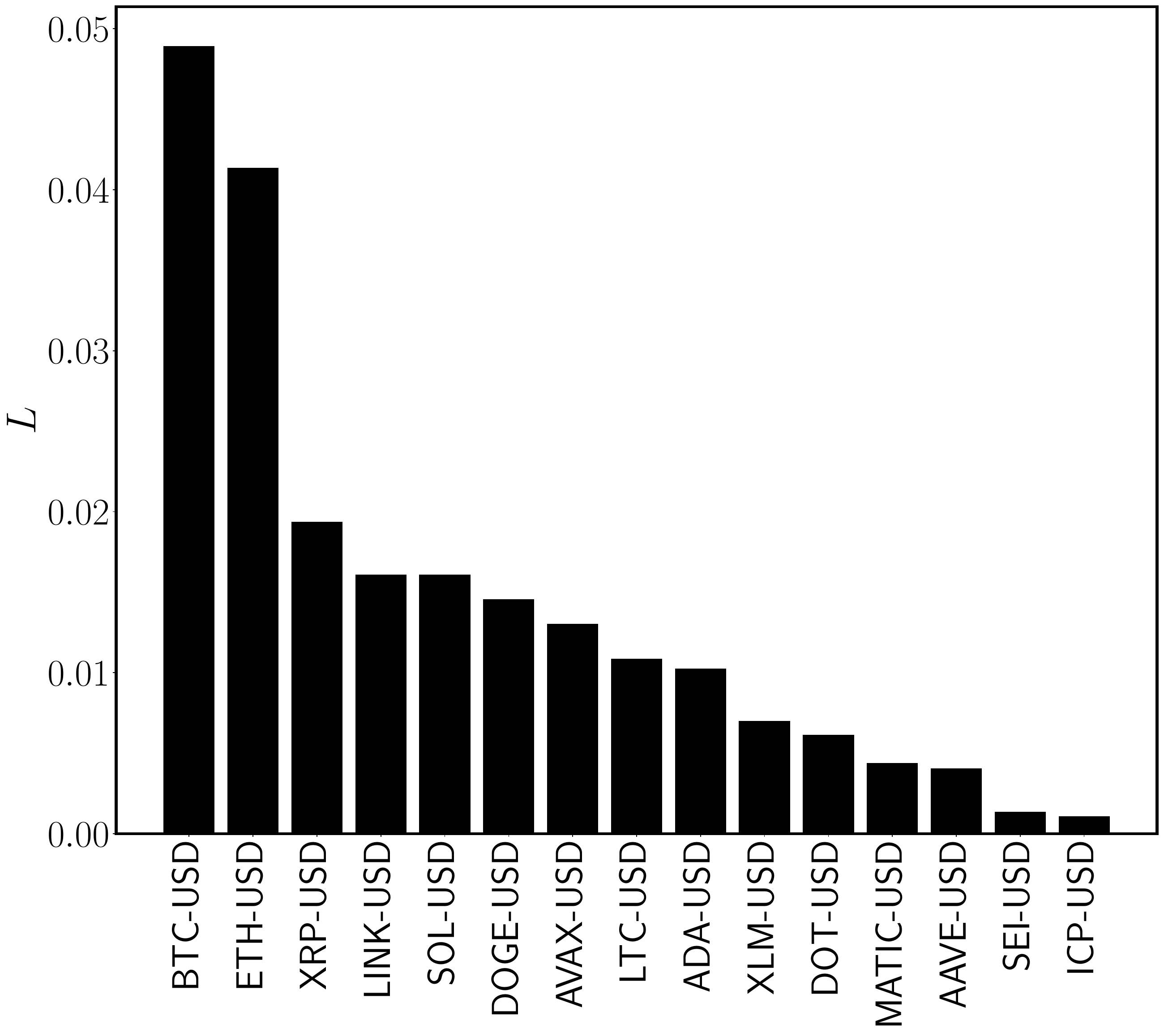}%
    }
    \subfloat[Receiver ratios \\\centering$(R^{i})_{1\leq i\leq D}$]{%
        \includegraphics[width=0.25\linewidth]{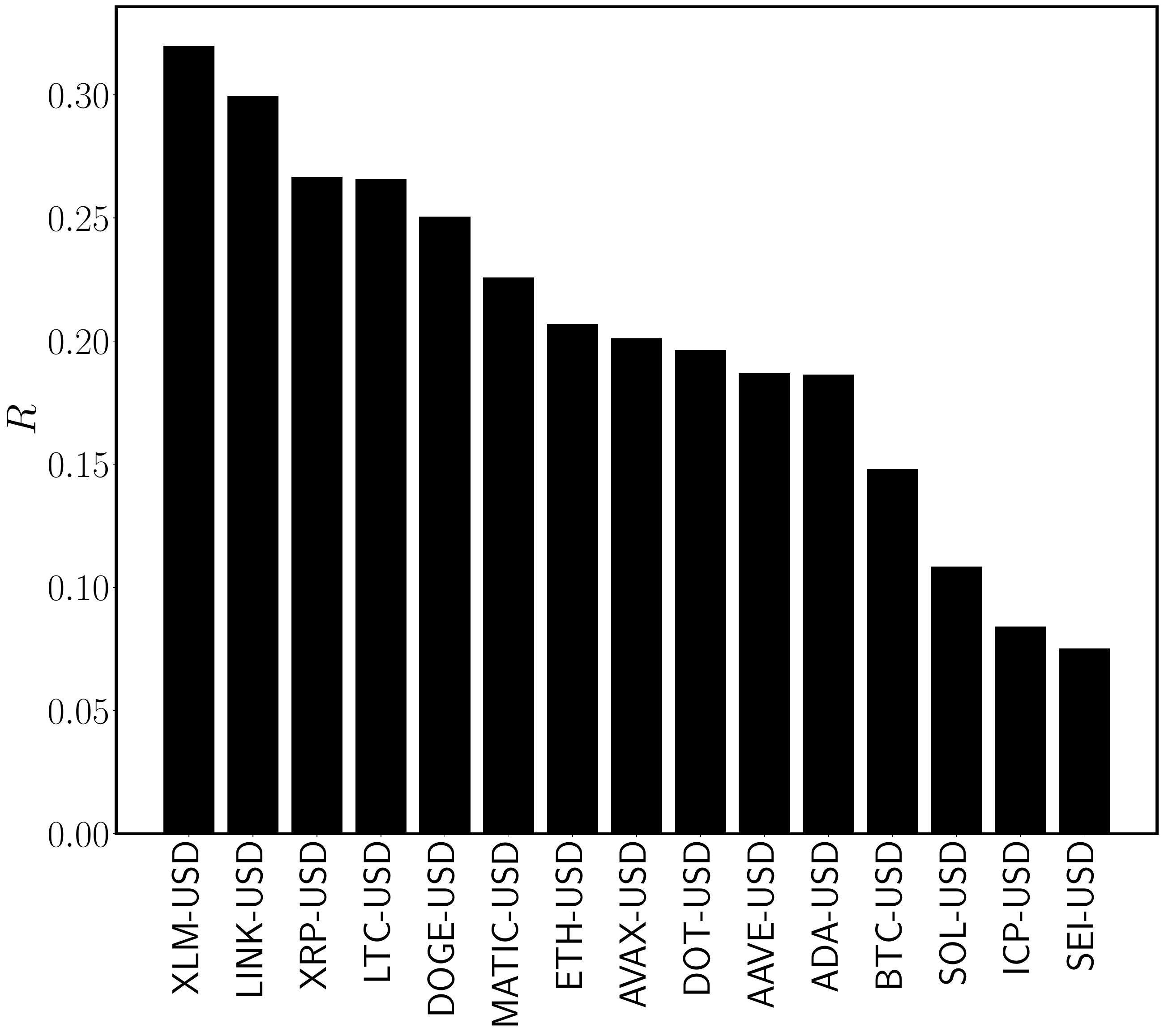}%
    }
    \subfloat[Participation rates $(\nu^{j})_{1\leq j\leq D}$]{%
        \includegraphics[width=0.25\linewidth]{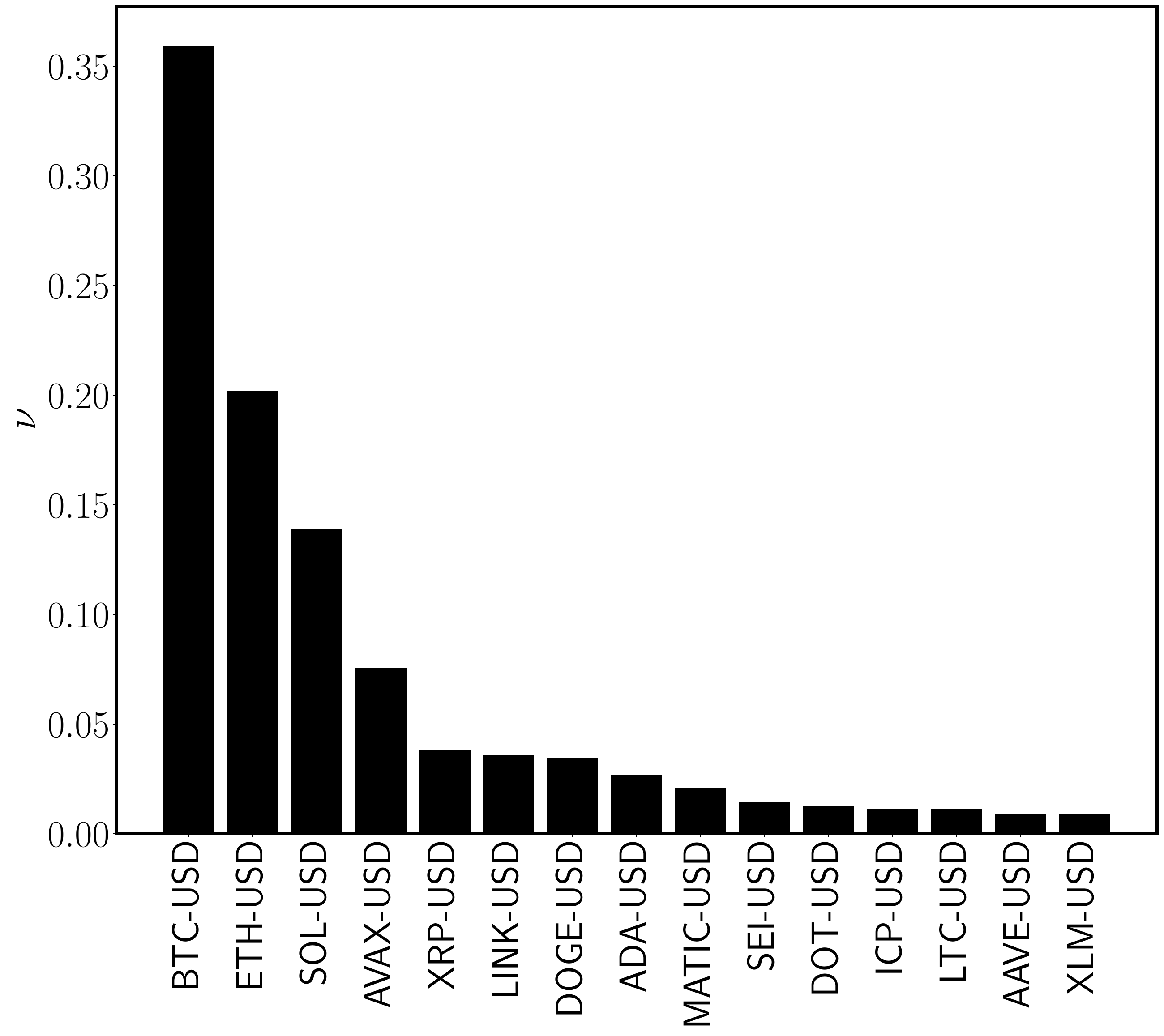}%
    }
    \caption{\textit{Cryptocurrency spillover} --- Spillover ratios and ranking of the leader, receiver measures and participation rates amongst the 15 cryptocurrency pairs. Estimation performed on the time windows spanning from 07:00 to 12:00 UTC.}
    \label{fig:coinbase_spillover_leadlag_measures_seasonality}
\end{figure}

\begin{figure}
    \centering
    \subfloat[Median spillover ratio matrix \centering$(S^{ij})_{1\leq i,j\leq D}$]{%
        \includegraphics[width=0.25\linewidth]{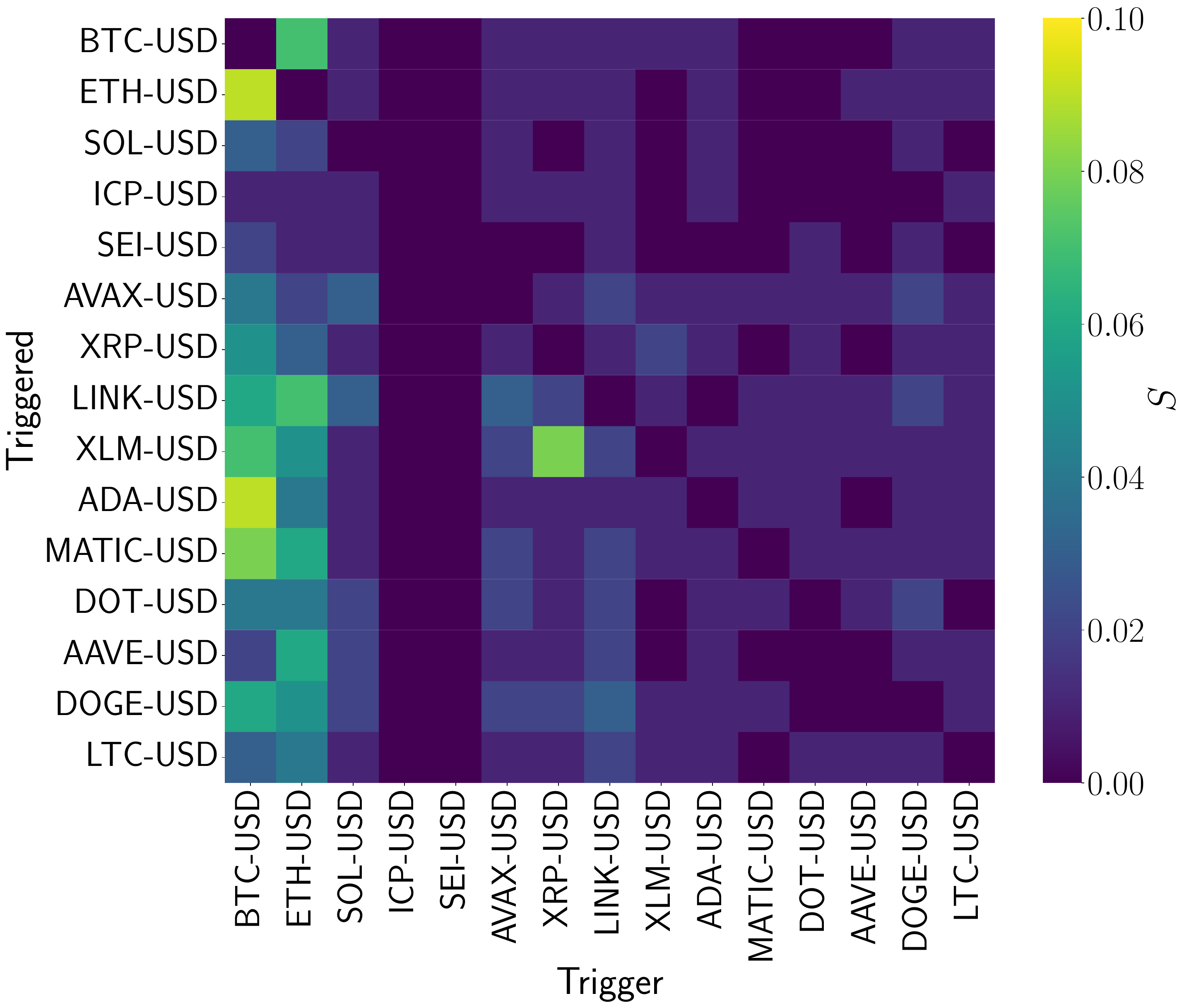}%
    }
    \subfloat[Box plot of leader ratios \centering$(L^{j})_{1\leq j\leq D}$]{%
        \includegraphics[width=0.25\linewidth]{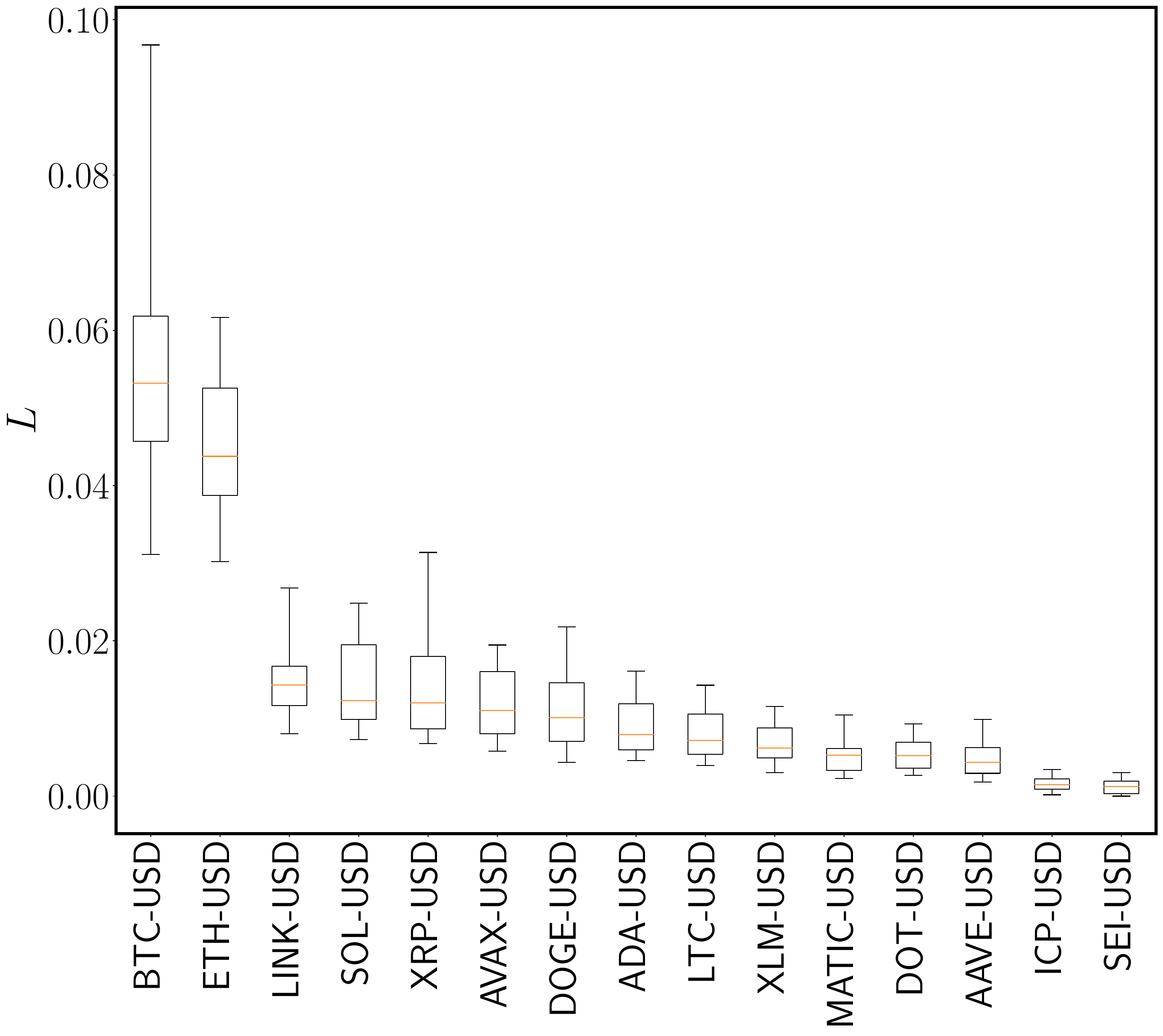}%
    }
    \subfloat[Box plot of receiver ratios \centering$(R^{i})_{1\leq i\leq D}$]{%
        \includegraphics[width=0.25\linewidth]{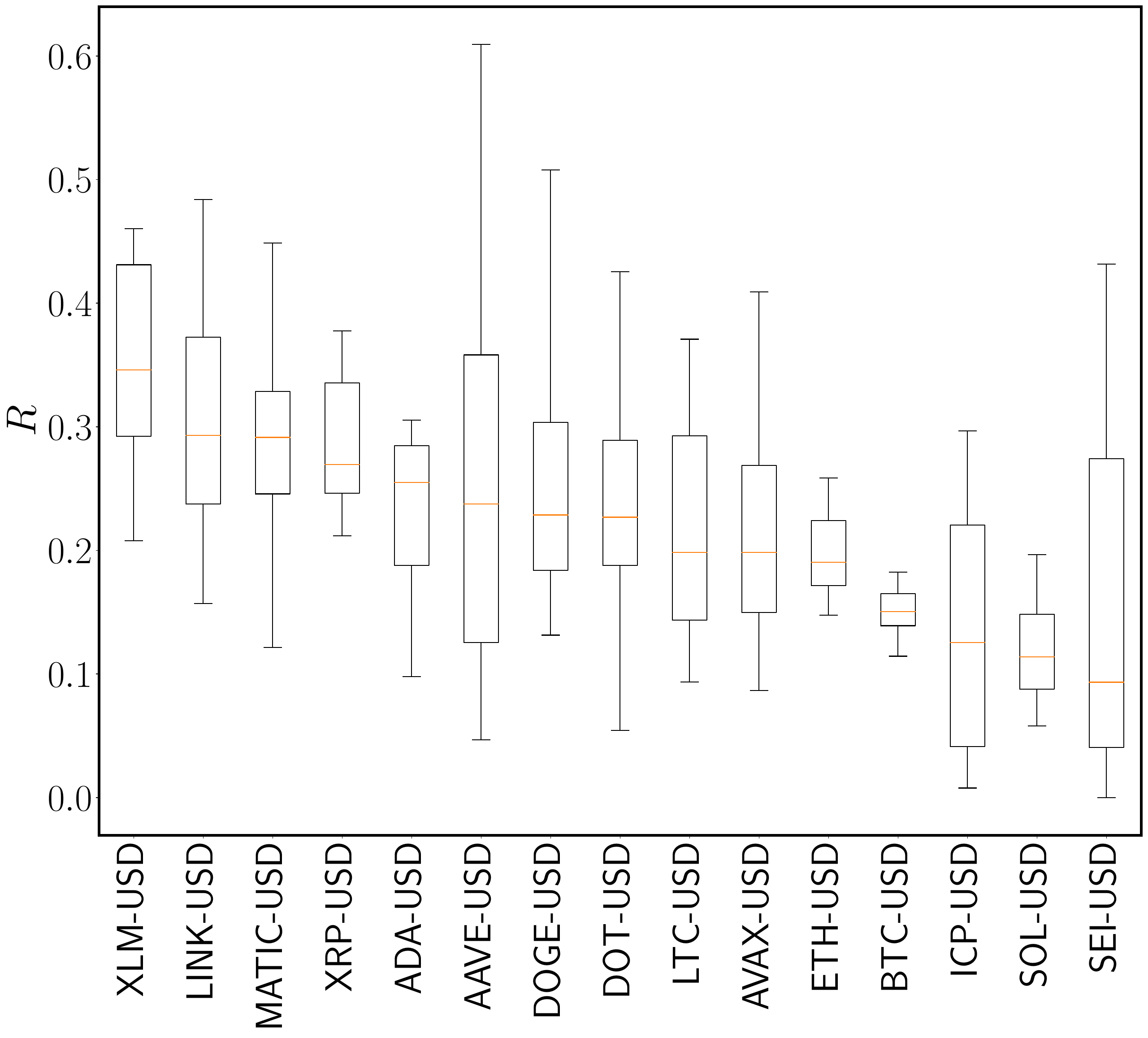}%
    }
    \subfloat[Box plot of participation rates \centering$(\nu^{j})_{1\leq j\leq D}$]{%
        \includegraphics[width=0.25\linewidth]{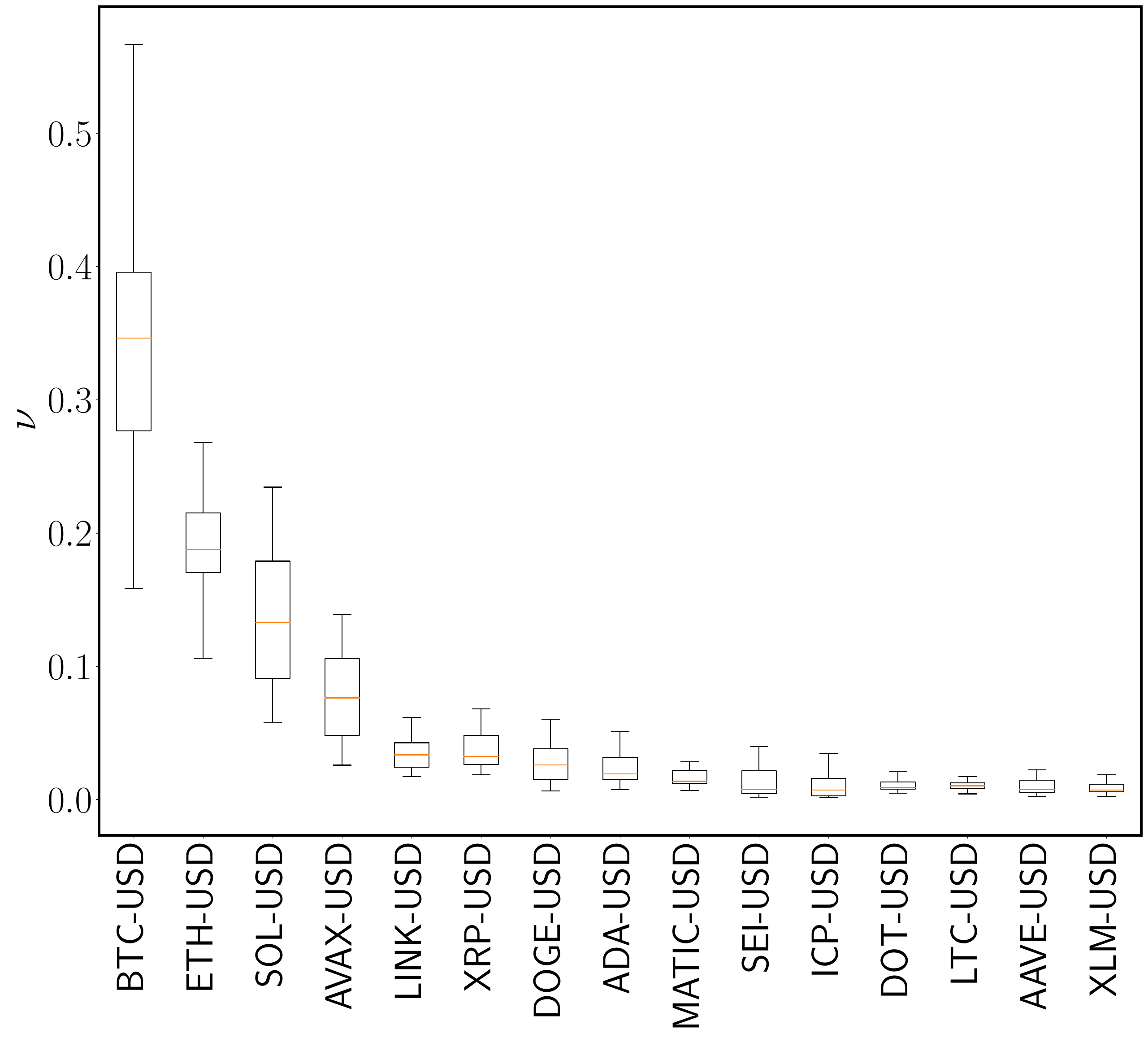}%
    }
    \caption{\textit{Cryptocurrency spillover} --- Median spillover ratios and ranking of the median leader, receiver measures and participation rates amongst the 15 cryptocurrency pairs. Estimation is performed daily on the time window spanning from 07:00 to 12:00 UTC, medians and box plots are computed using the 30 fits.}
    \label{fig:coinbase_spillover_leadlag_measures_seasonality_boxplot}
\end{figure}

\paragraph{Robustness with respect to intraday seasonality.}
\label{subsec:robustness_seasonality}
We provide here complementary figures concerning the robustness tests detailed in Section \ref{subsec:datastationarity}. Intraday seasonality is plotted on Figure \ref{fig:coinbase_intraday_intensity}.
Results of the first alternative estimation (single model on the 7am-12pm period) are given in Figure \ref{fig:coinbase_spillover_branching_ratio_seasonality} and \ref{fig:coinbase_spillover_leadlag_measures_seasonality}. 
 Results of the second alternative method (daily fits on the 7am-12pm period) are given in Figure \ref{fig:coinbase_spillover_leadlag_measures_seasonality_boxplot}.

\end{document}